\documentclass{elsart}
\usepackage{graphics}
\usepackage{amsmath}
\usepackage{amssymb}
\newcommand{\feyns}[1]{\slash \!\!\! #1}
\begin{document}
\begin{frontmatter}
\title{Effective Lagrangian Approach 
to pion photoproduction from the nucleon}
\author[IEM,US]{C. Fern\'andez-Ram\'{\i}rez\corauthref{cor}},
\corauth[cor]{Corresponding author.}
\ead{cesar@nuc2.fis.ucm.es}
\author[IEM,UCM]{E. Moya de Guerra},
\author[UCM]{J.M. Ud\'{\i}as}
\address[IEM]{Instituto de Estructura de la Materia, CSIC. \\ 
Serrano 123, E-28006, Madrid. Spain.}
\address[US]{Departamento de F\'{\i}sica At\'omica, Molecular 
y Nuclear. \\ Universidad de Sevilla. 
Apdo. 1065, E-41080, Sevilla. Spain}
\address[UCM]{Departamento de F\'{\i}sica At\'omica, Molecular 
y Nuclear. \\ Facultad de Ciencias F\'{\i}sicas. 
Universidad Complutense de Madrid. \\ Avda. Complutense s/n, 
E-28040, Madrid. Spain.}
\begin{abstract}
We present a pion photoproduction model on the free nucleon
based on an Effective Lagrangian Approach (ELA) which includes 
the nucleon resonances ($\Delta(1232)$, N(1440), N(1520), 
N(1535), $\Delta (1620)$, N(1650), and $\Delta (1700)$), 
in addition to Born and vector meson exchange terms.
The model incorporates a new theoretical treatment of 
spin-3/2 resonances, first introduced by Pascalutsa, 
avoiding pathologies present in previous models.
Other main features of the model are chiral symmetry,
gauge invariance, and crossing symmetry.
We use the model combined with modern optimization techniques
to assess the parameters of the nucleon resonances
on the basis of world data on electromagnetic multipoles.
We present results for electromagnetic multipoles,
differential cross sections, asymmetries, 
and total cross sections for all one pion photoproduction
processes on free nucleons. 
We find overall agreement with data from 
threshold up to 1 GeV in laboratory frame.
\end{abstract}
\end{frontmatter}

\section{Introduction} \label{sec:introduction}
In spite of the fact that
Quantum Chromodynamics (QCD) is regarded as the theory
of the strong interaction, in the energy regime of the mass 
of the nucleon and its resonances a perturbative approach
is not suitable.
Thus, we have to rely on an effective approach if
we are interested on the properties of nucleon resonances
and processes where they are involved
-- mainly meson production
which is the dominant decay channel.
This paper is devoted to 
pion photoproduction from the nucleon,
a classical topic within nuclear and particle physics, 
which has been proved as  
one of the best mechanisms to study the nucleon and its 
resonances as well as to study the role of the pion 
and resonances in nuclei \cite{Krusche}.

The experimental database 
\cite{Arndt90-2,SAID,SAIDdata} 
has been enormously increased
thanks to the experiments carried 
out at LEGS (Brookhaven) 
\cite{Shafi,Blanpied} and MAMI (Mainz) \cite{Mainz}
where photons are produced through laser backscattering 
and bremsstrahlung respectively. 
Because of this experimental 
effort, our knowledge of the $\Delta(1232)$ resonance 
region has been
largely increased, though several discrepancies 
between Mainz 
and Brookhaven analyses still remain \cite{Blanpied}. 
Polarization observables, 
differential cross sections and electromagnetic multipoles 
have been measured with 
a precision not possible a few years ago 
and a full description 
of the amplitudes in the $\Delta(1232)$ 
region is now available.
The database is expected to grow significantly 
once data from current experiments 
have been analyzed and when 
data from new laser backscattering facilities 
as GRAAL (Grenoble) and LEPS (Harima) 
become available. The last two facilities have started to 
run recently and operate at higher energies than LEGS. 
This situation opens a lot of 
possibilities for research on nucleon 
resonances.

In the last decades, pion photoproduction 
has been studied through many models and using 
various approaches to the description 
of the nucleon resonances.
Among them there are, 
Breit-Wigner models \cite{Walker,Drechsel}, 
K-matrix \cite{Olsson,Davidson91},
Effective Lagrangian Approach (ELA) 
\cite{EMoya,Feuster97,Vanderhaeghen,Scholten},
dynamical models \cite{Nozawa,Sato,Fuda,Pasc04},
Breit-Wigner plus a Regge-pole type background 
to take into account
the exchange of heavier mesons \cite{Aznauryan},
as well as quark models with pion 
treated as an elementary particle \cite{Zhao}.
Although in one way or another all models 
are phenomenological,
in this paper we adopt ELA method 
because we consider it
appealing in many respects and it is 
the most suitable approach in the energy range from 
threshold up to 1 GeV in laboratory frame, 
where the main low-lying resonances live.
This approach has 
proved to be a quite succesful tool 
to study pion photoproduction at 
low/threshold energy \cite{Bernard95,Bernard92,Thomas}
and provides
the most natural framework to extend the model to
pion electroproduction \cite{EMoya},
electromagnetic pion production 
in composite nuclei \cite{Garcilazo}
and halo nuclei \cite{Karataglidis},
two pion photoproduction \cite{Oset}, 
meson exchange currents \cite{MEC},
and exclusive X($\gamma$,$N \pi$)Y 
processes.

In the last years the Lagrangian 
description of spin-3/2 resonances has been
greatly improved and many
pathologies related 
to the pion-nucleon-resonance
and $\gamma$-nucleon-resonance  interactions
have been overcome \cite{Pasc98}.
This fact, combined with the substantial
enlargement of the pion photoproduction database,
demands to revisit the topic and 
to make the most of these advances in order to improve 
our knowledge on nucleon resonances and 
$\gamma$-nucleon-resonance 
vertices as well as on the 
pion photoproduction process itself. 

We focus on the analysis of 
pion photoproduction process on free nucleons 
with the aim of establishing 
a reliable set of coupling constants and 
achieving an accurate knowledge on 
nucleon resonances. The latter are 
needed for further studies
of resonances in nuclear medium
as well as to study the structure of the nucleon
through its excitations.
This requires to develop a pion
photoproduction model and to study the parameters of 
the nucleon resonances within the model 
for further implementation in the calculations 
previously mentioned. In this regard,
we consider this paper as a 
first step towards a deeper understanding of 
the role of the pion and the resonances 
in more complicated processes.
Our model is an improvement of the one in Ref. \cite{EMoya} 
where we have changed the spin-3/2 Lagrangians and 
explored other variations
which allow us to achieve crossing symmetry and
a better description of the resonance widths. 
The elements included in this model are nucleons, 
pions, photons, $\rho$ and $\omega$ mesons, as well as 
all  four stars status spin 1/2 and 3/2
nucleon resonances up to 1.7 GeV in 
Particle Data Group (PDG, in what follows) 
\cite{PDG2004}.
Spin-5/2 resonances are not expected to play an 
important role in the data analysis carried out 
in this paper and are left to future exploration.

The paper is organised as follows: 
in section \ref{sec:kinematics} 
we provide the basic features such as conventions 
and normalizations for cross 
sections and amplitudes which will be used throughout 
the article. In section \ref{sec:model} we describe 
the full model and its features in detail, 
stressing crossing symmetry and
the spin-3/2 treatment which 
avoids well-known pathologies of previous models. 
In section \ref{sec:results} we show results 
for multipoles, differential cross 
sections and remaining physical observables.
We also provide the reader with all 
the parameters
of the model explaining how they have been determined. 
In section \ref{sec:conclusions} we summarize 
the main conclusions and results.

\section{Kinematics, cross section and amplitude decomposition} 
\label{sec:kinematics}
Notation for kinematics is set to 
$k = (E_\pi,\vec{k} )$ for the outgoing pion, 
$q = (E_\gamma,\vec{q} )$ for the incoming photon, 
$p = ( E,\vec{p})$ for the incoming nucleon, and
$p' = (E',\vec{p'})$ for the outgoing nucleon. 
Mandelstam variables are defined as usual \cite{Halzen}

\begin{eqnarray}
s&=&\left(p+q\right)^2=\left(p'+k\right)^2 \: , \\
u&=&\left(p'-q\right)^2=\left(p-k\right)^2 \: , \\
t&=&\left(k-q\right)^2=\left(p-p'\right)^2 \: .
\end{eqnarray}

The photon
polarization vectors in spherical basis are
\begin{equation}
A^\mu_{\lambda_\gamma=\pm1}= \mp \frac{1}{\sqrt{2}} 
\left( 0,1,\pm i ,0 \right) \: .
\end{equation}

Following conventions and normalization of Ref. \cite{Halzen}, 
the differential cross section can be written in the center of 
mass (c.m.) reference system as

\begin{equation}
\sigma \left( \theta \right) \equiv \frac{d\sigma}{d\Omega_\pi^*}
=\frac{1}{64\pi^2}\frac{1}{s^*}\frac{k^*}{E_\gamma^*}\overline{
|{\mathcal M}|^2} \: . \label{eq:diff}
\end{equation}

Whenever a kinematical 
quantity appears starred it is defined in the c.m. 
reference frame. 
In particular the c.m. absolute values of the photon 
and the pion momenta are denoted by $q^*$ and $k^*$, 
which stand 
for $|\vec{q^*}|$ and $|\vec{k^*}|$ respectively. 
The transition probability is

\begin{equation}
\overline{|{\mathcal M}|^2}=\frac{1}{4}\sum_{\lambda_1 
\lambda_2 \lambda_\gamma}|{\mathcal A}_{\lambda_1 \lambda_2 
\lambda_\gamma}|^2 \: , \label{eq:amplitude}
\end{equation}
where ${\mathcal A}_{\lambda_1 \lambda_2 \lambda_\gamma}$ is the 
invariant amplitude, with photon polarization $\lambda_\gamma$, 
initial nucleon helicity $\lambda_1$, and final nucleon helicity 
$\lambda_2$. We use the following isospin decomposition 

\begin{equation}
{\mathcal A}=\chi_2^\dagger \left(A^0\tau_j +A^-\frac{1}{2}
[\tau_j,\tau_3]+A^+\delta_{j3} \right)\pi_j \chi_1 \: ,
\end{equation}
where for simplicity we have dropped helicity subindices
(another two isospin decompositions are used 
and are introduced in appendices).

The isospin decomposition can be related to the physical 
amplitudes
\begin{eqnarray}
{\mathcal A} \left(\gamma p \rightarrow p \pi^0 \right)
&=&A^++A^0 \: , \\
{\mathcal A} \left(\gamma n \rightarrow n \pi^0 \right)
&=&A^+-A^0 \: , \\
{\mathcal A} \left(\gamma n \rightarrow p \pi^- \right)
&=&\sqrt{2} \left(A^0-A^-\right) \: , \\
{\mathcal A} \left(\gamma p \rightarrow n \pi^+ \right)
&=&\sqrt{2}\left(A^0+A^-\right) \: .
\end{eqnarray}

For completeness we specify the conventions 
adopted throughout this article. Metric tensor: 
$g^{\mu \nu}\equiv diag(1,-1,-1,-1)$; 
Levi-Civit\'a tensor: 
$\epsilon_{0123}=1$, $\epsilon_{123}=1$; 
Pauli matrices: 
$\left[ \tau_j , \tau_k \right]= 2 \epsilon_{jkl}\tau_l$; 
Dirac-Pauli matrices:  
$\{ \gamma^\mu , \gamma^\nu \} = 2 g^{\mu \nu}$, 
$\gamma^{\mu \nu} = \frac{1}{2}
\left[ \gamma^\mu ,\gamma^\nu \right]$, 
$\gamma^{\mu \nu \alpha} = 
\frac{1}{2}\left( \gamma^\mu 
\gamma^\nu \gamma^\alpha - \gamma^\alpha 
\gamma^\nu \gamma^\mu \right)$, 
$\gamma_5=i\gamma_0 \gamma_1 \gamma_2 \gamma_3$; 
Electromagnetic field:  
$F^{\mu \nu}=\partial^\mu \hat{A}^\nu-
\partial^\nu \hat{A}^\mu$, 
$\tilde{F}^{\mu \nu}=\frac{1}{2}
\epsilon^{\mu \nu \alpha \beta} F_{\alpha \beta}$, 
where $\hat{A}^\mu$ is the photon field.

\section{The model} \label{sec:model}
In this section we present a complete description of the model and 
its features. 
Using as starting point Weinberg's theorem \cite{Weinberg79},
we construct a fully relativistic, chiral symmetric, 
gauge invariant, and 
crossing symmetric model based on 
suitable effective Lagrangians for 
particle couplings. From these Lagrangians we obtain the
invariant amplitudes and physical observables. 
This procedure has been adopted in many papers,
i.e. \cite{EMoya,Feuster97,Vanderhaeghen,Scholten},
and has been proved to be a succesful 
way to treat the pion photoproduction process. However, previous 
works had pathologies in the description of the spin-3/2 particles 
which are not present in our model. The basic idea 
is to build consistently the most general Lagrangians for 
vertices, taking into account all possible symmetries (crossing 
symmetry, gauge invariance, chiral symmetry), and to use Feynman 
rules to obtain invariant amplitudes which can be related to 
physical observables. The model can be split into three different 
types of contributions: Born terms (Fig. \ref{fig-diag1}), 
vector mesons exchange (Fig. \ref{fig-diag2}, diagram $(e)$), 
and spin-1/2 and spin-3/2
nucleon resonance excitations
(Fig. \ref{fig-diag2}, diagrams $(f)$ and $(g)$). 
There is no contribution 
from $\sigma$ meson exchange because of charge conjugation violation 
of the $\sigma \pi \gamma$ coupling \cite{Davidson91}. We consider that 
all the relevant degrees of freedom are taken into account except perhaps 
spin-5/2 resonances. 
Our choice of Lagrangians is explained and justified 
in the forthcoming subsections. 
All the invariant amplitudes can be found 
in appendix \ref{sec:invariantamplitudes}.

\subsection{Born terms} \label{sub:bornterms}

\begin{figure}[!t]
\begin{center}
\scalebox{1.0}{\includegraphics{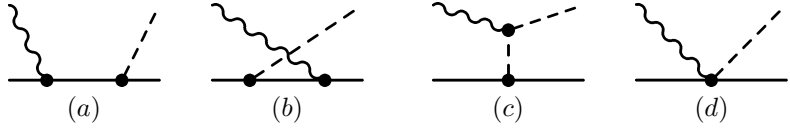}}
\end{center}
\caption{Feynman diagrams for Born terms: ($a$) direct or 
$s$-channel, ($b$) crossed or $u$-channel, ($c$) pion in flight 
or $t$-channel, and ($d$) Kroll-Rudermann (contact).} \label{fig-diag1}
\end{figure}

Born terms are the Feynman diagrams shown in Fig. \ref{fig-diag1} 
in which only pions, 
photons, and nucleons are involved. We start 
from the free Lagrangians for pions (Klein-Gordon) and nucleons 
(Dirac) and a phenomenological pion-nucleon interaction. 
This last interaction is chosen as a pseudovector (PV)  coupling to the 
pion because it is the lowest order in derivatives compatible with 
the low energy behavior of the pion and 
chiral symmetry \cite{Bernard95,Leutwyler}
\begin{equation}
{\mathcal L}_{\pi NN}=\frac{f_{\pi N}}{m_\pi} \bar{N} \gamma_\mu 
\gamma_5 \tau_j \left( \partial^\mu \pi_j \right) N \: , \label{eq:PV}
\end{equation}
where $m_\pi$ is the mass of the pion, $f_{\pi N}$ is the pseudovector 
coupling constant, and the sign is fixed by phenomenology. According 
to \cite{Arndt90-1} $f_{\pi N}$ is set to 
$f^2_{\pi N}/4 \pi = 0.0749 $. The use of the PV coupling for 
the pion in our effective Lagrangian grants that
low energy theorems of current 
algebra and partially conserved axial-vector current (PCAC) 
hypothesis are incorporated in the model.

The electromagnetic field is included in the usual way by minimal 
coupling to the photon field ($\partial^\alpha \to \partial^\alpha 
+ ie\hat{Q}\hat{A}^\mu$; where $\hat{Q}$ is the charge operator) 
and taking into account phenomenologically 
the anomalous magnetic moment of the nucleon

\begin{equation}
{\mathcal L}=-\frac{ie}{4M} F_2^V \bar{N} \frac{1}{2} 
\left( F_2^{S/V}+\tau_3 \right) \gamma_{\alpha \beta} N F^{\alpha \beta} \: .
\end{equation}

$F_2^{S/V}$ is defined as the ratio between isospin-scalar 
and isospin-vector form factors ($F_2^{S/V} \equiv F_2^S/F_2^V$).

The interacting Lagrangian for Born terms is
\begin{equation}
\begin{split}
{\mathcal L}_{Born}=&-ieF_\pi \hat{A}^\alpha \epsilon_{jk3}\pi_j 
\left( \partial_\alpha \pi_k  \right) 
-e\hat{A}^\alpha F_1^V \bar{N} \gamma_\alpha \frac{1}{2} 
\left( F_1^{S/V}+\tau_3 \right) N \\
&-ieF_1^V\frac{f_{\pi N}}{m_\pi} \hat{A}^\alpha \bar{N}\gamma_\alpha 
\gamma_5 \frac{1}{2}[\tau_j,\tau_3] \pi_j N \\
&-\frac{ie}{4M} F_2^V \bar{N} \frac{1}{2} \left( F_2^{S/V}+
\tau_3 \right) \gamma_{\alpha \beta} N F^{\alpha \beta} 
+\frac{f_{\pi N}}{m_\pi}\bar{N}\gamma_\alpha \gamma_5 \tau_j  N 
\left( \partial^\alpha \pi_j \right) \: ,
\end{split} \label{eq:PVlagrangian}
\end{equation}
where $e$ is the absolute value of the electron charge, $F_\pi$ 
is the pion form factor and $F_j^V = F_j^p - F_j^n$, 
$F_j^S = F_j^p + F_j^n$ 
are the isovector and isoscalar nucleon form 
factors which at the photon point ($q^2=0$) take the values
$F_1^S=F_1^V=1$, 
$F_2^S=\kappa^p+\kappa^n=-0.12$, $F_2^V=\kappa^p - \kappa^n=3.70$. 
We set $F_\pi = F_1^V$ in order to ensure gauge invariance. 
It is straightforward to check gauge invariance 
of the amplitudes in appendix \ref{sec:invariantamplitudes}
performing the replacement $A^\mu \to q^\mu$.

\subsection{Vector mesons} \label{sub:vectormesons}

\begin{figure}[!t]
\begin{center}
\scalebox{1.0}{\includegraphics{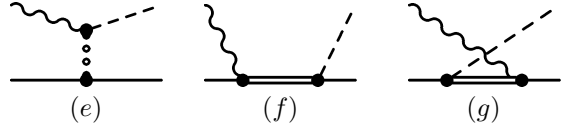}}
\end{center}
\caption{Feynman diagrams for vector meson exchange ($e$) and 
resonance excitations: ($f$) direct or $s$-channel and ($g$) 
crossed or $u$-channel.} \label{fig-diag2}
\end{figure}

The main contribution of mesons to pion photoproduction is 
given by $\rho$ (isospin-1 spin-1) and $\omega$ (isospin-0 spin-1) 
exchange. 
The phenomenological Lagrangians which describe 
vector mesons are \cite{Drechsel,EMoya}
\begin{eqnarray}
{\mathcal L}_\omega &=& -F_{\omega N N}\bar{N}\left[\gamma_\alpha 
-i\frac{K_\omega}{2M}\gamma_{\alpha \beta}\partial^\beta \right]
\omega^\alpha N +\frac{e G_{\omega \pi \gamma}}{m_\pi} 
\tilde{F}_{\mu \nu} \left( \partial^\mu \pi_j \right) \delta_{j3}
\omega^\nu \: ,\\
{\mathcal L}_\rho &=& -F_{\rho N N}\bar{N}\left[\gamma_\alpha 
- i\frac{K_\rho}{2M}\gamma_{\alpha \beta}\partial^\beta \right]
\tau_j \rho^\alpha_j N +\frac{e G_{\rho \pi \gamma}}{m_\pi} 
\tilde{F}_{\mu \nu} \left(\partial^\mu \pi_j \right)\rho^\nu_j \: .
\end{eqnarray}

Often the $\pi \gamma V$ coupling is written as 
${\mathcal L} =\frac{e G_{V \pi \gamma}}{2 m_\pi} 
\tilde{F}_{\mu \nu} V^{\mu \nu} \pi$ 
where $V^{\mu \nu} \equiv \partial^\mu V^\nu-\partial^\nu V^\mu$ 
and $V^\mu \equiv \rho^\mu,\omega^\mu$ \cite{Drechsel}. 
Both couplings yield the same amplitude.

\subsection{Spin-1/2 nucleon resonances} \label{sub:spin1/2}
In the model we deal with three different kinds of resonances 
with spin-$\frac{1}{2}$: S$_{11}$, S$_{31}$, and P$_{11}$; 
and we need Lagrangians and amplitudes to describe their behavior. 
The most simple isobar is 
isospin-$\frac{1}{2}$ spin-$\frac{1}{2}$ (S$_{11}$) 
which can be described by the following Lagrangian
\begin{equation}
{\mathcal L}_{\text{S}_{11}}= -\frac{h}{f_\pi}\bar{N}\gamma_\alpha 
\tau_j N^* \partial^\alpha \pi_j -\frac{ie}{4M}\bar{N}
\gamma_{\alpha \beta}\gamma_5\left(g_S+g_V\tau_3 \right) 
N^*F^{\alpha \beta} + HC \: ,
\end{equation}

where $HC$ stands for hermitian conjugate, $h$ is the 
strong coupling constant which can be related to the width of the resonance 
decay into a nucleon and a pion, 
and $f_{\pi}=92.3$ MeV is the leptonic 
decay constant of the pion. 
$g_V$ and $g_S$
stand for
the resonance
isovector and isoscalar 
form factors respectively. 
They are defined as $g_V=g_p-g_n$ and $g_S=g_p+g_n$, 
where subscripts $p$ and $n$ stand for the resonances
originating from the proton and the neutron, and
can be related to experimental helicity amplitudes
at the photon point 
as will be seen in the next sections.
The pion coupling has been chosen pseudovector 
in order to obtain the right low energy behavior and consistency 
with Born terms. The coupling to the photon used preserves 
gauge invariance.

The next isobar is isospin-$\frac{3}{2}$ spin-$\frac{1}{2}$ 
(S$_{31}$). Because of isospin we need to define isospinors as 
in Ref. \cite{Peccei}
\begin{eqnarray}
N_1^* &=& \sqrt{\frac{1}{2}}\begin{pmatrix} 
N^{*++}-\sqrt{\frac{1}{3}}N^{*0} \\
\sqrt{\frac{1}{3}}N^{*+}-N^{*-}
\end{pmatrix} \: , \\ 
N_2^* &=& i\sqrt{\frac{1}{2}}\begin{pmatrix}
N^{*++}+\sqrt{\frac{1}{3}} \\
\sqrt{\frac{1}{3}}N^{*+}+N^{*-}
\end{pmatrix} \: , \\ 
N_3^* &=& -\sqrt{\frac{2}{3}}\begin{pmatrix}
N^{*+} \\
N^{*0}
\end{pmatrix} \: .
\end{eqnarray}

In this basis, and under the same conditions as those for previous isobar, 
the S$_{31}$ Lagrangian is
\begin{equation}
{\mathcal L}_{\text{S}_{31}}=-\frac{h}{f_\pi}\bar{N}\gamma_\alpha 
N^*_j\partial^\alpha \pi_j -\frac{ieg}{2M} \bar{N}\gamma_{\alpha \beta}
\gamma_5N^*_3 F^{\alpha \beta} + HC \: .
\end{equation}
Just one electromagnetic coupling constant is needed here because
only the isovector part of the photon couples to the nucleon to produce
an isospin-$\frac{3}{2}$ field.

The P$_{11}$ Lagrangian is closely related to S$_{11}$ being parity the 
main change. This change is due to the angular momentum of the 
resonance, which implies a change in the parity of the coupling
\begin{equation}
{\mathcal L}_{\text{P}_{11}}
=-\frac{h}{f_\pi}\bar{N}\gamma_\alpha 
\gamma_5 \tau_j N^*\partial^\alpha \pi_j +\frac{ie}{4M} 
\bar{N}\gamma_{\alpha \beta}\left(g_S+g_V\tau_3 \right)N^* 
F^{\alpha \beta} + HC \: .
\end{equation}

\subsection{Spin-3/2 nucleon resonances} \label{sub:spin3/2}
The choice of spin-3/2 nucleon-resonance couplings is one of the
main improvements of the present model compared to former ones.
The choice that we use here is motivated by previous studies that
identified 
pathologies in former spin-3/2 couplings.
In what follows we provide a detailed comparison of both 
traditional (off-shell extension) and 
gauge invariant (GI) couplings, which exhibits 
the virtues of the choice adopted here. With regards 
to the traditional coupling, we restrict 
the discussion to the P$_{33}$ ($\Delta$)
resonance and its coupling to the pion and the nucleon, 
although it affects similarly to other spin-3/2 resonances.

\subsubsection{Traditional $\Delta$-nucleon-pion coupling} \label{sec:trad}
The basis of the traditional point of view is the seminal paper 
by Nath, Etemadi, and Kimel \cite{Nath}, based on the articles 
by Peccei \cite{Peccei} in the late sixties which dealt with this 
coupling. Peccei worked out a chiral Lagrangian with a pseudovector 
coupling to the pion, to ensure the low energy behavior, based upon 
the invariance of the $\Delta$ free field under the point 
transformation $\Delta_\mu \to \Delta_\mu- \frac{1}{4}\gamma_\mu 
\gamma_\beta \Delta^\beta$ and the ansatz 
$\gamma^\mu O_{\mu \nu} = 0$. Given the most general Lagrangian  
${\mathcal L}=h \bar{\Delta}^\mu_j O_{\mu \nu}N\partial^\nu \pi_j$ 
we obtain the well known Peccei Lagrangian \cite{Peccei}
\begin{equation}
{\mathcal L}_{Peccei}=ih \bar{\Delta}^\alpha_j 
\left( 4 g_{\alpha \beta}-\gamma_\alpha \gamma_\beta \right)N 
\partial^\beta \pi_j + HC \: .
\end{equation}

Restrictions such as Peccei's ansatz are needed in order to reduce 
the number of degrees of freedom (DOF) of the spin-3/2 field. 
When a massive spin-3/2 particle is 
described within the Bargmann and Wigner equations \cite{Greiner}, 
a problem of extra DOF arises because a vector-spinor has sixteen 
componentes whilst only four are needed. These constraints naturally 
emerge in the free theory thanks to the Euler-Lagrange \cite{Nath} 
or the Hamiltonian formalism \cite{Pasc98}, but for interacting 
particles the picture is not so straightforward and additional 
restrictions have to be imposed.

Nath et al. \cite{Nath}
proved that Peccei's ansatz was too restrictive, 
developing a generalization which -- despite of its many pathologies 
\cite{Pasc98,Nath,Benmerrouche,Pasc01} -- has become the traditional 
and most popular approach to interacting spin-3/2 particles for the 
last thirty years.

The starting point of Nath et al. is the massive spin-3/2 
free theory, which can be found in Refs. 
\cite{Peccei,Nath,Benmerrouche}. 
The following Lagrangian is defined
\begin{equation}
\begin{split}
{\mathcal L}_\Delta=&\bar{\Delta}^\alpha\Big{[} \left( i\partial_\mu 
\gamma^\mu-M^* \right)g_{\alpha \beta}+i\omega \left(\gamma_\alpha 
\partial_\beta + \gamma_\beta \partial_\alpha \right) \\
&+ \frac{i}{2}\left( 3\omega^2+2\omega+1 \right)\gamma_\alpha 
\partial^\mu \gamma_\mu \gamma_\beta + M^* \left( 3\omega^2+3\omega
+1 \right)\gamma_\alpha \gamma_\beta \Big{]}\Delta^\beta \: ,
\end{split} \label{freefield}
\end{equation}
where $\omega \neq -\frac{1}{2}$ and the 
Lagrangian is invariant under the point 
transformation
\begin{eqnarray}
\Delta^\mu & \to & \Delta^\mu+a\gamma^\mu \gamma^\nu \Delta_\nu \: , 
\label{pointtransformation}\\
\omega   & \to & \frac{\omega-2a}{1+4a} \: ; \label{pointtransformation2}
\end{eqnarray}
with $a \neq -\frac{1}{4}$. Subsidiary 
constraints $\gamma_\mu \Delta^\mu=0$ and $\partial_\mu \Delta^\mu=0$ 
appear in order to reduce the number of DOF to four, as expected 
for a spin-3/2 particle. A detailed description of the DOF 
counting technique is given in reference \cite{Pasc99}.
The parameter 
$\omega$ does not affect physical quantities, so that one is free to set 
it to the most convenient value, usually $\omega=-1$ which recovers 
the Rarita-Schwinger theory \cite{Rarita}.

The point transformation of Eq. (\ref{pointtransformation}) 
does not affect the spin-3/2 content of the free field because of 
the constraint $\gamma_\mu \Delta^\mu=0$, but for interacting 
$\Delta$ particles this constraint does not apply, and the excess 
of DOF shows up as a contribution to the spin-1/2 sector.

The most general interacting Lagrangian containing only  
first-order derivatives of the pion field and consistent with 
(\ref{freefield}), (\ref{pointtransformation}),
and (\ref{pointtransformation2}) 
is given by
\begin{equation}
{\mathcal L}_{int}= \varkappa \bar{\Delta}^\alpha \left( g_{\alpha \beta} 
+ a\gamma_\alpha \gamma_\beta \right)N \partial^\beta \pi + HC \: ,
\label{offshell}
\end{equation}
where $\varkappa$ is a coupling constant and $a$ is called the
off-shell parameter, which can be set to different values. 
This is named 
off-shell extension framework.
If $a=-\frac{1}{4}$ we recover Peccei theory.
This family of Lagrangians 
has been widely used in pion-nucleon scattering 
\cite{Scholten,Feuster98}, pion photoproduction 
\cite{EMoya,Feuster97,Vanderhaeghen,Scholten,Nozawa,Sato,Peccei} 
and compton scattering \cite{Scholten,Pasc95} 
in the $\Delta$-region, as well 
as for the description of meson exchange currents \cite{MEC}. The 
off-shell parameter can be set to a fixed value, $a=-1$ \cite{Nath}, 
$a=-\frac{1}{4}$ \cite{EMoya,Peccei} or just let it run 
freely \cite{Feuster97,Pasc95} in order to get the best possible fit.

However, it is not possible to remove the spin-1/2 
sector from the amplitude for any value of $a$ \cite{Benmerrouche}. 
The physical meaning of the 
off-shell parameter is unclear and could be considered just as a 
free parameter with a fuzzy physical meaning set only for 
fitting 
improvement. The disadvantage is that there is
a heavy dependence of the 
coupling constants on the off-shell parameter, as was proved by 
Feuster and Mosel \cite{Feuster97}. Other pathologies related to 
Eq. (\ref{offshell}) coupling are: quantization 
anomalies (except for $a=-1$), 
so that the naive Feynman rules we \textit{read} 
from the Lagrangian are no longer valid 
\cite{Pasc98,Nath}; Johnson-Sudarshan (JS) problem  
(nonpositive definite commutators) \cite{Johnson,Hagen} and 
Velo-Zwanziger (VZ) problem (acausal propagations) \cite{Hagen,Velo}.

A consistent theory for interacting spin-3/2 particles is 
expected to be free of such problems. This theory has been 
developed in recent years and will be detailed in the next paragraphs.

\subsubsection{Gauge invariant couplings} 
\label{gaugeinvariantcouplings}
A different approach to massless fields of arbitrary spin 
$\lambda$ was developed in the seventies. It was proved that 
the massless theory obtained from the massive one has a 
simple structure for both integer \cite{Fronsdal} and 
half-integer \cite{Fang} spin fields, even if the massive 
theory is rather complicated. The free massless Lagrangians 
for half-integer spin fields can be obtained just from 
first principles requiring the action to be invariant under 
the gauge transformation $\psi \to \psi + \delta \psi$, 
where $\delta \psi = \partial \eta$ \cite{Curtright,Weinberg95}, 
$\psi$ is a tensor-spinor with rank $\ell$ which stands 
for the particle and $\eta$ a complex tensor-spinor field with 
rank $\ell-1$. For a spin-3/2 field 
$\delta \psi_\mu = \partial_\mu \eta$, with $\psi_\mu$ a 
vector spinor and $\eta$ a spinor field. This gauge condition 
reduces the number of DOF of the spin-$\lambda$ field to 2 -- 
helicity states $-\lambda$ and $+\lambda$ -- as it is required for 
a massless particle. In this framework, it is quite simple to 
build consistent interactions for half-integer spin fields as 
suggested by Weinberg and Witten \cite{Weinberg80} just enforcing 
them to fulfill this gauge invariance condition. For example, 
the spin-3/2 $\psi_\mu$ field should appear in the interaction 
as $\partial_\mu \psi_\nu - \partial_\nu \psi_\mu$, the 
spin-5/2 $\psi_{\mu \nu}$ as $\partial_\mu \partial_\nu 
\psi_{\rho \sigma} - \partial_\mu \partial_\sigma 
\psi_{\rho \sigma} - \partial_\rho \partial_\nu 
\psi_{\mu \sigma}+\partial_\rho \partial_\sigma \psi_{\mu \nu}$, 
and, more generally, an arbitrary spin-$\lambda$ tensor-spinor 
field as the antisymetrization of $\partial_{\alpha_1} 
\partial_{\alpha_2} \cdots \partial_{\alpha_{\left( \lambda-1/2 
\right)}} \psi_{\beta_1 \beta_2 \cdots \beta_{
\left( \lambda-1/2 \right)}}$. Thus the vertices 
${\mathcal O}^{\mu ...}$ of the Feynman diagrams for 
massless spin-3/2 particles will fulfill the condition 
$p_\mu {\mathcal O}^{\mu ...}=0$ where $p$ is the four-momentum 
of the spin-3/2 particle, $\mu$ the vertex index which couples 
to the spin-3/2 field, and the dots stand for other possible 
indices. This is what is called GI coupling scheme.

We apply this procedure to the $\Delta$ case. We start from 
Lagrangian (\ref{freefield}) for a free massless spin-3/2 particle.
For $\omega=-1$ it can be written as
\begin{equation}
{\mathcal L}_{3/2,massless} = \bar{\psi}_\mu \gamma^{\mu \nu 
\alpha} \partial_\alpha \psi_\nu \: .
\end{equation}
The inclusion of the mass term 
\begin{equation}
{\mathcal L}_{3/2,massive} ={\mathcal L}_{3/2,massless}-M^* 
\bar{\psi}_\mu \gamma^{\mu \nu} \psi_\nu \: , \label{deltamassive}
\end{equation}
breaks gauge symmetry, raising the number of DOF from 2 to 4 as it should be.

Let us now consider the interaction. For an interacting 
massless spin-3/2 particle we write the Lagrangian
\begin{equation}
{\mathcal L} = {\mathcal L}_{3/2,massles}+{\mathcal L}_{int} \: .
\end{equation}
The interaction has been built within the GI coupling scheme 
and can be written as \cite{Pasc98}
\begin{equation}
{\mathcal L}_{int} = \psi^\dagger_\mu J^\mu + HC \: ,
\end{equation}
where $J^\mu$ has no dependence on $\psi^\mu$ and  gauge 
invariance imposes $\partial_\mu J^\mu = 0$. The inclusion of 
the mass term -- if it is properly done as in 
(\ref{deltamassive}) -- breaks  gauge symmetry increasing the 
number of DOF of the spin-3/2 field from 2 to 4 and does not 
affect ${\mathcal L}_{int}$ \cite{Pasc98,Weinberg95}. Hence 
the number of DOF in the interacting massive field is the right 
one and no unphysical components are present. 
Focusing on our photoproduction model, we are 
interested in two couplings: the $\Delta$ to the pion and the 
nucleon, and the $\Delta$ to the photon and the nucleon. The 
simplest consistent $\Delta N \pi$-coupling is \cite{Pasc98}
\begin{equation}
{\mathcal L}_{int}=-\frac{h}{f_\pi M^*} \bar{N} \epsilon_{\mu 
\nu \lambda \beta} \gamma^\beta \gamma^5 \left( \partial^\mu 
N^{*\nu}_j \right) \left( \partial^\lambda \pi_j  \right) + 
HC \: .\label{GIcoupling}
\end{equation}
We have to clarify that the vector coupling to the pion 
is a consequence of GI prescription. 
Whithin this prescription, the scalar coupling 
to the pion gives no contribution to the amplitude \cite{Pasc98}.

Concerning the $\Delta$N$\gamma$ coupling, Jones and Scadron 
\cite{Jones} suggestion has been widely used in the ($G_1,G_2$) 
decomposition with 
\cite{Davidson91,EMoya,Feuster97,Vanderhaeghen,Pasc95}
or without \cite{Nozawa,Sato} off-shell extension. Another 
decomposition ($G_E$, $G_M$), based upon the same idea as 
the Sachs form factors for the nucleon \cite{Ernst}, 
is also possible. This decomposition is
directly connected to physical quantities, as 
electric and magnetic multipoles, in particular to the E2/M1 
ratio which is of great interest 
from both experimental and theoretical points of view 
\cite{Blanpied,Pasc04,Thomas}. 
This second decomposition is consistent 
with the GI approach and can be written as
\cite{Pasc03}
\begin{equation}
{\mathcal L}=\frac{3e}{2M \left( M+M^* \right)} \bar{N} 
\left[ ig_1 \tilde{F}_{\mu \nu}+g_2 \gamma^5 F_{\mu \nu} 
\right] \left( \partial^\mu N^{*\nu}_3\right)+ HC \: ,
\end{equation}
where $g_1$ and $g_2$ can be easily related to $G_E$ and 
$G_M$ \cite{Pasc99} by
\begin{eqnarray}
G_E&=&-\frac{1}{2}\frac{M^* - M}{M^* + M}g_2 \: ,\\
G_M&=& g_1 + \frac{1}{2}\frac{M^* - M}{M^* + M}g_2 \: .
\end{eqnarray}

Other possible consistent choices can be found in Refs.
\cite{Pasc99,Kondratyuk01}.

GI couplings have been proved to be free of the pathologies 
which are inherent 
to the traditional scheme. No anomalies are found in the 
quantization; neither JS nor VZ problems appear; and no 
spin-1/2 sector arises when the invariant amplitudes are 
calculated \cite{Pasc98}. Moreover, Pascalutsa and Timmermans 
\cite{Pasc99} claim that DOF counting is
the reason why GI couplings are 
consistent while the off-shell extension couplings 
of Nath et al. are not. They 
blame the unphysical extra components for the appearance of 
pathologies. Both, GI (\ref{GIcoupling}) and 
traditional (\ref{offshell}) couplings, provide the same 
result on-shell (if we set properly the coupling constants). 
However, their off-shell behavior is completely different.

Based on the previous discussion, the P$_{33}$ 
Lagrangian that will be used in this
work is
\begin{equation}
\begin{split}
{\mathcal L}_{\text{P}_{33}}=&-\frac{h}{f_\pi M^*} \bar{N} 
\epsilon_{\mu \nu \lambda \beta} \gamma^\beta \gamma^5 
\left( \partial^\mu N^{*\nu}_j \right) \left( \partial^\lambda 
\pi_j  \right) \\
&+ \frac{3e}{2M \left( M+M^* \right)} \bar{N} \left[ ig_1 
\tilde{F}_{\mu \nu}+g_2 \gamma^5 F_{\mu \nu} \right] 
\left( \partial^\mu N^{*\nu}_3\right) \\ 
&+ HC \: .
\end{split} \label{P33}
\end{equation}

From this Lagrangian it is straightforward to obtain 
phenomenological Lagrangians for other spin-3/2 resonances. 
To obtain the P$_{13}$ resonance Lagrangian from (\ref{P33}) 
only an isospin change is needed
\begin{equation}
N^{* \alpha}_j \to \tau_j N^{* \alpha}, \quad j=1,2,3 \: ;
\end{equation}
for the strong vertex, and
\begin{equation}
N^{* \alpha}_3 \to N^{* \alpha}, \quad g_j \to \frac{1}{2} 
\left[ g_j^S+g_j^V \tau_3 \right], \quad j=1,2 \: ;
\end{equation}
for the photon vertex.

Thus the Lagrangian is
\begin{equation}
\begin{split}
{\mathcal L}_{\text{P}_{13}}=&-\frac{h}{f_\pi M^*} \bar{N} 
\epsilon_{\mu \nu \lambda \beta} \gamma^\beta \gamma^5 \tau_j 
\left( \partial^\mu N^{*\nu} \right) \left( \partial^\lambda 
\pi_j  \right) \\
&+ \frac{3e}{4M \left( M+M^* \right)} \bar{N} \Big{[} i 
\left( g_1^S+g^V_1 \tau_3 \right) \tilde{F}_{\mu \nu} \\
&+\left( g_2^S+g_2^V \tau_3 \right) \gamma^5 F_{\mu \nu} 
\Big{]} \left( \partial^\mu N^{*\nu} \right) + HC \: .
\end{split}
\end{equation}

Lagrangians for D$_{33}$ and D$_{13}$ resonances are obtained 
easily from P$_{33}$ and P$_{13}$. We only need to change the 
parity of the coupling placing an overall $\gamma_5$
\begin{equation}
\begin{split}
{\mathcal L}_{\text{D}_{33}}=&-\frac{h}{f_\pi M^*} \bar{N} 
\epsilon_{\mu \nu \lambda \beta} \gamma^\beta \left( 
\partial^\mu N^{*\nu}_j \right) \left( \partial^\lambda 
\pi_j  \right)  \\
&+ \frac{3e}{2M \left( M+M^* \right)} \bar{N} \left[ ig_1 
\tilde{F}_{\mu \nu} \gamma_5 +g_2  F_{\mu \nu} \right] \left( 
\partial^\mu N^{*\nu}_3\right) \\
&+ HC \: ,
\end{split}
\end{equation}

\begin{equation}
\begin{split}
{\mathcal L}_{\text{D}_{13}}=&-\frac{h}{f_\pi M^*} \bar{N} 
\epsilon_{\mu \nu \lambda \beta} \gamma^\beta \tau_j 
\left( \partial^\mu N^{*\nu} \right) 
\left( \partial^\lambda \pi_j  \right)  \\
&+ \frac{3e}{4M \left( M+M^* \right)} \bar{N} \Big{[} i 
\left( g_1^S+g_1^V \tau_3\right) \tilde{F}_{\mu \nu} \gamma_5 \\
&+ \left( g_2^S+g_2^V \tau_3 \right) F_{\mu \nu} \Big{]} 
\left( \partial^\mu N^{*\nu}\right) + HC \: .
\end{split}
\end{equation}

Although we restrict ourselves to spin-3/2, it is clear 
that higher spin interactions can be built within the same theoretical 
framework. This is left to future works.

\subsection{Propagators and widths} \label{sub:propagatorsandwidths}
With regards to the propagators of the resonances, for a spin-1/2
resonance we use
\begin{equation}
iG(v)=i\frac{\feyns{v}+M^*}{v^2-M^{*2}+iM^*\Gamma(s,u)} \: ,
\end{equation}
and for the spin-3/2 propagator we use the Rarita-Schwinger propagator
\begin{equation}
\begin{split}
iG_{\alpha \beta}(v)= & i\frac{\feyns{v}+M^*}{v^2-M^{*2}
+iM^*\Gamma(s,u)} \\ 
\times & \left[ -g_{\alpha \beta}+\frac{1}{3}\gamma_\alpha 
\gamma_\beta+\frac{2}{3M^{*2}}v_\alpha v_\beta -\frac{1}{3M^*}
\left(v_\alpha \gamma_\beta -\gamma_\alpha v_\beta \right) \right] \: ,
\end{split}
\end{equation}
where $v$ is the resonance four-momentum. A phenomenological 
width $\Gamma \left( s,u \right)$ is included in the propagator 
denominator consistently with what is obtained if we dress it 
with pions \cite{Pasc03,Kondratyuk00}.

The energy dependence of the width is chosen phenomenologically as
\begin{equation}
\Gamma \left(s,u \right) 
= \sum_j \Gamma_j X_j \left( s , u \right) \: ,\label{eq:width}
\end{equation}
where $j = \pi , \pi \pi , \eta$ stands for the different decay 
channels and
\begin{equation}
X_j \left( s , u \right) \equiv X_j \left( s \right) +  X_j 
\left( u \right) - X_j \left( s \right)  X_j \left( u \right) \: ,
\label{eq:Xj}
\end{equation}
with $X_j\left( l \right)$ given by
\begin{equation}
X_j \left( l \right) = 2 \frac{\left(\frac{|\vec{k}_j|}{|\vec{k}_{j0}|} 
\right)^{2L+1}}{1+\left( \frac{|\vec{k}_j|}{|\vec{k}_{j0}|}
\right)^{2L+3}} \: \Theta \left( l - \left( M + m_j \right)^2 \right) \: ,
\end{equation}
where $L$ is the angular momentum of the resonance, 
$\Theta$ is the Heaviside step function, and 

\begin{equation}
|\vec{k}_j|=\sqrt{\left(l-M^2-m_j^2 \right)^2-4m_j^2M^2}/
\left( 2 \sqrt{l}\right) \: ,
\end{equation}
with $m_{\pi \pi} \equiv 2m_\pi$ 
and $|\vec{k}_{j0}| = |\vec{k}_{j}|$ when $l=M^{*2}$.

This parametrization has been built in order to fulfill 
the following conditions
\begin{enumerate}
\item [(i)]$\Gamma = \Gamma_0$ at $\sqrt{s}=M^*$, 
\item [(ii)]$\Gamma \to 0$ when $|\vec{k}_j| \to 0$, 
\item [(iii)]a correct angular momentum barrier at 
threshold $|\vec{k}_j|^{2L+1}$,
\item [(iv)]crossing symmetry.
\end{enumerate}

This parametrization of the width is an improvement over the one used 
in Ref. \cite{EMoya} and includes decays to $\eta$ and $2\pi$ 
which take into account inelastic channels \cite{Drechsel} 
and condition (iv). 
The width  
contributes to both $s$ and $u$-channels, 
so that crossing symmetry 
is preserved due to Eq. (\ref{eq:Xj}). 
In \cite{Feuster97} 
the authors made
an analysis of the  energy dependence of the width.
It was concluded that, as long as it provides a decrease 
of the width beyond the resonance position, the specific 
way in which $X_j$ is parametrized is not so important.

\subsection{Form factors} \label{sub:formfactors}
For the numerical calculations
we include form factors for Born terms and vector mesons, 
in order to regularize the high energy behavior of these terms. 
We choose form factors as suggested by Davidson and Workman 
\cite{Davidson01-1} that allow to fulfill gauge invariance and 
crossing symmetry. Actually, $X_j(s,u)$ in Eq. (\ref{eq:Xj}) also 
follows this choice. Thus for Born terms
\begin{equation}
\begin{split}
\hat{F}_B(s,u,t)=& F_1(s)+F_2(u)+F_3(t)-F_1(s)F_2(u) \\
&- F_1(s)F_3(t)-F_2(u)F_3(t)+F_1(s)F_2(u)F_3(t) \: ,
\end{split}
\end{equation}
where,
\begin{eqnarray}
F_1(s)&=& \left[1+ \left( s-M^2 \right)^2/\Lambda_B^4 \right]^{-1} \: , \\
F_2(u)&=& \left[1+ \left( u-M^2 \right)^2/\Lambda_B^4 \right]^{-1} \: , \\
F_3(t)&=& \left[1+ \left( t-m_\pi^2 \right)^2/\Lambda_B^4 \right]^{-1} \: .
\end{eqnarray}
For vector mesons we adopt $\hat{F}_V(t) = F_3(t)$ with 
the changes $m_\pi \to m_V$ and $\Lambda_B \to \Lambda_V$. 
In order to have as few free parameters as possible in the numerical
calculations we use the same $\Lambda \equiv \Lambda_B=\Lambda_V$ 
for both vector 
mesons and Born terms.
For the resonance-pion-nucleon vertex, the form factor
$\sqrt{X_\pi\left( s,u \right)}$ has to be used for consistency
with the width used in the propagator discussed previously.

Models like the ones by Garcilazo and 
Moya de Guerra \cite{EMoya} 
and Feuster and Mosel \cite{Feuster97} needed a cutoff 
in the $u$-channels of spin-3/2 resonances to obtain a good 
description of observables. This cutoff was needed
because the high-energy contributions 
of these diagrams are not reduced by the denominator of the 
propagator. In Ref. \cite{EMoya} it was argued that 
the need of this 
cutoff could be justified by
the two possible interpretations of the 
resonance excitation. 
From an effective field theory point 
of view, $u$-channels should be introduced with their full 
strength. On the other hand if we consider 
resonances as pure $\pi N$ 
rescattering states (Chew-Low description), the $u$-channels 
contributions should be dropped.
Hence, the cutoff was interpreted in Ref. \cite{EMoya}
as a way to have an interplay between both descriptions.
However, in this way crossing symmetry was broken.
Our present model relies entirely on effective field theory,
and we preserve crossing symmetry and there is no need for that
cutoff in the $u$-channel amplitudes.
Ought to the vector 
couplig to the resonance the $u$-channel amplitudes are 
suppresed by themselves which is a strong point in favor of the
GI coupling.

\section{Results} \label{sec:results}

\subsection{Study of the parameters of the model}
The first choice that has to be made is the 
nucleon resonances to be taken into account. 
We have included seven resonances: 
$\Delta(1232)$, N(1440), N(1520), N(1535), 
$\Delta(1620)$, N(1650), and $\Delta(1700)$ 
which are all the four star nucleon resonances 
in PDG up to 1.7 GeV and up to 
spin-3/2. Among four star resonances only spin-5/2 
N(1675) and N(1680) resonances are left aside 
for future work.

In a Lagrangian model, the determination 
of the parameters of a single resonance
is affected by the determination of 
the parameters of the other resonances. 
Thus,
we have decided not to include three star
resonances because their contribution would 
be very small and would introduce a sort of 
noise in the determination of 
the parameters. 

There are quite a number of parameters to be set in the model. 
Some of them are well known and established 
independently of the photoproduction data,
such as nucleon and pion masses 
($M=938.9175$ MeV,
$m_{\pi^0}= 134.9766$ MeV, 
$m_{\pi^\pm}=139.5673$ MeV), but 
some others have to be established from fits to the pion 
photoproduction data, 
namely electromagnetic coupling constants. 
In the forthcoming paragraphs we give the values of
every  parameter of the model 
as well as the procedures employed to establish them. 

\subsubsection{Vector meson coupling constants}
Vector meson contributions
are characterized by eleven parameters: 
$m_\omega$, $F_{\omega NN}$, $K_\omega$, 
$G_{\omega \pi \gamma}$, $m_{\rho^0}$, 
$m_{\rho^\pm}$, $F_{\rho NN}$, $K_\rho$, 
$G_{\rho^0 \pi \gamma}$, $G_{\rho^\pm \pi \gamma}$,
and cutoff $\Lambda$. Masses are given 
by PDG and the $\pi \gamma V$ couplings are related to the 
decay widths $\Gamma_{\pi \gamma V}$ of PDG \cite{PDG2004} 
through the equation 

\begin{equation}
\Gamma_{V \to \pi \gamma} = \frac{e^2 G^2_{V \pi \gamma}}
{96 \pi}\frac{m^3_V}{m_\pi^2}\left( 1-\frac{m_\pi^2}{m^2_V} 
\right)^3 \: . \label{eq:decayVM}
\end{equation}

We take from PDG the following values: 
$m_\omega=782.57$ MeV, $m_{\rho^0}=768.5$ MeV, 
$m_{\rho^\pm}=766.5$ MeV, $\Gamma_{\rho^0 \pi \gamma}=0.121$ 
MeV ($G_{\rho^0 \pi \gamma} = 0.1161$), 
$\Gamma_{\rho^\pm \pi \gamma}= 0.068$ MeV 
($G_{\rho^\pm \pi \gamma} = 0.0906$), and 
$\Gamma_{\omega \pi \gamma}=0.70476$ MeV 
($G_{\omega \pi \gamma} = 0.2804$). 
Thus, only five constants 
remain unknown. One of them is the cutoff $\Lambda$ 
which will be discussed later.
The four remaining constants are taken from the 
analysis of
nucleon electromagnetic form factors by Mergell, 
Mei\ss ner, and Drechsel \cite{MMD}: $F_{\rho NN}=2.6$, 
$K_\rho =6.1 \pm 0.2$, $F_{\omega NN}=20.86 \pm 0.25$, and 
$K_\omega=-0.16 \pm 0.01$, 
which compare well to the data, including the 
latest experiments at Jefferson Lab \cite{Jlab}.

\subsubsection{Masses and widths of 
the nucleon resonances} \label{sec:speedplot}
\begin{table}
\caption{Masses, widths, and branching ratios from Refs. 
\cite{PDG2004,Vrana} and from the speed plot calculation 
(see text). 
Masses and widths in MeV.
We have taken 
$\Gamma_{\pi \pi}/\Gamma=1-\Gamma_\pi/\Gamma-\Gamma_\eta/\Gamma$. 
Subscripts $P$, $V$ and $SP$ stand for PDG \cite{PDG2004}, 
Vrana \cite{Vrana} and Speed Plot 
respectively. PDG masses and widths are mean values.} 
\label{tab:massesandwidths}
\begin{tabular}{lccccccc}

\hline

&$\Delta(1232)$ 
&     N(1440) 
&     N(1520) 
&     N(1535) 
&$\Delta(1620)$ 
&     N(1650) 
&$\Delta(1700)$ \\

\hline \hline

$M^*_P$ 
& 1210 & 1365 & 1510 & 1505 & 1607 & 1660 & 1660 \\
$M^*_V$ 
& 1217 & 1383 & 1504 & 1525 & 1607 & 1663 & 1726 \\
$M^*_{_{SP}}$
& 1211 & 1372 & 1516 & 1540 & 1608 & 1664 & 1641 \\
$\Gamma_P$
&  100 &  210 &  115 &  170 &  115 &  160 &  200 \\
$\Gamma_V$
&   96 &  316 &  112 &  102 &  148 &  240 &  118 \\
$\Gamma_{_{SP}}$
&   98 &  290 &   48 &  107 &  141 &  159 &  955 \\
$\frac{\Gamma_\pi}{\Gamma_P}$
& 1.00 & 0.65 & 0.55 & 0.45 & 0.25 & 0.72 & 0.15 \\ 
$\frac{\Gamma_\eta}{\Gamma_P}$
&  --  & 0.00 & 0.00 & 0.51 &  --  & 0.06 &  --  \\
$\frac{\Gamma_{\pi \pi}}{\Gamma_P}$
& 0.00 & 0.35 & 0.45 & 0.04 & 0.75 & 0.22 & 0.85 \\
$\frac{\Gamma_\pi}{\Gamma_V}$
& 1.00 & 0.72 & 0.63 & 0.35 & 0.45 & 0.74 & 0.05 \\
$\frac{\Gamma_\eta}{\Gamma_V}$
&  --  & 0.00 & 0.00 & 0.51 &  --  & 0.06 &  --  \\
$\frac{\Gamma_{\pi \pi}}{\Gamma_V}$
& 0.00 & 0.28 & 0.37 & 0.14 & 0.55 & 0.20 & 0.95 \\

\hline

\end{tabular}
\end{table}

We have used three different sets of masses and widths of the 
nucleon resonances (Table \ref{tab:massesandwidths}): 
First, the PDG values \cite{PDG2004}; second, 
the multichannel analysis of Vrana 
et al. \cite{Vrana}; and third, the 
speed plot (SP) calculation that we explain below. 
For the partial decay widths we have two different sets, 
one from PDG and one from Vrana et al. 
which lies within 
the PDG error bars. The 
Vrana et al. set of partial decay widths 
has been chosen for the SP calculation.

Masses and widths of nucleon resonances can be obtained 
from $\pi N$ partial wave analysis using the 
speed plot technique
\cite{Hoehler}. First we define the 
speed by
\begin{equation}
SP\left( W \right) = |dT\left( W \right)/dW| \: ,
\end{equation}
with $W=\sqrt{s}$ and
\begin{equation}
T\left( W \right)=\frac{1}{2i}\left[ \exp \left( 2i\delta
\left( W \right)\right)-1 \right] \: ,
\end{equation}
where $T(W)$ is the dimensionless resonance partial wave 
amplitude and $\delta \left( W \right)$ its phase.

\begin{figure}
\begin{center}
\rotatebox{-90}{\scalebox{0.35}[0.4]{\includegraphics{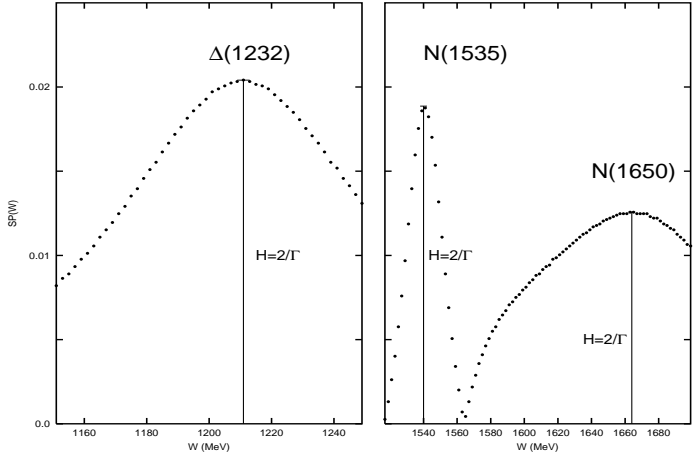}}}
\end{center}
\caption{Speed plot examples. 
Left figure shows the $\Delta(1232)$ 
speed plot. Right figure shows the speed plot 
for the S$_{11}$ region. 
Data have been taken from SAID database 
for $\pi N$ scattering 
\cite{SAIDdata}.} \label{fig-speed}
\end{figure}

To calculate masses and widths we have taken phases from 
the current solution of the SAID $\pi N$ partial wave analysis 
\cite{SAIDdata}. In figure \ref{fig-speed} we show $SP(W)$ 
for  P$_{33}$ -- in the $\Delta (1232)$ region -- and S$_{11}$ 
multipoles. The position of the peak provides the pole mass, 
and the height provides the width: $H=2/\Gamma$.

The baryon resonances show up clearly and the 
calculation is straightforward. The only problem is related 
to the existence of a background which induces a phase shift 
in the $\pi N$ phases. In the region of the peak this phase 
shift can be considered approximately constant and its effect 
in $SP(W)$ is negligible.

Our fits to photoproduction data shown in the next subsections
are clasified according to six sets
of parameters which are given in table \ref{tab:fits}.
Sets \#1 and \#4 are based on PDG values for masses and widths;
set \#2 and set \#4 are based on Vrana et al. \cite{Vrana}; and
sets \#3 and \#6 are based on the SP calcualtion.

The strong coupling constants ($h$'s) of the resonances are 
obtained from equations in appendix \ref{sec:decaywidths}, 
using the partial decay widths into one pion of the resonances. 
We choose all the strong coupling constants to be positive, thus 
the overall sign of the amplitude of 
each resonance depends on the 
sign of the electromagnetic coupling constants.

\subsubsection{Electromagnetic coupling 
constants of the nucleon resonances}\label{sec:emcoupling}

\begin{table}
\caption{Specifications of the parameter sets.
Masses, widths, and $\Lambda$ are in GeV. 
The coupling constants for the 
vector mesons are dimensionless.
We provide also the $\chi^2/\chi^2_{\text{PDG}}$
in order to compare fits.} \label{tab:fits}
\begin{tabular}{lcccccc}

\hline

Set
&~~~\#1~~~
&~~~\#2~~~
&~~~\#3~~~
&~~~\#4~~~
&~~~\#5~~~
&~~~\#6~~~ \\

\hline\hline

Masses \& Widths~~~~& PDG&Vrana&SP&PDG&Vrana&SP\\
$\delta_{FSI}$   & Yes  & Yes   & Yes  & No   & No   & No \\
$\chi^2/\chi^2_{\text{PDG}}$&1&0.53&0.60&9.30&5.57&4.56 \\
\hline
$\Lambda$        &1.121&1.050&1.040&1.494&0.951&0.962 \\
\hline
$K_\rho$         &6.30 &6.30 &6.30&6.30 &5.90 &5.90  \\
$F_{\omega NN}$  &21.11&21.11&21.11&20.61&21.11&21.11 \\
$K_\omega$       &-0.17&-0.17&-0.17&-0.15&-0.17&-0.17 \\
\hline
$M^* \left[ \Delta(1232) \right]$& 
1.209&1.215&1.209&1.210&1.215&1.209\\
$\Gamma \left[ \Delta(1232) \right]$&
0.102&0.098&0.100&0.102&0.094&0.099\\
\hline
$M^* \left[ \text{N(1440)} \right]$ &
1.385&1.381&1.370&1.385&1.381&1.370\\
$\Gamma \left[ \text{N(1440)} \right]$&
0.160&0.318&0.292&0.260&0.314&0.288\\
\hline
$M^* \left[ \text{N(1520)} \right]$&
1.505&1.502&1.514&1.505&1.502&1.514\\
$\Gamma \left[ \text{N(1520)} \right]$&
0.110&0.110&0.050&0.110&0.110&0.050\\
\hline
$M^* \left[ \text{N(1535)} \right]$&
1.495&1.527&1.542&1.495&1.523&1.538\\
$\Gamma \left[ \text{N(1535)} \right]$&
0.250&0.104&0.109&0.099&0.100&0.109\\
\hline
$M^* \left[ \Delta(1620) \right]$&
1.590&1.605&1.606&1.620&1.605&1.606\\
$\Gamma \left[ \Delta(1620) \right]$&
0.100&0.150&0.143&0.100&0.150&0.143\\
\hline
$M^* \left[ \text{N(1650)} \right]$&
1.680&1.665&1.666&1.640&1.665&1.666\\
$\Gamma \left[ \text{N(1650)} \right]$&
0.150&0.238&0.157&0.150&0.242&0.157\\
\hline
$M^* \left[ \Delta(1700) \right]$&
1.620&1.728&1.639&1.620&1.728&1.639\\
$\Gamma \left[ \Delta(1700) \right]$&
0.250&0.120&0.957&0.250&0.120&0.957\\
\hline

\end{tabular}
\end{table}

At this point, only the electromagnetic coupling constants and 
the cutoff $\Lambda$ remain undetermined. The best
way to establish them is by fitting to 
pion photoproduction experimental data. Among 
all the observables (cross section, asymmetries, etc.) 
for pion photoproduction, the set of data we have chosen
is the one given by the current SAID multipole energy 
independent solution \cite{Arndt90-2,SAID,SAIDdata}. 
There are two main  reasons for this choice. 
First, electromagnetic multipoles are directly related to 
the amplitudes and are more sensitive to coupling properties
than are other observables.
Deficiencies in the model 
show up much more clearly in multipoles than in any other observable. 
Second, all the observables can be expressed in terms of the 
multipoles, thus, if the multipoles are properly fitted by the 
model, so should be the other observables. The explicit expressions 
for the multipoles in terms of the amplitudes can be found 
in appendix \ref{sec:multipole}.

Another issue to take into account is unitarity.
Models below the two pion production threshold fulfill 
Watson's theorem \cite{Watson} to achieve unitarity
using either 
$\pi N$ scattering phases \cite{Drechsel}, dynamical models 
\cite{Nozawa,Sato,Fuda,Pasc04} or 
K-matrix \cite{Davidson91}. 
Beyond the two pion production limit,
implementation of unitarity is unclear and usually relies 
on experimental data and/or extensions of the methods applied 
below the two pion threshold.

We would like to note that,
although our calculation seems to be at tree-level, it is not quite so 
due to the inclusion of the width and the 
form factors, which take into account higher order diagrams 
and structure effects. If we perform a truly tree-level 
calculation -- straightforwardly from amplitudes of appendix 
\ref{sec:invariantamplitudes} -- we would find out that all the 
amplitudes are real and that it would be impossible to fulfill the 
unitarity condition $SS^\dagger=1$, where $S$ is the scattering 
matrix. 
In an effective Lagrangian perturbative model, unitarity should 
be restored by the inclusion of higher order diagrams. 
To avoid this tedious and difficult task, we adopt a 
phenomenological point of view. The main higher order effects can 
be taken into account including a width in the propagator, as 
we do in section \ref{sub:propagatorsandwidths} 
(which amounts to dress the propagator), and including also effective 
final state interactions (FSI). 
Once the width is included, unitarity restoration 
may be achieved through FSI. 
We can assume that it is possible 
to isolate the FSI effects factorizing
the multipoles ${\mathcal M}$ in the following way:
\begin{equation}
{\mathcal M}^{I,\ell,\Pi}=|{\mathcal M}^{I, \ell,\Pi}|\exp 
\left[ \delta_{width}\right] F^{I,\ell,\Pi} \: ,
\end{equation}
where $F^{I, \ell, \Pi}$ is a phase factor that
takes into account FSI, and $\ell$ 
stands for orbital angular momentum, 
$\Pi$ for parity, and $I$ for 
isospin:
\begin{equation}
F^{I,\ell, \Pi}= \exp \left[ i \delta_{FSI}^{ I,\ell, 
\Pi} \right] \: .
\end{equation}

Then, the absolute value of the multipoles 
must be well reproduced by the model 
and only the phases of the multipoles 
remain unknown. We are interested in the 
bare values of the coupling constants, so the best choice is 
to use directly the experimental phases. Hence, the multipole 
phase can be written as
\begin{equation}
\delta^{I, \ell, \Pi} = \delta_{width} + \delta_{FSI}^{I, \ell, \Pi}\: , 
\end{equation}
where we call 
$\delta_{width}$ to the phase given by the calculated amplitudes
and comparison 
with experimental phase shifts ($\delta^{I, \ell, \Pi}$) provides 
us with the unknown final state interaction phase shifts 
$\delta_{FSI}^{I, \ell, \Pi}$. 
Phases $\delta^{I, \ell, \Pi}$ 
are taken from the current energy dependent multipole solution
of SAID analysis \cite{Arndt90-2,SAID,SAIDdata}. 
For each set of masses and widths we obtain two types of fits,
one with and one without SAID phases.

\begin{table}
\caption{Coupling constants of the resonances.
The E2/M1 Ratio (EMR) of $\Delta (1232)$ is also given. 
All magnitudes are dimensionless.} \label{tab:couplings}
\begin{tabular}{lllcccccc}

\hline

 & &  & \#1 & \#2 & \#3 & \#4 & \#5 & \#6 \\

\hline \hline

$\Delta$(1232) & P$_{33}$ 
  & $h$   &0.764&0.721&0.757&0.759&0.706&0.753 \\ 
& & $g_1$ &6.061&5.574&5.630&6.254&5.382&4.984 \\
& & $g_2$ &2.414&1.187&1.123&4.032&7.253&7.696 \\
& & $G_E$ &-0.152&-0.076&-0.071&-0.255&-0.466&-0.485 \\
& & $G_M$ &6.213&5.650&5.701&6.509&5.848&5.469 \\
& & EMR&-2.45\%&-1.35\%&-1.24\%&-3.92\%&-7.97\%&-8.87\% \\
\hline
N(1440)  & P$_{11}$ 
  & $h$  &0.213&0.304&0.303&0.272&0.302&0.300 \\ 
& &$g^p$ &0.255&-0.269&-0.247&0.255&-0.164&0.017 \\
& &$g^n$ &-0.125&0.273&0.234&-0.125&0.096&-0.128 \\
\hline
N(1520)    & D$_{13}$ 
  & $h$    &0.560&0.567&0.366&0.560&0.567&0.360 \\ 
& &$g_1^p$ &-5.753&-4.848&-5.607&-5.498&-0.580&-2.348 \\
& &$g_1^n$ &1.217&2.829&1.982&0.301&-1.503&0.105 \\
& &$g_2^p$ &-0.861&-0.645&-0.520&-0.920&-0.986&-0.691 \\
& &$g_2^n$ &1.462&0.960&0.979&1.674&2.731&2.174 \\
\hline
N(1535)  & S$_{11}$ 
  & $h$  &0.132&0.079&0.078&0.083&0.078&0.079 \\ 
& &$g^p$ &0.219&0.078&0.028&0.435&0.230&0.084 \\
& &$g^n$ &-0.102&-0.127&-0.080&-0.164&-0.195&-0.129 \\
\hline
$\Delta$(1620) & S$_{31}$ 
  &$h$ &0.133&0.159&0.155&0.126&0.159&0.155\\ 
& &$g$ &-0.154&-0.324&-0.308&-0.063&0.008&0.044 \\
\hline
N(1650)  & S$_{11}$ 
  & $h$  &0.102&0.132&0.107&0.110&0.134&0.107\\ 
& &$g^p$ &0.113&-0.167&0.025&0.117&0.074&0.127\\
& &$g^n$ &0.018&0.411&0.324&0.019&0.281&0.056\\
\hline
$\Delta$(1700) & D$_{33}$ 
  & $h$    &0.285&0.149&0.528&0.285&0.149&0.528\\ 
& & $g_1$  &-3.513&0.663&-11.875&-3.996&-19.642&-26.531 \\
& & $g_2$  &1.871&0.548&-2.392&2.000&3.701&7.293 \\

\hline

\end{tabular}
\end{table}

In order to fit the data and determine 
the best parameters of the resonances 
we have written a genetic algorithm combined with the 
\texttt{E04FCF} routine from NAG libraries \cite{NAG}. Although 
genetic algorithms are computationally more expensive 
than other algorithms, in a minimization problem it is much 
less likely for them to get stuck at local minima than for other
methods, namely 
gradient based minimization methods. Thus, in a 
multiparameter minimization like the one we face here
it is probably 
the best possibility to search for the minimum. 
It is out of the scope 
of this paper to go through an explanation on genetic algorithms 
and details on them can be found elsewhere 
\cite{genetic}.

The function to minimize is the $\chi^2$ defined as 
\begin{equation}
\chi^2 = \sum_j \frac{\left( \mathcal{M}^{exp}_j-
\mathcal{M}^{th}_j 
\right)^2}{\left( \Delta \mathcal{M}^{exp}_j \right)^2} \: ,
\end{equation}
where ${\mathcal M}^{exp}$ stands for the current energy 
independent extraction of the multipole analysis of SAID up 
to 1 GeV for $E_{0+}$, $M_ {1-}$, $E_{1+}$, $M_{1+}$, $E_{2-}$, 
and $M_{2-}$ multipoles in the three isospin channels 
$I=\frac{3}{2},p,n$ for the $\gamma p \to \pi^0 p$ process.
$\Delta \mathcal{M}^{exp}$ is the error 
and $\mathcal{M}^{th}$ is the multipole given by the model 
which depends on the parameters.
These parameters are the 
cutoff $\Lambda$ and the  
electromagnetic coupling constants in Table \ref{tab:couplings}, 
which are related to the 
helicity amplitudes $A_{\lambda}^I$ in Table \ref{tab:helicities}
through equations given in appendix \ref{sec:helicityamplitudes}.

The minimization procedure applied is as follows: First 
the genetic algorithm has been run and when the convergence 
conditions were accomplished the \texttt{E04FCF} routine was 
used for fine tunning. The program has been run many times 
with different seeds in order to ensure that the minimum 
was not local.
We have taken into account 763 data for the real part of 
the multipoles and the same amount for the imaginary part. 
Thus, 1526 data points have been used in the fits.

In Tables \ref{tab:fits}, \ref{tab:couplings}, 
and \ref{tab:helicities} 
we show results for the six different sets 
and provide the reader with all the parameters of the model. 
Table \ref{tab:fits} shows masses and widths, the cutoff 
$\Lambda$, as well as the vector meson parameters $K_\rho$, 
$F_{\omega NN}$, and $K_\omega$, 
for each set.
Table
\ref{tab:couplings} provides all the 
coupling constants of the resonances as well as the 
the E2/M1 Ratio (EMR) of the 
$\Delta(1232)$ resonance.
Table \ref{tab:helicities} contains the helicity 
amplitudes of the resonances which can be compared to those in
other
references such as \cite{Davidson91,EMoya,Feuster97,PDG2004}. 

\begin{table}
\caption{Helicity amplitudes in GeV$^{-1/2}$ 
for the different sets.} \label{tab:helicities}
\begin{tabular}{lllcccccc}

\hline

 & & & \#1 & \#2 & \#3 & \#4 & \#5 & \#6 \\

\hline \hline

$\Delta$(1232) & P$_{33}$ ~~& 
$A^\Delta_{1/2}$  ~~&
-0.129 ~&
-0.123 ~&
-0.123 ~&
-0.129 ~&
-0.101 ~&
-0.090 ~\\ 
& & $A^\Delta_{3/2}$&-0.247&-0.225&-0.224&-0.263&-0.248&-0.231 \\
\hline
N(1440) & P$_{11}$
&  $A^p_{1/2}$ &-0.061&0.064&0.058&-0.061&0.039& -0.004\\ 
& &$A^n_{1/2}$ &0.030&-0.065&-0.055&0.030&-0.023&0.030 \\
\hline
N(1520) & D$_{13}$
& $A^p_{1/2}$  &-0.020&-0.020&-0.034&-0.015&0.027& 0.006\\ 
& &$A^n_{1/2}$ &-0.050&-0.013&-0.022&-0.068&-0.121& -0.092\\
& &$A^p_{3/2}$ &0.161& 0.129&0.136&0.161&0.095& 0.092\\
& &$A^n_{3/2}$ &-0.128&-0.118&-0.107&-0.128&-0.190& -0.163\\
\hline
N(1535) & S$_{11}$
& $A^p_{1/2}$  &0.060&0.022&0.008&0.119&0.065&0.024\\ 
& &$A^n_{1/2}$ &-0.028&-0.036&-0.023&-0.045&-0.055&-0.037\\
\hline
$\Delta$(1620) & S$_{31}$
& $A^\Delta_{1/2}$ &0.038&0.081&0.077&0.016&-0.002&-0.011 \\ 
\hline
N(1650) & S$_{11}$
& $A^p_{1/2}$  &0.037&-0.054&0.008&0.037&0.024& 0.041\\ 
& &$A^n_{1/2}$ &0.006& 0.133&0.105&0.006&0.091& 0.018\\
\hline
$\Delta$(1700) & D$_{33}$
& $A^\Delta_{1/2}$   &0.109&0.015&0.222&0.119&0.406& 0.573\\ 
& & $A^\Delta_{3/2}$ &0.063&0.055&0.057&0.063&-0.156& 0.006\\

\hline

\end{tabular}
\end{table}

\subsection{Multipole analysis}
As has been previously explained, 
in order to determine the parameters of the resonances 
and the cutoff we have used the data for electromagnetic multipoles.
In this section we discuss the results
obtained for multipoles as well as the 
quality of the different fits.

Lagrangian models like the one presented here are more
complicated than Breit-Wigner models such as MAID
\cite{Drechsel}. The latter are simple and describe accurately 
experimental observables but do not provide much information
about properties of the resonances such as the strength of 
the couplings. Breit-Wigner treatment of resonances 
can be considered 
naive because each resonance contributes only to the 
multipole with its same angular momentum quantum number 
In this way there is no background from resonances, 
which is very different from Lagrangian models where,
for a given resonance, the direct term contributes
only to a single spin-isospin channel, while the crossed term
contributes to different spin-isospin channels as background, 
and then one resonance does indeed affect to the determination of the
parameters of other resonances.
Contributions from crossed terms to the background cannot be 
neglected and 
there are resonant contributions to several 
multipoles.
For instance, N(1520) contributes to $E_{2-}^{p,n}$ and 
$M_{2-}^{p,n}$, 
as expected for a D$_{13}$ isobar, but also contributes
strongly to $M_{1+}^p$. 
Thus, the background of Breit-Wigner models is much simpler
because it only has
contributions from Born terms and vector mesons 
($\rho$ and $\omega$).

\begin{figure}
\begin{center}
\rotatebox{0}{\scalebox{0.75}[0.8]{
\includegraphics{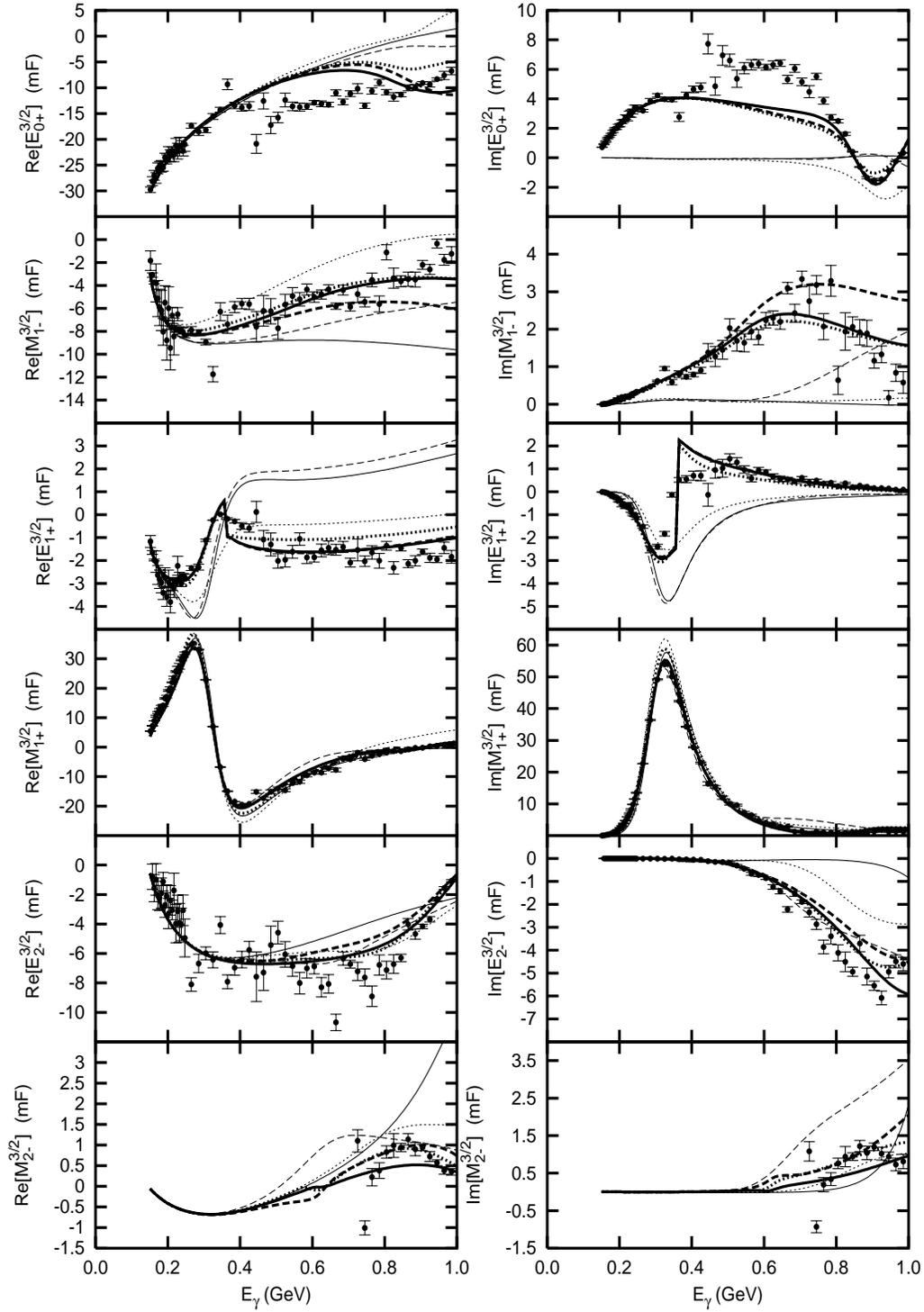}}}
\caption{Electromagnetic multipoles for the isospin-3/2 channel. 
Data have been taken 
from Ref. \cite{SAIDdata}. 
Photon energy is given in the 
laboratoy frame. Curves conventions: 
thick dotted set \#1;
thick solid set \#2;
thick dashed set \#3;
thin dotted set \#4;
thin solid set \#5;
thin dashed set \#6.}
\label{fig:multipole_delta}
\end{center}
\end{figure}

\begin{figure}
\begin{center}
\rotatebox{0}{\scalebox{0.75}[0.8]{
\includegraphics{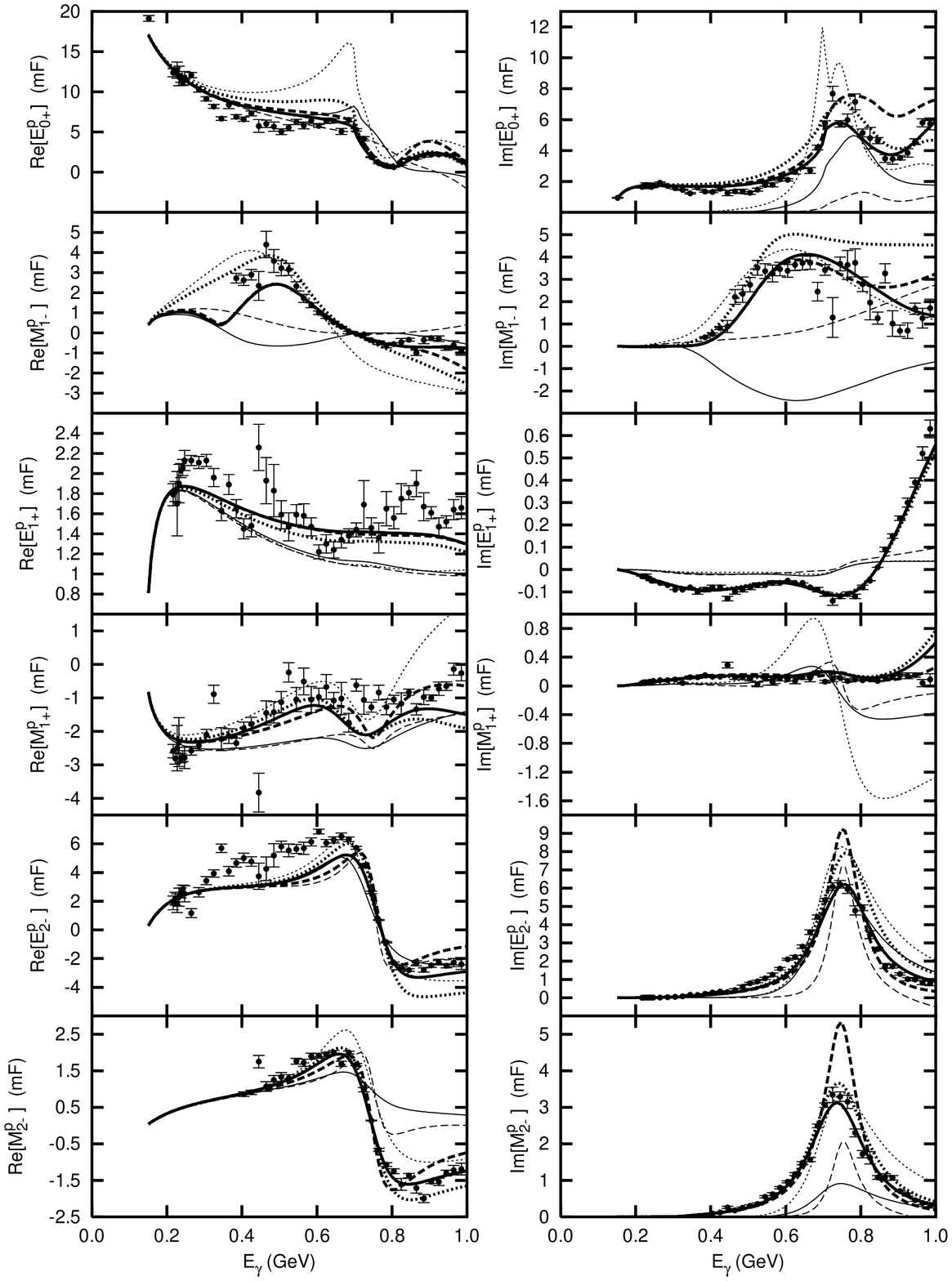}}}
\caption{Electromagnetic multipoles 
for the isospin-1/2 proton channel. 
Same conventions as in Fig. 
\ref{fig:multipole_delta} apply.
Data have been taken 
from Ref. \cite{SAIDdata}} \label{fig:multipole_p}
\end{center}
\end{figure}

\begin{figure}
\begin{center}
\rotatebox{0}{\scalebox{0.75}[0.8]{
\includegraphics{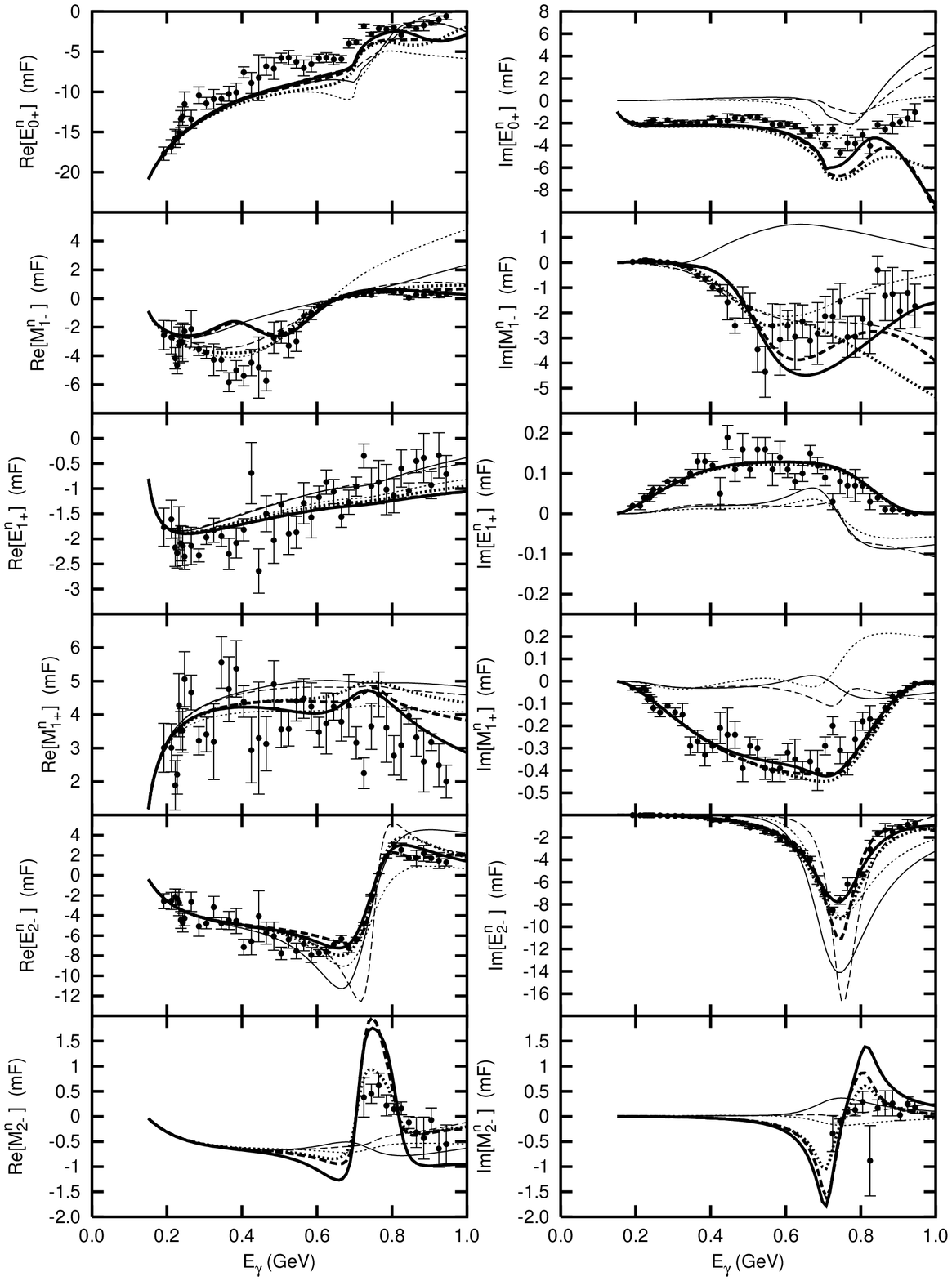}}}
\caption{Electromagnetic multipoles 
for the isospin-1/2 neutron channel. 
Same conventions as in Fig. 
\ref{fig:multipole_delta} apply.
Data have been taken 
from Ref. \cite{SAIDdata}} \label{fig:multipole_n}
\end{center}
\end{figure}

Figures \ref{fig:multipole_delta},  
\ref{fig:multipole_p}, and
\ref{fig:multipole_n} show the comparison of the six different sets 
of Table \ref{tab:fits} to
experimental data from SAID database \cite{SAIDdata}. 
Without FSI, at low energies, we get nice fits 
to some of the 
multipoles: $M_{1+}^{3/2}$, $E_{2-}^{3/2}$, and
$E_{2-}^p$.
With increasing energy there is a breakdown of the model
which calls for further improvements.
The major ingredient that lacks the model is FSI, which we
introduce phenomenologically as described in section 
\ref{sec:emcoupling}.
Indeed, the fits are greatly improved 
-- specially the fits of the imaginary parts of the multipoles -- 
when FSI are included, as it stems from the comparison of the
$\chi^2$ (Table \ref{tab:fits}). 
The 
experimental data are quite well reproduced by theory
with better quality for the low energy region 
than for the high energy (900 MeV and further), where some 
of the fits start to diverge 
(i.e. Im$M_{1+}^p$ and Im$E_{0+}^n$). 
In this section we focus on fits that include FSI,
except in the case in which comparison 
with non-FSI sets provides relevant information.

Despite of the difference between SP and Vrana et al. 
masses and widths, the curves that we obtain for sets \#2 and \#3 
are very close to each other
(so are their $\chi^2$, see Table \ref{tab:fits}), 
sometimes undistinguishable, 
except for some high order multipoles as Im$M_{2-}^p$. 
Curves from set \#1 do not reproduce data as well as \#2 and \#3 do
and the
$\chi^2$ is almost twice as large
due to the additional restrictions in the values of the parameters.

\begin{figure}
\begin{center}
\rotatebox{-90}{\scalebox{0.5}[0.5]{
\includegraphics{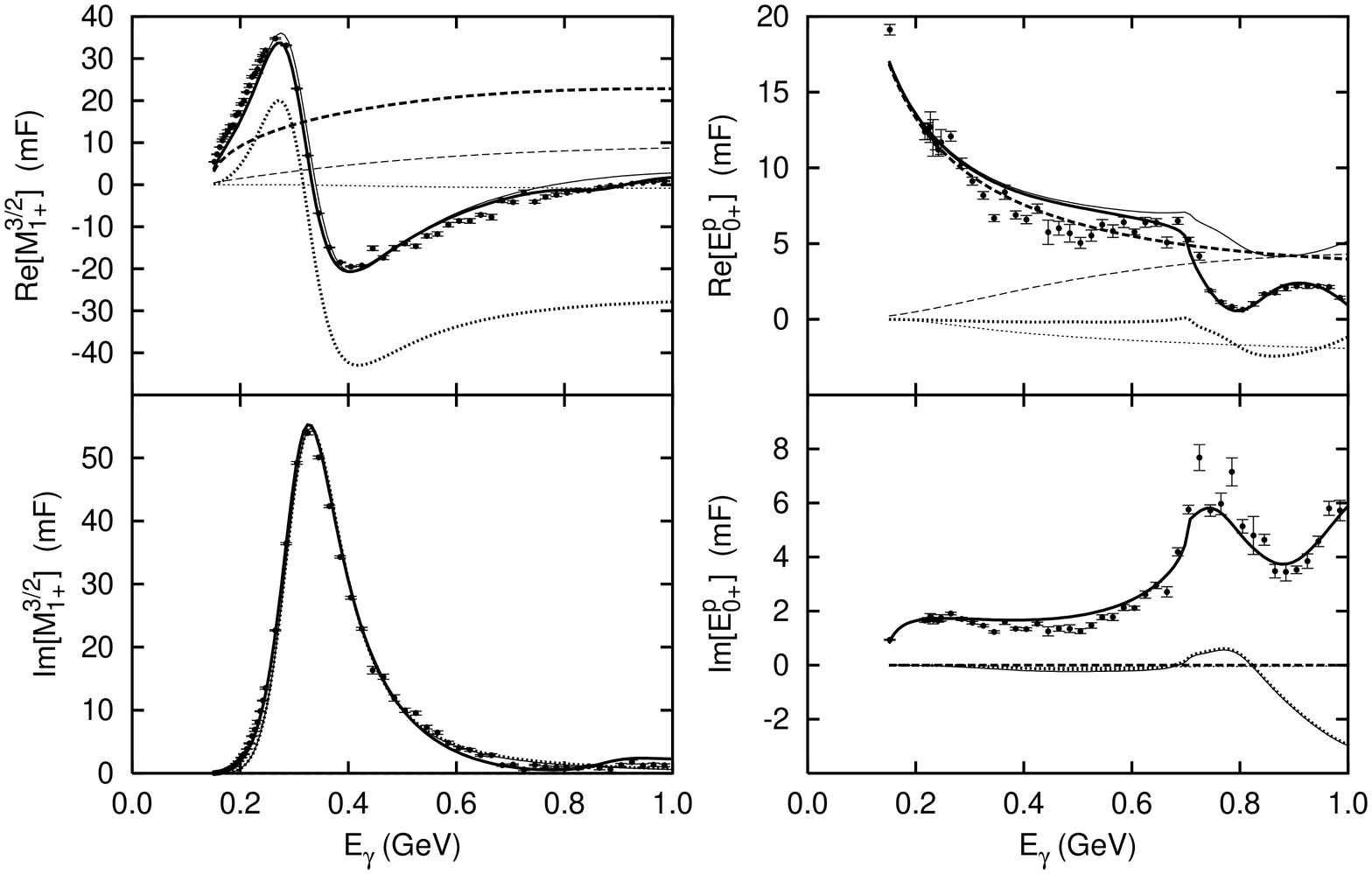}}}
\caption{Examples of various contributions 
to the multipoles. Left panel shows the 
$M_{1+}^{3/2}$ multipole, 
right panel the $E_{0+}^p$ multipole.
Data have been taken 
from Ref. \cite{SAIDdata}. All the curves have been
obtained using 
set \#2 parameters. Thick dashed: Born terms contribution;
thin dashed: vector meson contributions;
thick dotted: direct terms contribution from resonances;
thin dotted: crossed terms contribution from resonances;
thin solid: full calculation without FSI; 
thick solid: full calculation with FSI.} 
\label{fig:multipole_evol}
\end{center}
\end{figure}

If we go through the multipoles in detail, 
it is convenient to start with
$E_{1+}^{3/2}$ and $M_{1+}^{3/2}$ 
(both in Fig. \ref{fig:multipole_delta})
which provide information 
about the most important low-lying nucleon resonance, 
the $\Delta (1232)$.
These multipoles are of great interest at present 
and a lot of experimental effort has been put in 
the study of the $\Delta (1232)$ in the last years 
\cite{Blanpied,Bellini}.
The $M_{1+}^{3/2}$ presents a quite simple 
structure which is very well reproduced by 
all our sets and it is not affected by FSI. 
That is why all sets are quite similar.
Sets \#1, \#4, and \#5 overestimate the multipole 
peaks which will cause 
an overestimation of the cross section 
as will be seen in section \ref{sec:crosssections}.
The situation is much more complicated for the 
$E_{1+}^{3/2}$, where the FSI are critical, 
as can be inferred when we compare data to sets 
with and without FSI and check the strong differences
among them.
For these multipoles, 
data cannot be well reproduced without the inclusion of 
$\delta_{FSI}$.
When the latter is included the multipoles show 
a discontinuity 
at $\sqrt{s}=1.249$ GeV 
due to an abrupt change in SAID phases 
at that energy.

An important quantity  related to 
$\Delta (1232)$ is the E2/M1 Ratio (EMR)
which is related to the deformation 
of the nucleon \cite{Blanpied,Pasc04,Thomas}.
This quantity can be defined as the
$G_E/G_M$ ratio
\begin{equation}
\text{EMR} =\frac{G_E}{G_M}= 
-\frac{\left( M_\Delta - M \right) g_2}
{2\left( M_\Delta + M \right) g_1 
+\left( M_\Delta - M \right)g_2} 
\times 100 \% \: . \label{eq:EMR1}
\end{equation}

We obtain a negative value for this ratio, 
which according to Ref. \cite{Vanderhaeghen}
corresponds to an oblate deformation.
The values from the most reliable fits
(sets \#2 and \#3) are very similar, around $-1.3\%$.
This result compares well to some other analysis:
$-1.45\%$ (K-matrix) \cite{Davidson91};
$-1.42\%$ (ELA) \cite{EMoya};
$-1.43\%$ (ELA) \cite{Vanderhaeghen};
$-1.3\%$ (dynamical model) \cite{Sato};
$-2.09\%$ (dynamical model) \cite{Fuda}.
However,it is quite different from the result recently
obtained by Pascalutsa and Tjon ($\text{EMR}=3.8\%$)
\cite{Pasc04} within a dynamical model.
This ratio is discussed in more detail in  
Ref. \cite{fernandez}.

The multipoles $M_{1-}^p$ and $M_{1-}^n$ are 
closely related to the N(1440) resonance.
If we focus on sets \#2 and \#3, 
when $\delta_{FSI}$ is included the fits look quite 
well except for the real part of the $M_{1-}^n$
(second figure of the left panel in Fig. \ref{fig:multipole_n})
where a serious discrepancy between theory and data 
is found in the $0.2-0.5$ GeV energy range. 
Also,a rather odd behavior in the $M_{1-}^p$ 
is found between $0.3$ and $0.4$ GeV
(see Fig. \ref{fig:multipole_p}), where  
no experimental data are available. 
For these multipoles related to N(1440) resonance, background
and resonant contributions are not well established.
As a consequence, 
the parameters of the P$_{11}$
resonance cannot be well determined.
These multipoles also show the importance of 
FSI in the model in order to determine the properties of the
resonances
because of the large 
discrepancies among fits with and 
without $\delta_{FSI}$.
However, if we focus
on sets \#1 and \#4, FSI do not seem so 
important if the PDG values are used.
Actually,
set \#4  provides better results
than set \#1 except for 
the high energy region of 
Re$M_{1-}^p$ and Re$M_{1-}^n$. 
More research on the properties of this resonance 
(and of its role in nuclear medium) has 
to be done in forthcoming years \cite{Krusche}.

Resonance N(1520) contributes mainly to 
$E_{2-}^{p,n}$ and $M_{2-}^{p,n}$ due to its angular 
momentum and isospin. It also  
contributes sizeably
to other multipoles. The $s$-channel
contributes to $M_{1+}^p$ and its crossed term to
Im$M_{1-}^{3/2}$ as background. 
It also has small
contributions to the background of other multipoles.
Considering set \#2 and multipoles
$E_{2-}^{p,n}$ and $M_{2-}^{p,n}$,
the agreement is excellent except where 
there are few experimental data.
Set \#3 overestimates the peak of the resonance  
in the multipoles and so will do for the cross section.

$E_{0+}^{3/2,p,n}$ multipoles get contributions from 
Born terms and vector mesons mainly.
Resonances N(1535), $\Delta (1620)$, and N(1650)
only contribute in the high energy region, 
but in that region they acquire great importance 
defining the shape of the multipoles.
For example, the cusp peak that shows up 
in Im$E_{0+}^{p,n}$ (Figs. \ref{fig:multipole_p} 
and \ref{fig:multipole_n})
is due to the structure of the 
phenomenological width -- Eq. (\ref{eq:width}) -- 
and to the inclusion of the partial decay width 
$\Gamma_\eta / \Gamma$ in N(1535) resonance.
Multipoles $E_{0+}^{p,n}$ are  well reproduced
by sets \#2 and \#3, except in the high energy region
of Im$E_{0+}^n$. 
The multipole $E_{0+}^{3/2}$ (Fig. \ref{fig:multipole_delta})
is not so well reproduced in the 
intermediate energy region ($0.4-0.8$ GeV), 
with an overestimation of the real part 
and an underestimation of the imaginary part.
This indicates that the prediction of the model is correct
for the absolute values of the multipoles and that
there may be a problem with the phases.

Only one resonance remains, $\Delta (1700)$, 
which is associated mainly to multipoles
$E_{2-}^{3/2}$ and $M_{2-}^{3/2}$.
As one can see in Fig. \ref{fig:multipole_delta},
when enough data points are available the 
fits are good, yet
the large ambiguities in the 
mass and width of this resonance
make somewhat unreliable the determination of 
its coupling constants and its contribution to the
observables (see Table \ref{tab:couplings}).
Further research on the properties of this 
resonance is necessary.

In figure \ref{fig:multipole_evol} we show 
two examples of the various contributions 
to the multipoles using the coupling 
constants of set \#2. It is clear that, without 
FSI, the Born terms and vector mesons 
do not contribute to the imaginary part of the 
multipoles and represent a background 
--- it has to be noticed that when the FSI are included, 
they do contribute to both real and imaginary 
parts of the multipole.
Left panel shows the multipole $M_{1+}^{3/2}$, 
whose main contribution is the $\Delta (1232)$. 
In this multipole, 
FSI are not important and curves with and without 
SAID phases differ little. 
Thus, the phenomenological width 
included is enough to describe accurately the 
multipole and its structure is quite simple.
However, the situation is different for the multipole 
$E_{0+}^p$ which presents a more complex structure
because its dominant contribution comes from Born terms
and vector mesons. In the absence of FSI 
the imaginary part of this multipole
is practically zero up to $0.8$ GeV.
While inclusion of FSI makes Born and vector mesons 
contribute to the imaginary part too, improving agreement with data.

We have not considered spin-5/2 resonances in the model.
This will be required in order to extend the model to 
multipoles of higher
angular momentum. For the energy range considered 
here, their contribution is expected not to be important, 
although their contribution to the background
could improve the agreement with data.

\subsection{Results at threshold energy}

\begin{table}
\caption{Reduced cross section at threshold
$\frac{q^*}{k^*}\frac{d \sigma}{d \Omega}$ in $\mu$b/sr. 
Experimental data 
have been taken from reference \cite{EMoya}.} \label{tab:redxsec}
\begin{tabular}{lccccccc}

\hline

Sets &\#1&\#2&\#3&\#4&\#5&\#6& Experiment \\

\hline \hline

$\gamma p \to p \pi^0$
&0.0984&0.0998&0.0998&0.0949&0.1023&0.1020& $0.094 \pm 0.017$\\
$\gamma n \to n \pi^0$
&0.0046&0.0045&0.0045&0.0049&0.0044&0.0044& \\
$\gamma n \to p \pi^-$
&18.92&18.93&18.93&18.91&18.95&18.95& $20.4 \pm 0.7$ \\
& & & & & & & $20.0 \pm 0.3$ \\ 
& & & & & & & $19.7 \pm 1.4$ \\
$\gamma p \to n \pi^+$
&14.51&14.50&14.50&14.52&14.48&14.49& $15.4 \pm 0.5$ \\
& & & & & & & $15.6 \pm 0.5$ \\

\hline

\end{tabular}
\end{table}

Special attention has to be paid to the behavior of 
the model at low/threshold energy, because
cross sections and multipoles are well predicted 
by Low Energy Theorems (LET) \cite{Naus} and Chiral 
Perturbation Theory  (ChPT) \cite{Bernard95,Bernard92}.
Owing to the change in the spin-3/2 coupling scheme, 
the threshold energy results change substantially 
when compared to previous works. 
In particular, in Ref. \cite{EMoya},
using the off-shell formalism, 
it was found that the contributions from
resonances, direct and 
crossed terms, were of great importance 
in order to explain the reduced cross section 
at threshold and the low energy behavior of 
the cross section. These contributions were specially 
important in neutral processes, mainly 
because of $\Delta (1232)$, and were almost the 
total contribution to the $n \pi^0$ 
production channel \cite{cefera}.
However, in the present calculation we obtain
a zero contribution to the 
reduced cross section at threshold from both  
direct and crossed resonance terms.
The reason for such a change is the spin-3/2
coupling scheme used in the present article,
which has no spurious spin-1/2 sector.
The reduced differential cross section 
at threshold is 
proportional to the $E_{0+}$ multipole 
\cite{Bernard92} which is a spin-1/2 multipole. 
Thus, at threshold, any contribution of the
direct channel from spin-3/2
resonances is a contamination which unveils a pathology in the model.
This is the case of models based upon the traditional 
spin-3/2 formalism
explained in section \ref{sec:trad},
as the one used in Ref. \cite{EMoya}.
This result is 
independent on the phenomenology 
of the decay width and on the form factors.
Therefore, we conclude that, at threshold, 
only Born 
terms (Fig. \ref{fig-diag1}) 
and vector mesons 
(Fig. \ref{fig-diag2}, diagram $(e)$)
contribute, as the spin-1/2 resonances 
are at much higher energy.
In Table \ref{tab:redxsec} we present results 
for the reduced cross section at threshold 
for the four different processes and we find 
a good agreement with experimental values.
Tiny differencies are found with various parameter sets
(\#1 to \#6) that use different cutoff $\Lambda$ and small
variations in the vector meson parameters.

\subsection{Differential cross sections and asymmetries}

\begin{figure}
\begin{center}
\rotatebox{-90}{\scalebox{0.52}[0.52]
{\includegraphics{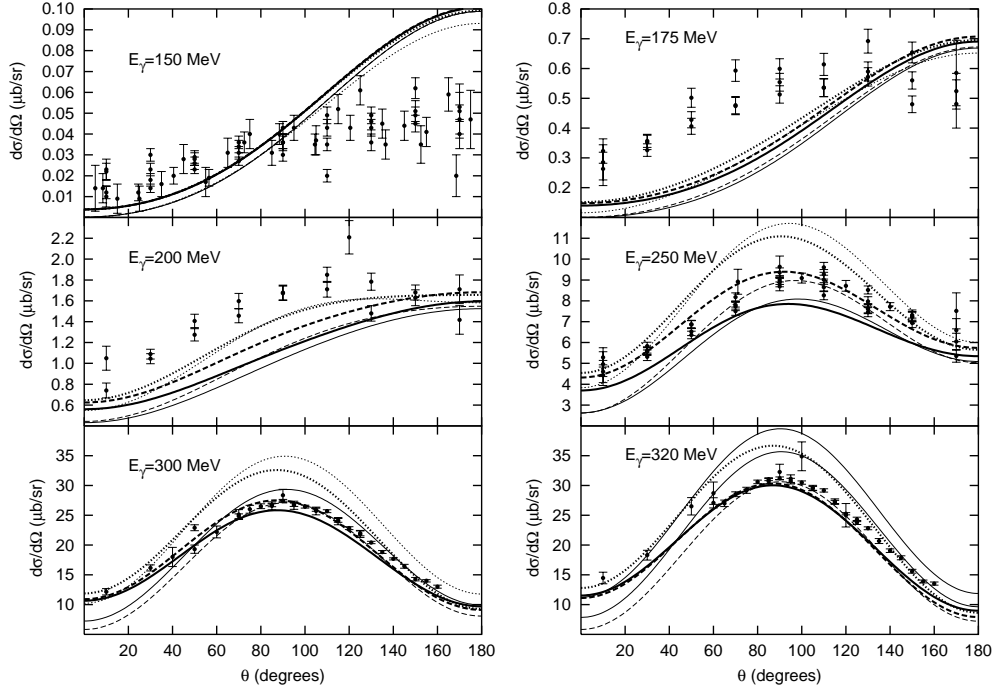}}}
\end{center}
\caption{Differential cross section in $\mu$b/sr
of the $\gamma p \to \pi^0 p$ reaction for
different photon energies in the laboratory frame. 
Pion scattering angle $\theta$ in the center of mass 
reference system.
Experimental data  have been taken
from reference \cite{SAIDdata} and are 
within the range $E_\gamma \pm 1$ MeV.
Curve conventions: 
thick dotted set \#1; 
thick solid  set \#2; 
thick dashed set \#3; 
thin dotted  set \#4; 
thin solid   set \#5; 
thin dashed  set \#6.} 
\label{fig:diffpi0p}
\end{figure}

\begin{figure}
\begin{center}
\rotatebox{-90}{\scalebox{0.52}[0.52]
{\includegraphics{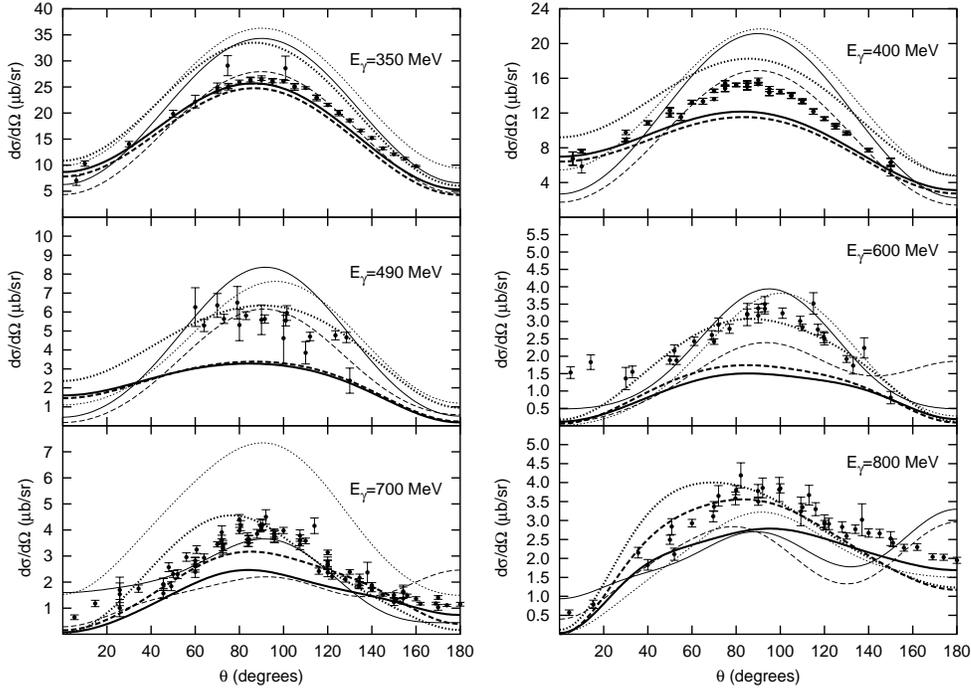}}}
\end{center}
\caption{Same as in Fig. \ref{fig:diffpi0p}. 
Experimental data are 
within the range $E_\gamma \pm 3$ MeV.} 
\label{fig:diffpi0p2}
\end{figure}

In this section we show results for the differential 
cross sections together with results for
five asymmetries: 
the recoil nucleon polarization, $P$; 
the polarized target asymmetry, $T$; 
the polarized beam asymmetry, $\Sigma$;
and the double polarization parameters $G$ and $H$. 
Details on the  
definition of these quantities 
can be found in Refs. \cite{Walker,Arndt90-1,Bussey}.
The asymmetries are of great interest in the search of 
\textit{missing} resonances which do not show up
so clearly in other observables \cite{Dutta}. 
The formulae which relate the amplitudes 
with the asymmetries 
will be presented in forthcoming paragraphs.
We provide the reader with a wide sample of figures 
in order to have a broad outlook of the model 
compared to data whenever available.

\begin{figure}
\begin{center}
\rotatebox{-90}{\scalebox{0.52}[0.52]
{\includegraphics{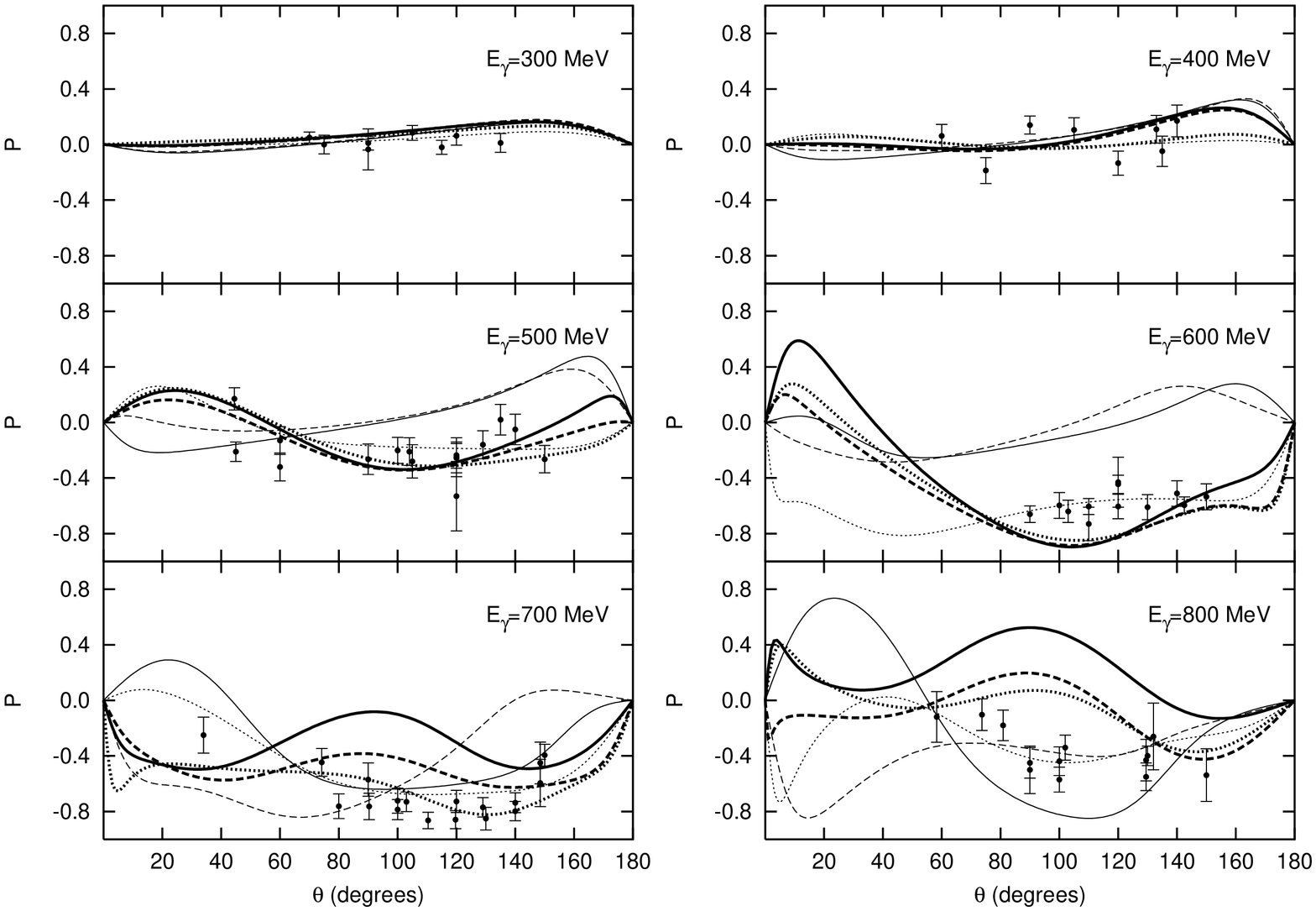}}}
\end{center}
\caption{Recoil nucleon polarization of the 
$\gamma p \to \pi^0 p$. Photon energy in the 
laboratory frame. Pion angle in the 
center of mass reference system. 
Experimental data are
within the range $E_\gamma \pm 3$ MeV.
Conventions for the curves are as in Fig. 
\ref{fig:diffpi0p}.}
\label{fig:Ppi0p}
\end{figure}

\begin{figure}
\begin{center}
\rotatebox{-90}{\scalebox{0.52}[0.52]
{\includegraphics{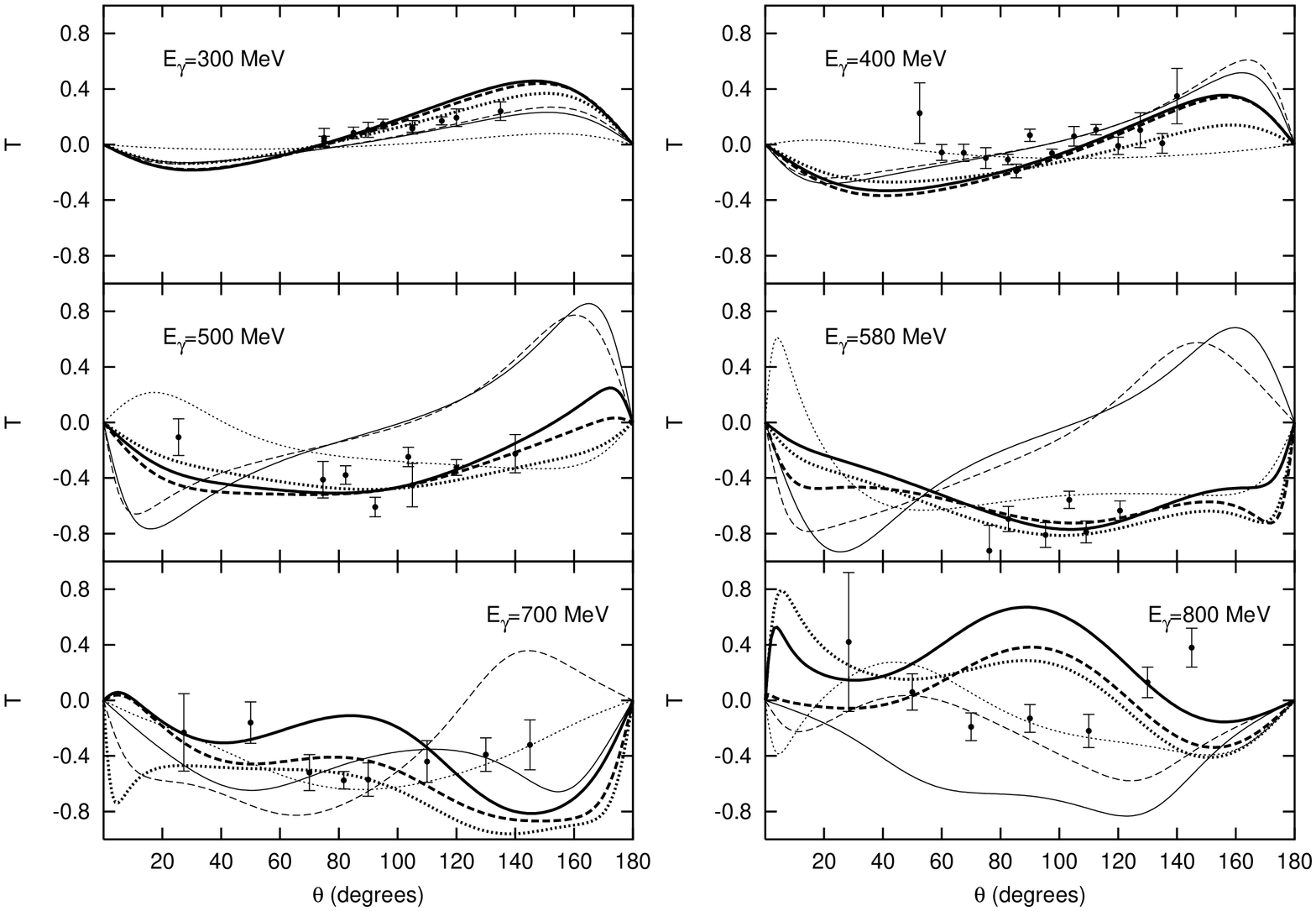}}}
\end{center}
\caption{Polarized target asymmetry of the 
$\gamma p \to \pi^0 p$ reaction.
Experimental data are 
within the range $E_\gamma \pm 3$ MeV.
Same conventions as in Fig. \ref{fig:Ppi0p}.} 
\label{fig:Tpi0p}
\end{figure}

The FSI treatment described in section \ref{sec:emcoupling}
has been applied only to the $\gamma p \to \pi^0 p$ process.
For the other three pion production processes, no FSI 
phases have been included because we have no means to determine 
them from the data available. 
We calculate the observables for these processes for the six sets
of coupling constants obtained by fitting $\gamma p \to \pi^0 p$
multipoles, given in Tables \ref{tab:fits} and \ref{tab:couplings}.
Thus, these calculations have no adjustable parameters.
As we shall see in what follows an overall good 
agreement with data has been found.
As the energy increases, differences among 
the curves obtained with the
different sets of parameters show up more
Data favour the sets of coupling 
constants obtained using FSI.

\begin{figure}
\begin{center}
\rotatebox{-90}{\scalebox{0.52}[0.52]
{\includegraphics{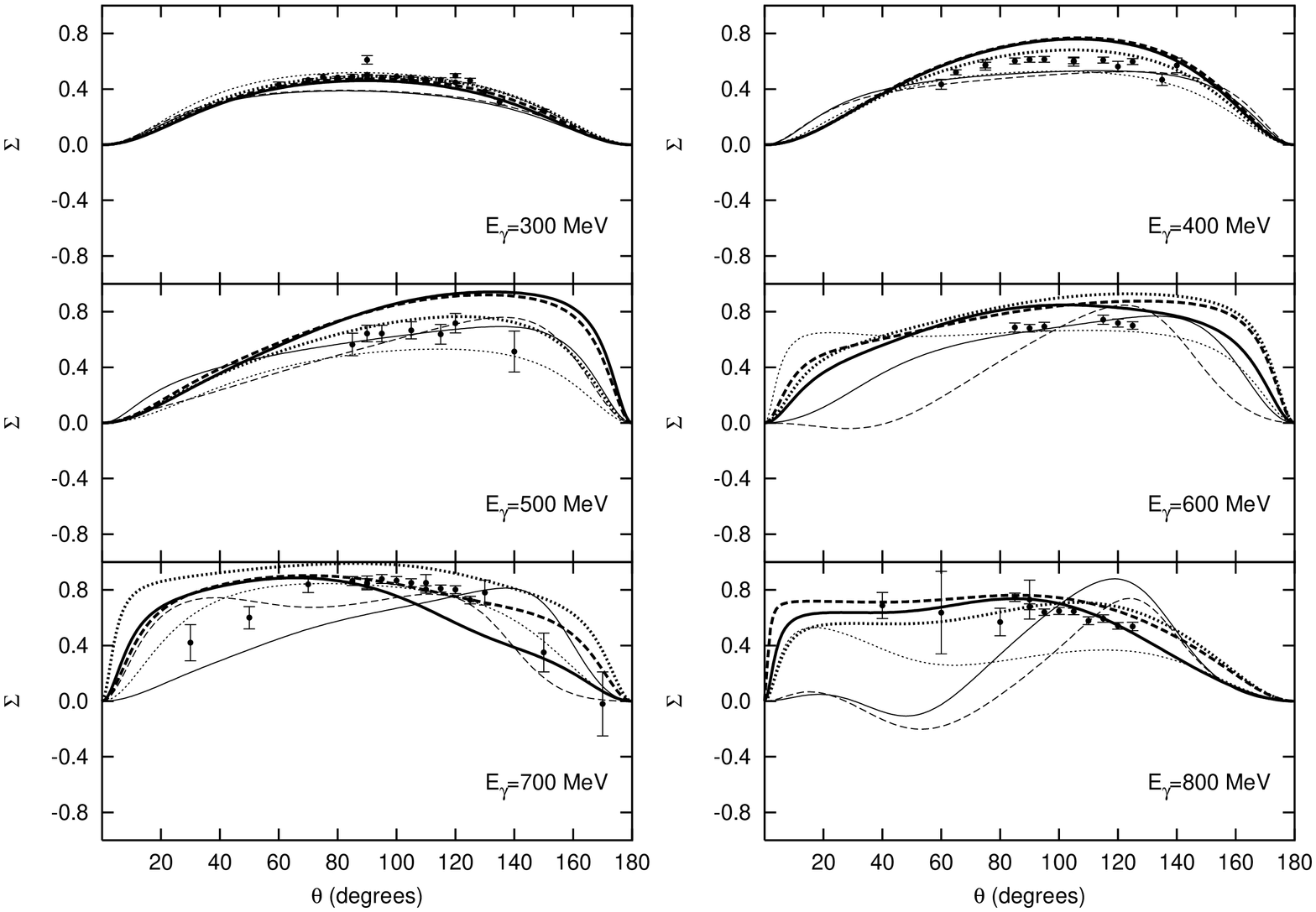}}}
\end{center}
\caption{Photon beam asymmetry of the 
$\gamma p \to \pi^0 p$ reaction.
Experimental data are 
within the range $E_\gamma \pm 3$ MeV.
Same conventions as in Fig. \ref{fig:Ppi0p}.} 
\label{fig:Spi0p}
\end{figure}

\begin{figure}
\begin{center}
\rotatebox{-90}{\scalebox{0.52}[0.52]
{\includegraphics{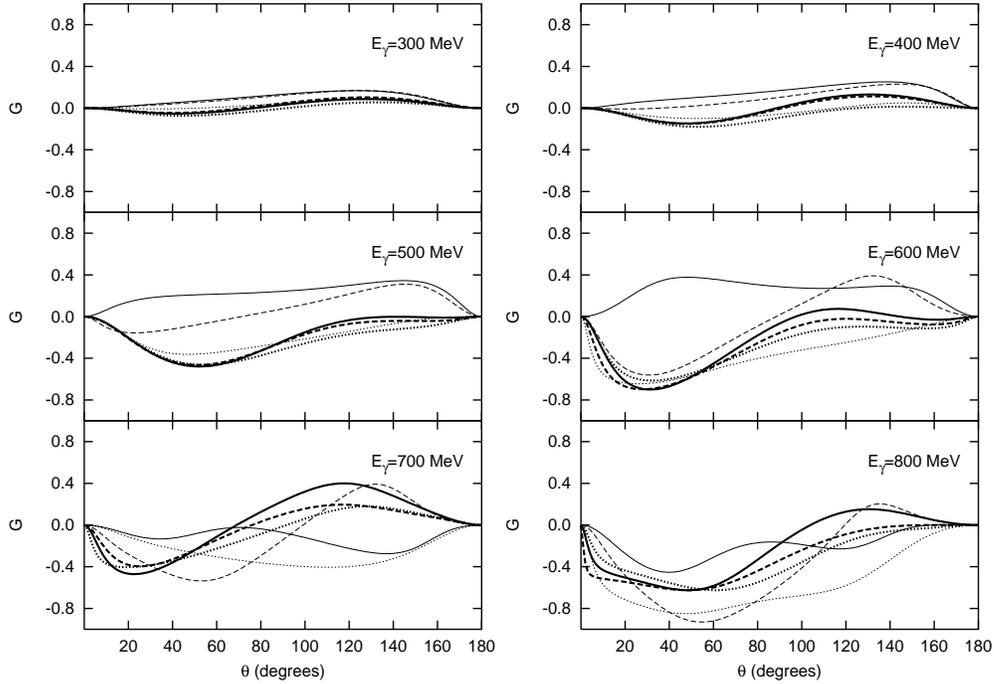}}}
\end{center}
\caption{$G$ asymmetry of the 
$\gamma p \to \pi^0 p$ reaction. 
Same conventions as in Fig. \ref{fig:Ppi0p}.} 
\label{fig:Gpi0p}
\end{figure}

\subsubsection{$\gamma p \to \pi^0 p$}

Let us first consider the process $\gamma p \to \pi^0 p$,
for which the experimental database  
has been largely increased in the last ten years 
mainly thanks to the experimental
programs developed at Mainz (MAMI) and Brookhaven (LEGS).
For this process 
the amount of experimental information is much larger 
than for any other pion photoproduction process. 
Even so,
the database on asymmetries is not yet large enough and more
measurements are needed in order to fill in the existing 
gaps. Figures \ref{fig:diffpi0p} and 
\ref{fig:diffpi0p2} show
theoretical curves for the differential cross sections 
compared to experimental data.
Differential cross sections have been calculated using 
equations from section \ref{sec:kinematics} and
amplitudes from
appendix \ref{sec:invariantamplitudes}.

\begin{figure}
\begin{center}
\rotatebox{-90}{\scalebox{0.52}[0.52]
{\includegraphics{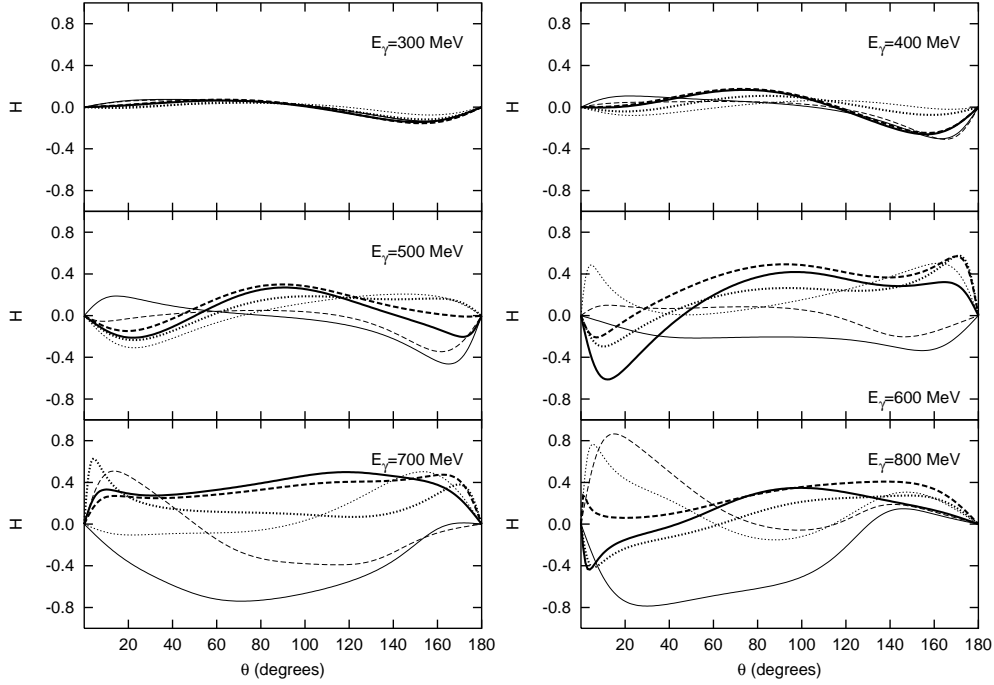}}}
\end{center}
\caption{$H$ asymmetry of the 
$\gamma p \to \pi^0 p$ reaction.
Same conventions as in Fig. \ref{fig:Ppi0p}.} 
\label{fig:Hpi0p}
\end{figure}

\begin{figure}
\begin{center}
\rotatebox{-90}{\scalebox{0.52}[0.52]
{\includegraphics{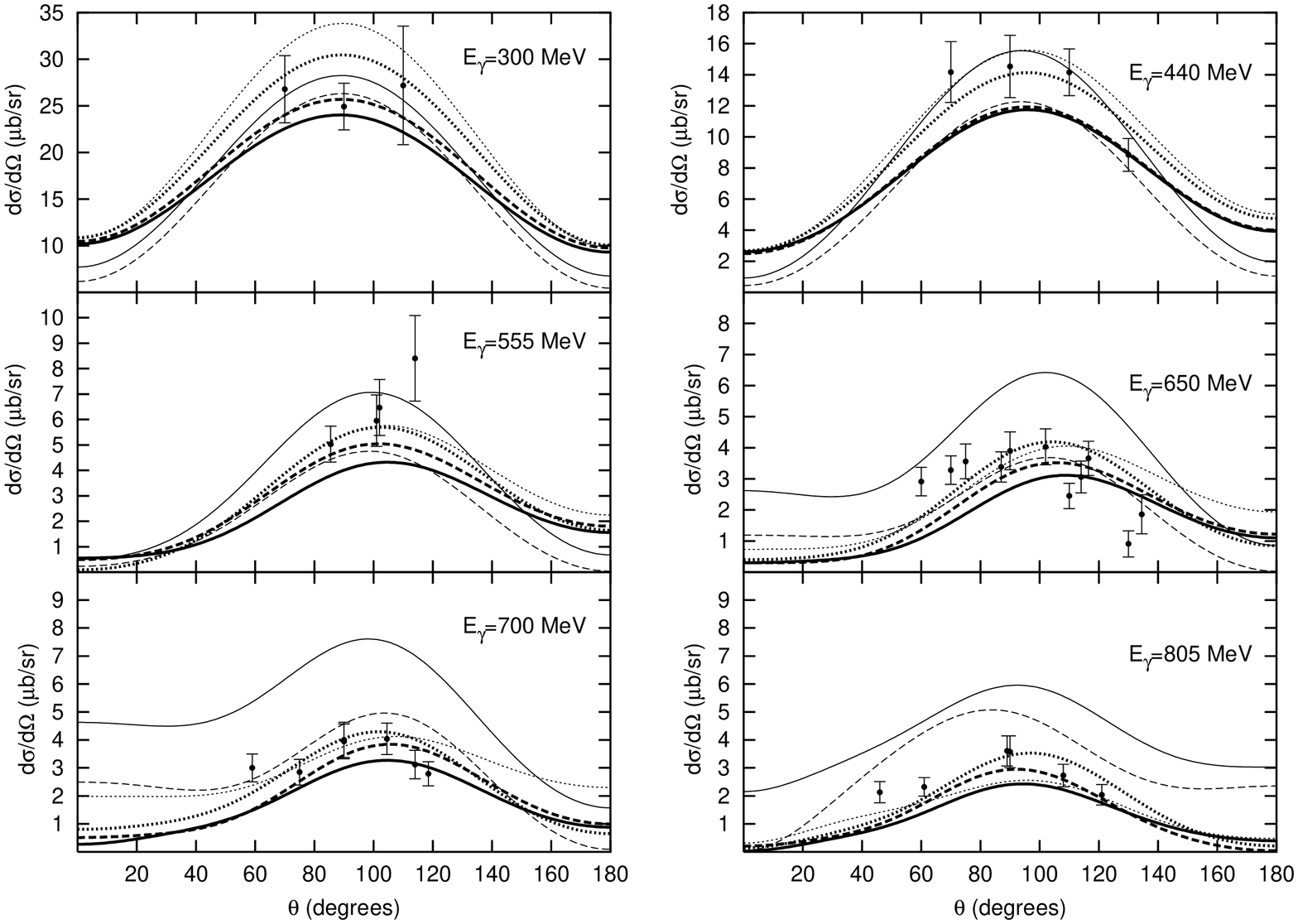}}}
\end{center}
\caption{Differential cross section of the 
$\gamma n \to \pi^0 n$ reaction.
Experimental data are 
within the range $E_\gamma \pm 5$ MeV.
Same conventions as in Fig. \ref{fig:diffpi0p}.}
\label{fig:diffpi0n}
\end{figure}

\begin{figure}
\begin{center}
\rotatebox{-90}{\scalebox{0.52}[0.52]
{\includegraphics{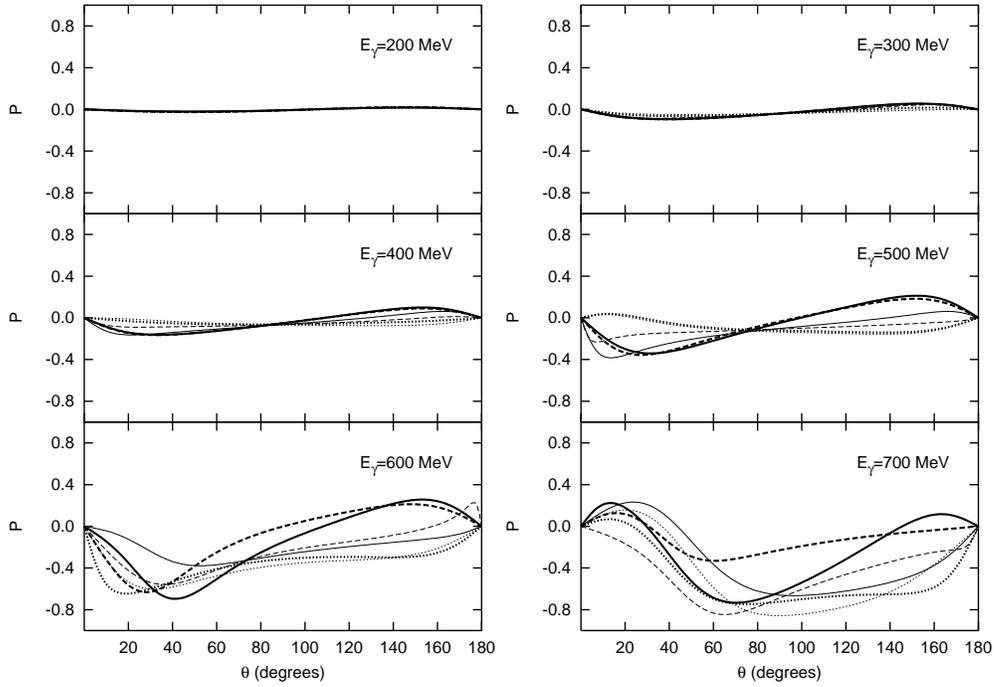}}}
\end{center}
\caption{Recoil nucleon polarization of the 
$\gamma n \to \pi^0 n$.
Same conventions as in Fig. \ref{fig:Ppi0p}.} 
\label{fig:Ppi0n}
\end{figure}

\begin{figure}
\begin{center}
\rotatebox{-90}{\scalebox{0.52}[0.52]
{\includegraphics{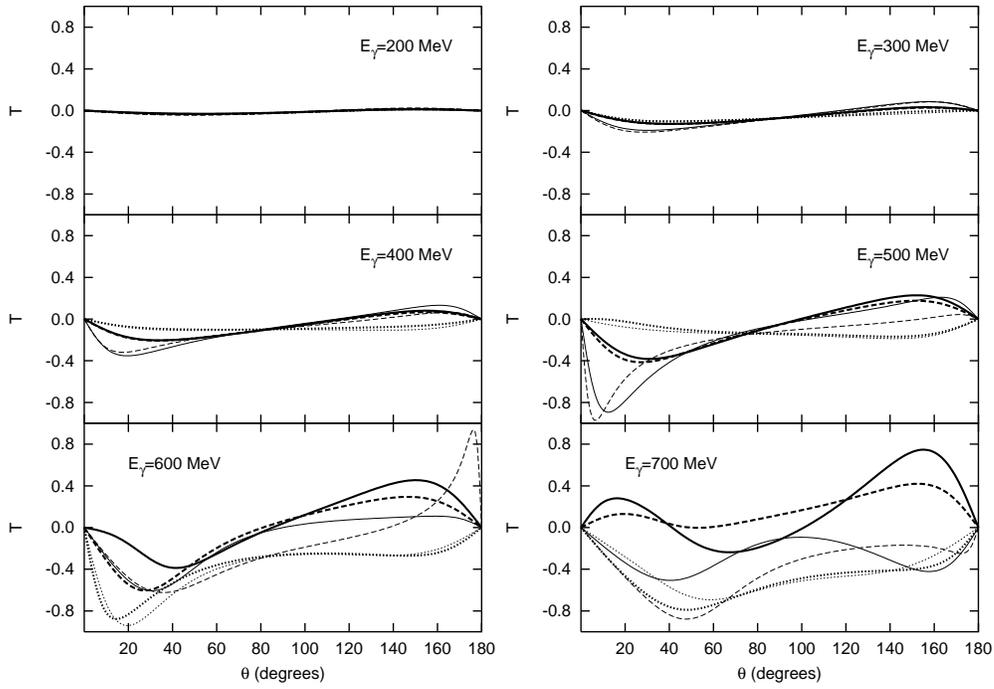}}}
\end{center}
\caption{Polarized target asymmetry of the 
$\gamma n \to \pi^0 n$ reaction.
Same conventions as in Fig. \ref{fig:Ppi0p}.} 
\label{fig:Tpi0n}
\end{figure}

\begin{figure}
\begin{center}
\rotatebox{-90}{\scalebox{0.52}[0.52]
{\includegraphics{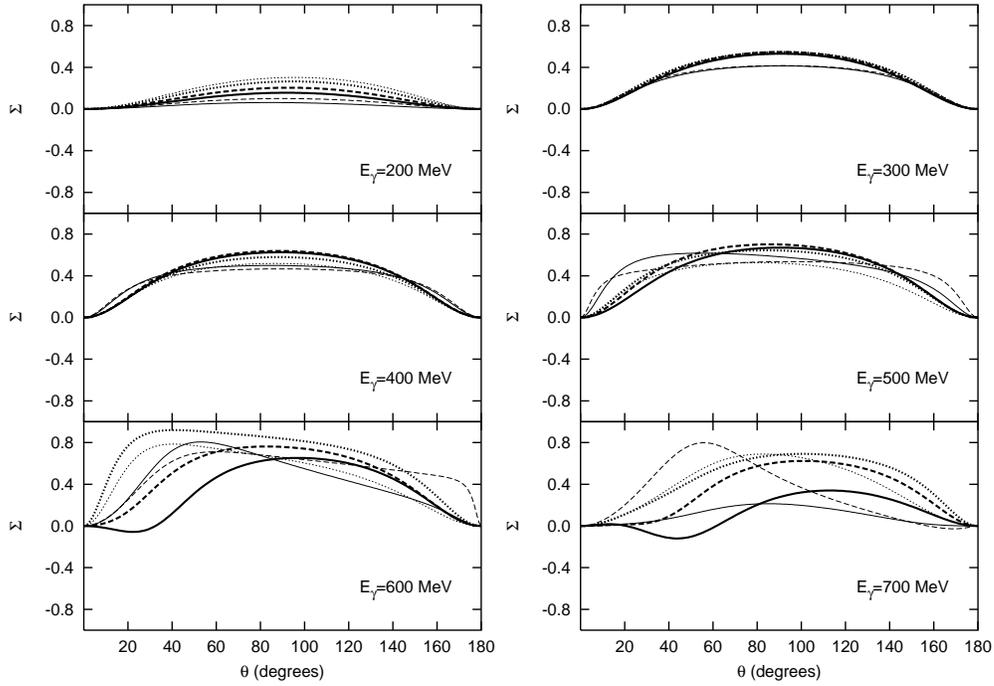}}}
\end{center}
\caption{Photon beam asymmetry of the 
$\gamma n \to \pi^0 n$ reaction.
Same conventions as in Fig. \ref{fig:Ppi0p}.} 
\label{fig:Spi0n}
\end{figure}

Because of parity, among the eight helicity amplitudes, 
only four of them are independent
\begin{eqnarray}
H_1    &=& \mathcal{A}_{1/2,-1/2,1}
= - \mathcal{A}_{-1/2,1/2,-1} \: , \\
H_2 &=& \mathcal{A}_{-1/2,-1/2,1}
= \mathcal{A}_{1/2,1/2,-1} \: ,\\
H_3 &=& \mathcal{A}_{1/2,1/2,1}
= \mathcal{A}_{-1/2,-1/2,-1} \: , \\
H_4    &=& \mathcal{A}_{-1/2,1/2,1}
= - \mathcal{A}_{1/2,-1/2,-1} \: .
\end{eqnarray}

In terms of these four independent helicity amplitudes
(see section \ref{sec:kinematics}
for $\mathcal{A}_{\lambda_1,\lambda_2,\lambda_\gamma}$ definition),
it is possible 
to define all the physical observables \cite{Arndt90-2}. In particular, 
the five asymmetries previously mentioned.

\begin{figure}
\begin{center}
\rotatebox{-90}{\scalebox{0.52}[0.52]
{\includegraphics{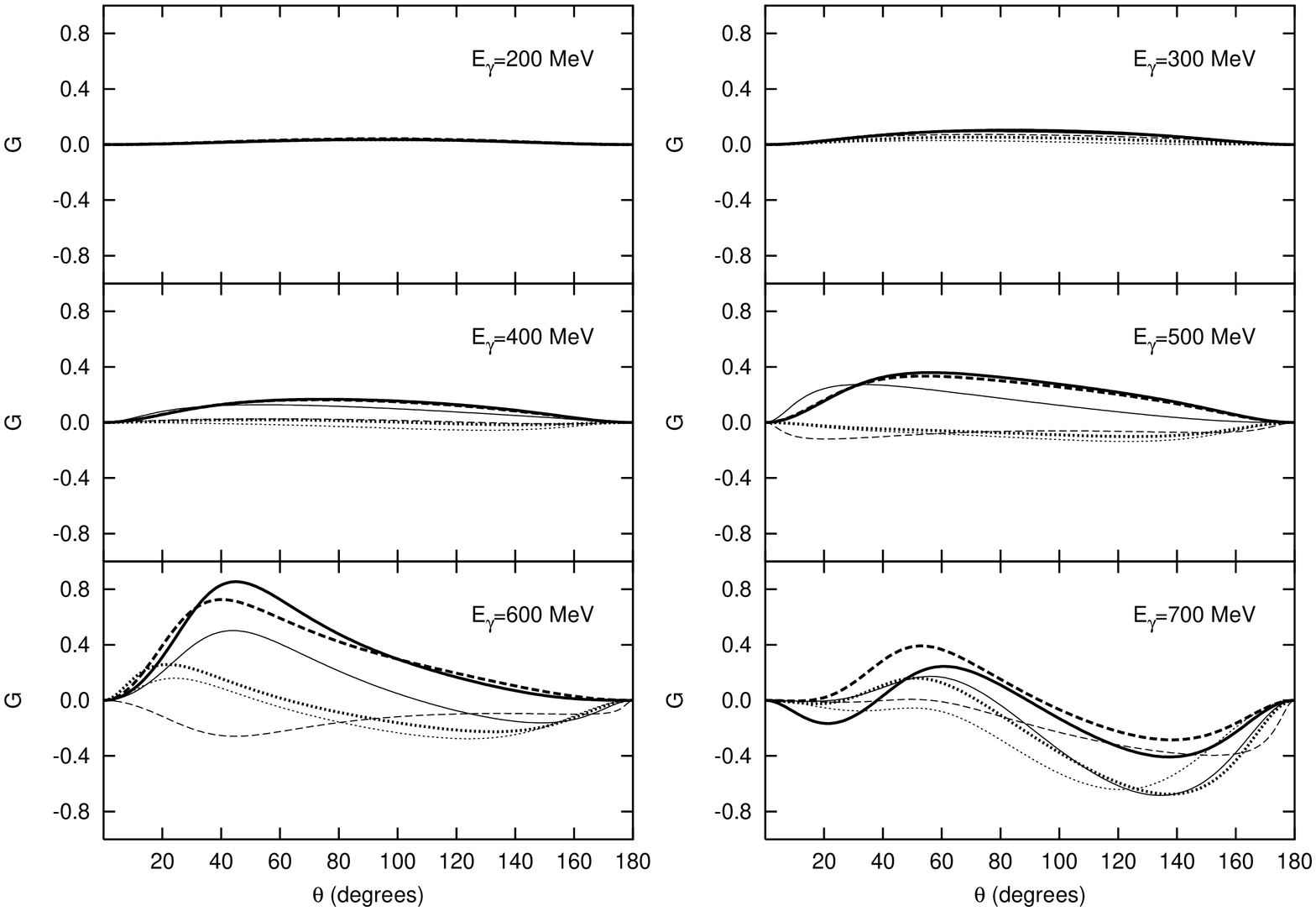}}}
\end{center}
\caption{$G$ asymmetry of the 
$\gamma n \to \pi^0 n$ reaction.
Same conventions as in Fig. \ref{fig:Ppi0p}.} 
\label{fig:Gpi0n}
\end{figure}

Focusing on sets with $\delta_{FSI}$ phases,
the fits 
are qualitatively good
in the whole energy region, and even quantitatively
so in the range 
$250-400$ MeV. Asymmetries are 
well predicted in almost the 
whole energy range.

\begin{figure}
\begin{center}
\rotatebox{-90}{\scalebox{0.52}[0.52]
{\includegraphics{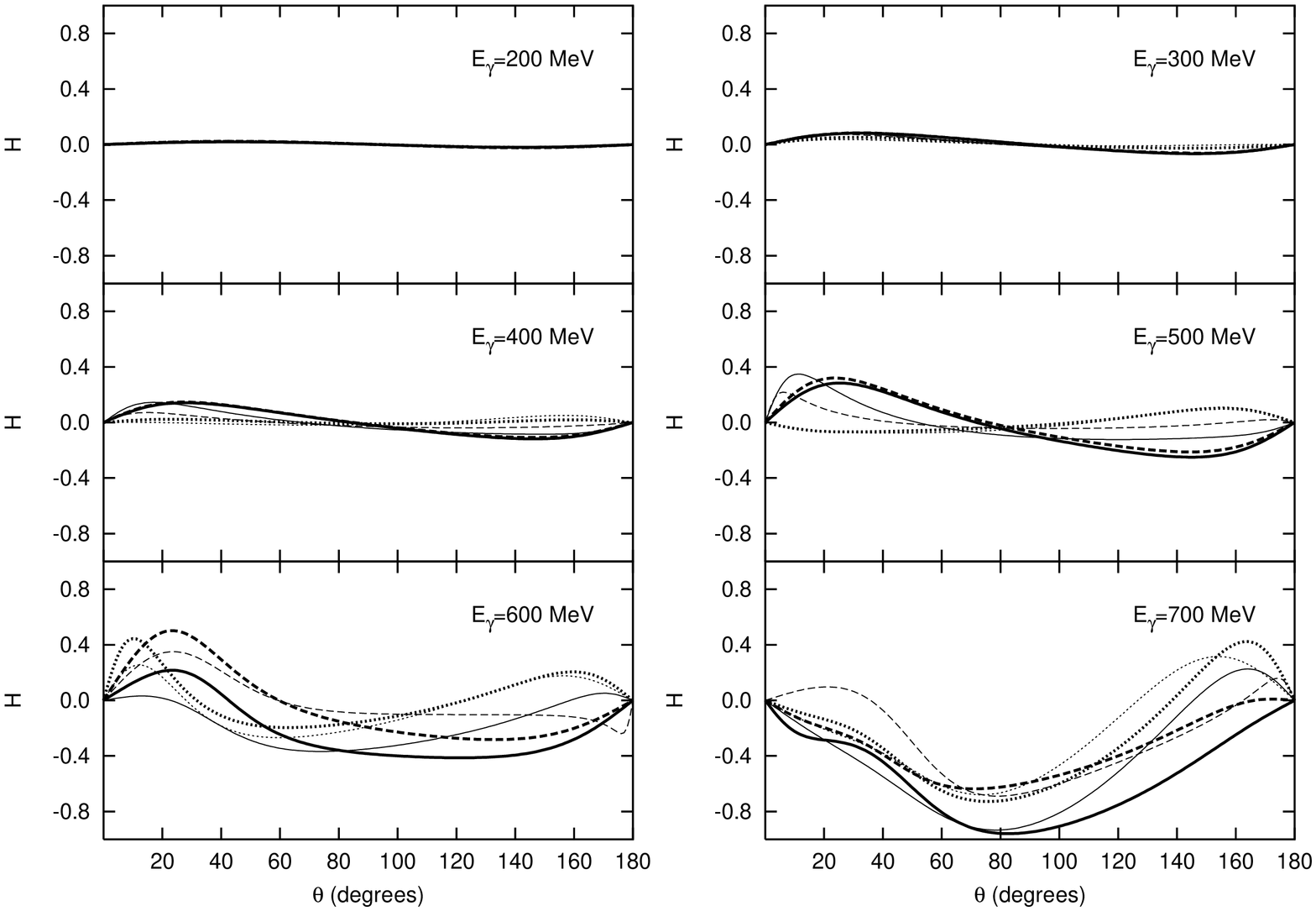}}}
\end{center}
\caption{$H$ asymmetry of the 
$\gamma n \to \pi^0 n$ reaction.
Same conventions as in Fig. \ref{fig:Ppi0p}.} 
\label{fig:Hpi0n}
\end{figure}

In Fig. \ref{fig:Ppi0p} we provide 
recoil nucleon polarization asymmetries ($P$) defined by
\begin{equation}
\sigma \left( \theta \right) P = 
- \frac{1}{64 \pi^2 s^*}\frac{k^*}{E_\gamma^*}
\text{Im}\left[ H_2 \bar{H}_4    
+ H_1 \bar{H}_3 \right] \: ,\label{Pasymmetry}
\end{equation}
where the bar over the helicity amplitudes $H_j$ stands for 
complex conjugate and
$\sigma \left( \theta \right)$ for the differential cross section 
given by Eq. (\ref{eq:diff}). Up to $600$ MeV, data are well
reproduced by sets with FSI. Above this energy, data are reproduced
qualitatively but not quantitatively.

\begin{figure}
\begin{center}
\rotatebox{-90}{\scalebox{0.52}[0.52]
{\includegraphics{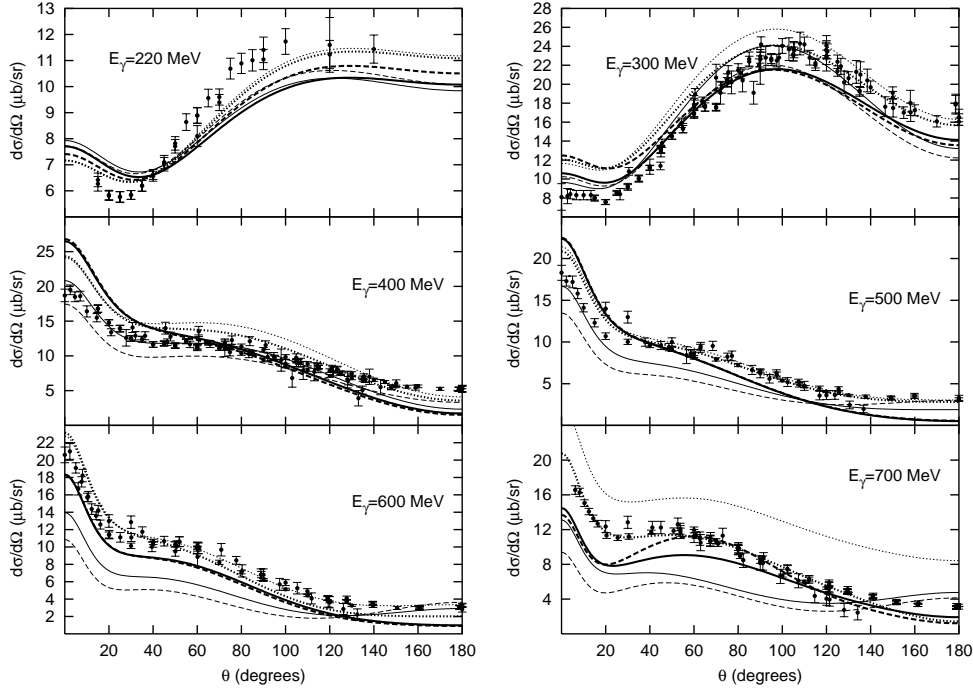}}}
\end{center}
\caption{Differential cross section of the 
$\gamma p \to \pi^+ n$ reaction.
Experimental data are 
within the range $E_\gamma \pm 5$ MeV.
Same conventions as in Fig. \ref{fig:diffpi0p}.} 
\label{fig:diffpipn}
\end{figure}

\begin{figure}
\begin{center}
\rotatebox{-90}{\scalebox{0.52}[0.52]
{\includegraphics{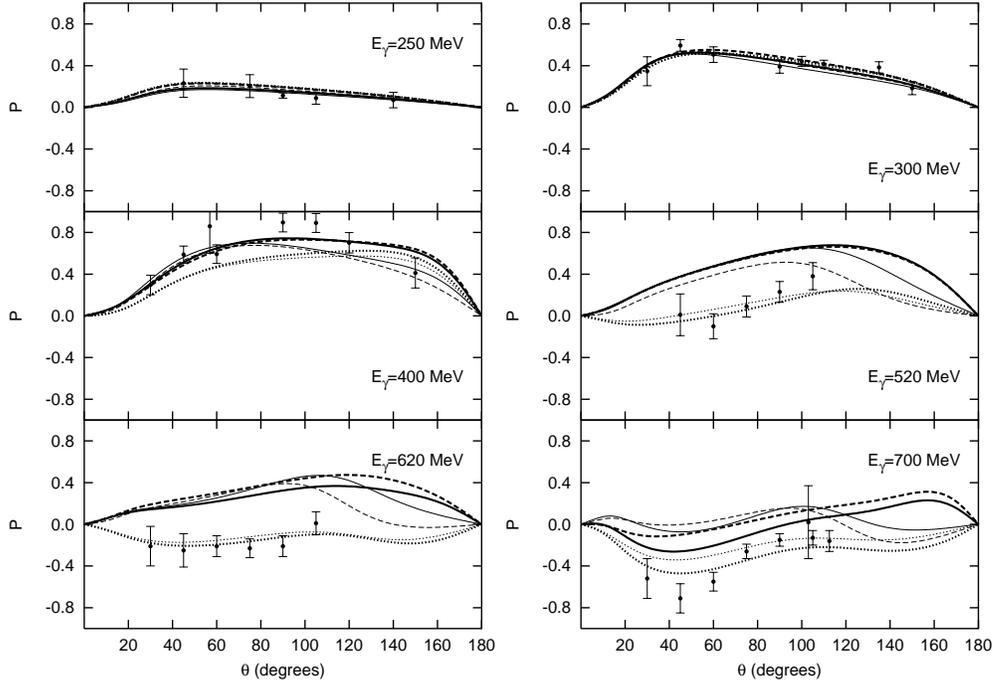}}}
\end{center}
\caption{Recoil nucleon polarization of the 
$\gamma p \to \pi^+ n$ reaction.
Experimental data are 
within the range $E_\gamma \pm 3$ MeV.
Same conventions as in Fig. \ref{fig:Ppi0p}.} 
\label{fig:Ppipn}
\end{figure}

In Fig. \ref{fig:Tpi0p} we present 
polarized target asymmetry ($T$) given by equation
\begin{equation}
\sigma \left( \theta \right) T 
= \frac{1}{64 \pi^2 s^*}\frac{k^*}{E_\gamma^*}
\text{Im}\left[ H_2 \bar{H}_1    
+ H_4 \bar{H}_3 \right] \: .\label{Tasymmetry}
\end{equation}
Up to $400$ MeV the six curves are very similar.
For $500$ and $580$ MeV the sets with phases provide good results 
and the sets without phases do not.
The high energy region ($700$ and $800$ MeV) is not 
well reproduced in general.

Polarized beam asymmetry ($\Sigma$) is well predicted 
in the whole energy range by sets with FSI (Fig. \ref{fig:Spi0p}).
Even sets without FSI provide good results except 
in the very high energy region ($800$ MeV).
Helicity amplitudes are related to $\Sigma$ through
\begin{equation}
\sigma \left( \theta \right) \Sigma 
= \frac{1}{64 \pi^2 s^*}\frac{k^*}{E_\gamma^*}
 \text{Re}\left[ H_2 \bar{H}_3 
- H_1 \bar{H}_4 \right] \: .\label{Sasymmetry}
\end{equation}

In short, compared to data, good agreement is obtained 
for energies below $800$ MeV. Beyond that energy some observables
(v.g. $\Sigma$) are also reasonably well described

In the energy region considered here there are no experimental data
on the other two asymmetries $G$ and $H$.
These asymmetries are expressed in terms of 
helicity amplitudes by means of the following equations
\begin{equation}
\sigma \left( \theta \right) G 
= - \frac{1}{64 \pi^2 s^*}\frac{k^*}{E_\gamma^*}
\text{Im}\left[ H_2 \bar{H}_3 
+ H_1 \bar{H}_4 \right] \: ,\label{Gasymmetry}
\end{equation}

\begin{equation}
\sigma \left( \theta \right) H 
= - \frac{1}{64 \pi^2 s^*}\frac{k^*}{E_\gamma^*}
\text{Im}\left[ H_2 \bar{H}_4    
+ H_3 \bar{H}_1 \right] \: . \label{Hasymmetry}
\end{equation}
We have also calculated these asymmetries and our results are presented
in Figs. \ref{fig:Gpi0p} and \ref{fig:Hpi0p}.

\begin{figure}
\begin{center}
\rotatebox{-90}{\scalebox{0.52}[0.52]
{\includegraphics{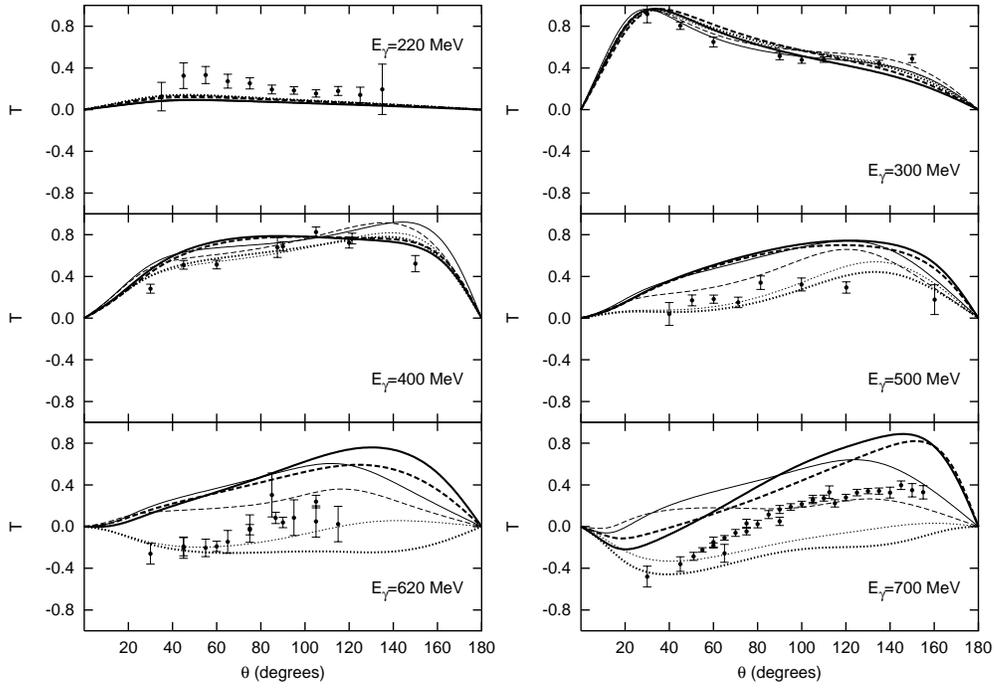}}}
\end{center}
\caption{Polarized target asymmetry of the 
$\gamma p \to \pi^+ n$ reaction.
Experimental data are 
within the range $E_\gamma \pm 4$ MeV.
Same conventions as in Fig. \ref{fig:Ppi0p}.} 
\label{fig:Tpipn}
\end{figure}

\subsubsection{$\gamma n \to \pi^0 n$}

The 
situation for the $\gamma n \to \pi^0 n$ 
process is quite different from the previous case. 
The amount of experimental information is very small:
No asymmetry data are available and the differential cross
section data are scant. 
In figure \ref{fig:diffpi0n} we show 
differential cross sections and in figures
\ref{fig:Ppi0n}, \ref{fig:Tpi0n}, 
\ref{fig:Spi0n}, \ref{fig:Gpi0n}, 
and \ref{fig:Hpi0n} the predicted 
asymmetries 
($P$, $T$, $\Sigma$, $G$, and $H$ respectively)
obtained with sets of parameters.
There is 
a reasonable
agreement with data, and sets \#1 and \#4 
(PDG values) provide the best results globally.

\subsubsection{Charged pion production}

\begin{figure}
\begin{center}
\rotatebox{-90}{\scalebox{0.52}[0.52]
{\includegraphics{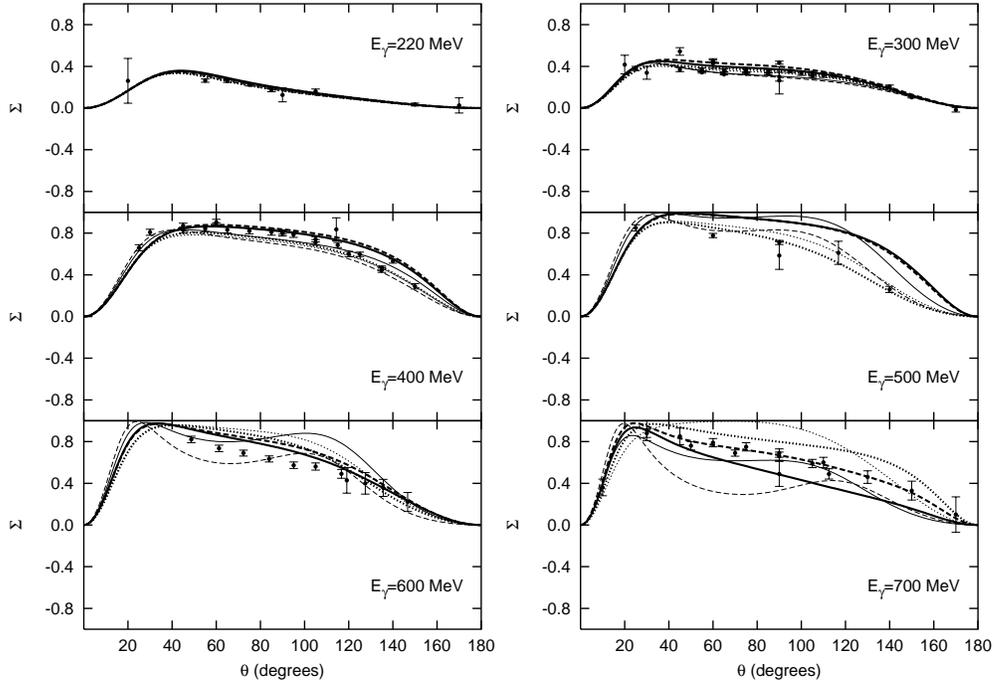}}}
\end{center}
\caption{Photon beam asymmetry of the 
$\gamma p \to \pi^+ n$ reaction.
Experimental data are 
within the range $E_\gamma \pm 4$ MeV.
Same conventions as in Fig. \ref{fig:Ppi0p}.} 
\label{fig:Spipn}
\end{figure}

\begin{figure}
\begin{center}
\rotatebox{-90}{\scalebox{0.52}[0.52]
{\includegraphics{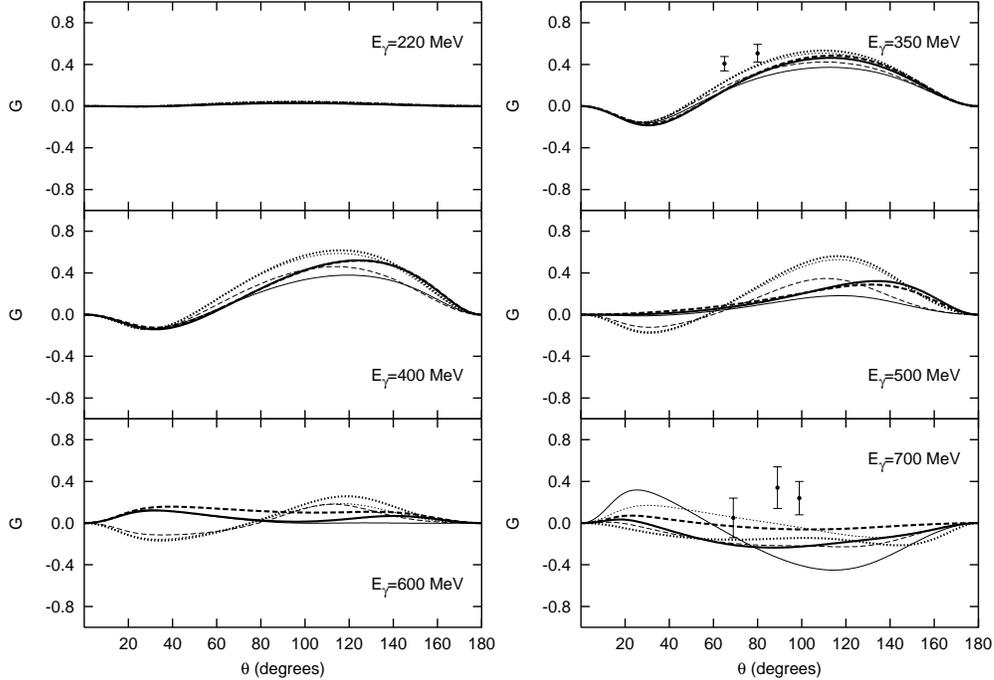}}}
\end{center}
\caption{$G$ asymmetry of the 
$\gamma p \to \pi^+ n$ reaction.
Experimental data are 
within the range $E_\gamma \pm 3$ MeV.
Same conventions as in Fig. \ref{fig:Ppi0p}.} 
\label{fig:Gpipn}
\end{figure}

\begin{figure}
\begin{center}
\rotatebox{-90}{\scalebox{0.52}[0.52]
{\includegraphics{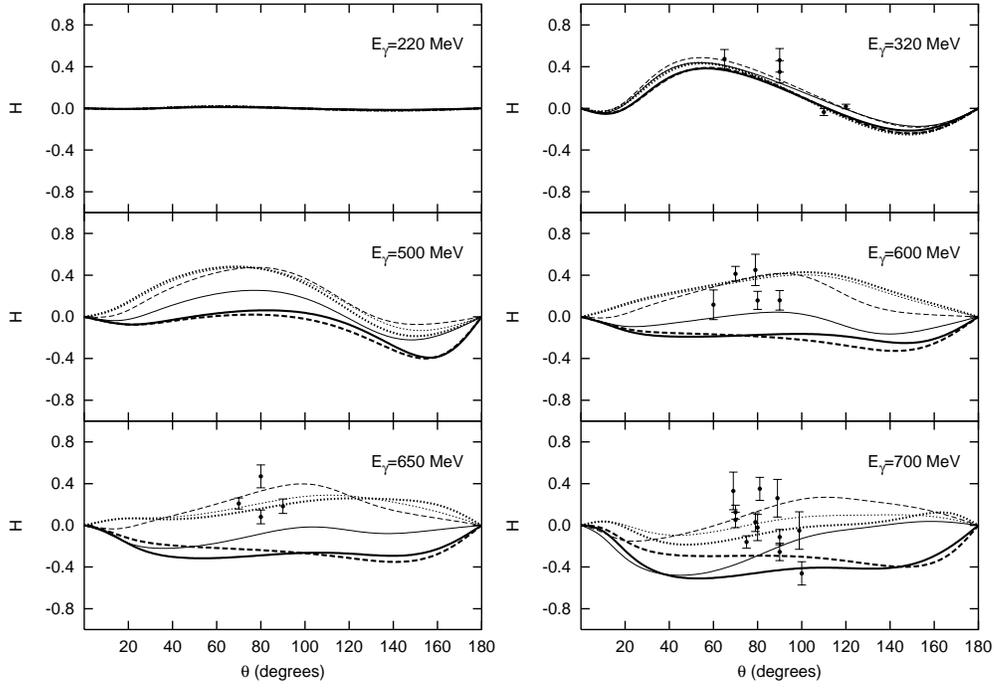}}}
\end{center}
\caption{$H$ asymmetry of the 
$\gamma p \to \pi^+ n$ reaction.
Experimental data are 
within the range $E_\gamma \pm 3$ MeV.
Same conventions as in Fig. \ref{fig:Ppi0p}.} 
\label{fig:Hpipn}
\end{figure}

\begin{figure}
\begin{center}
\rotatebox{-90}{\scalebox{0.52}[0.52]
{\includegraphics{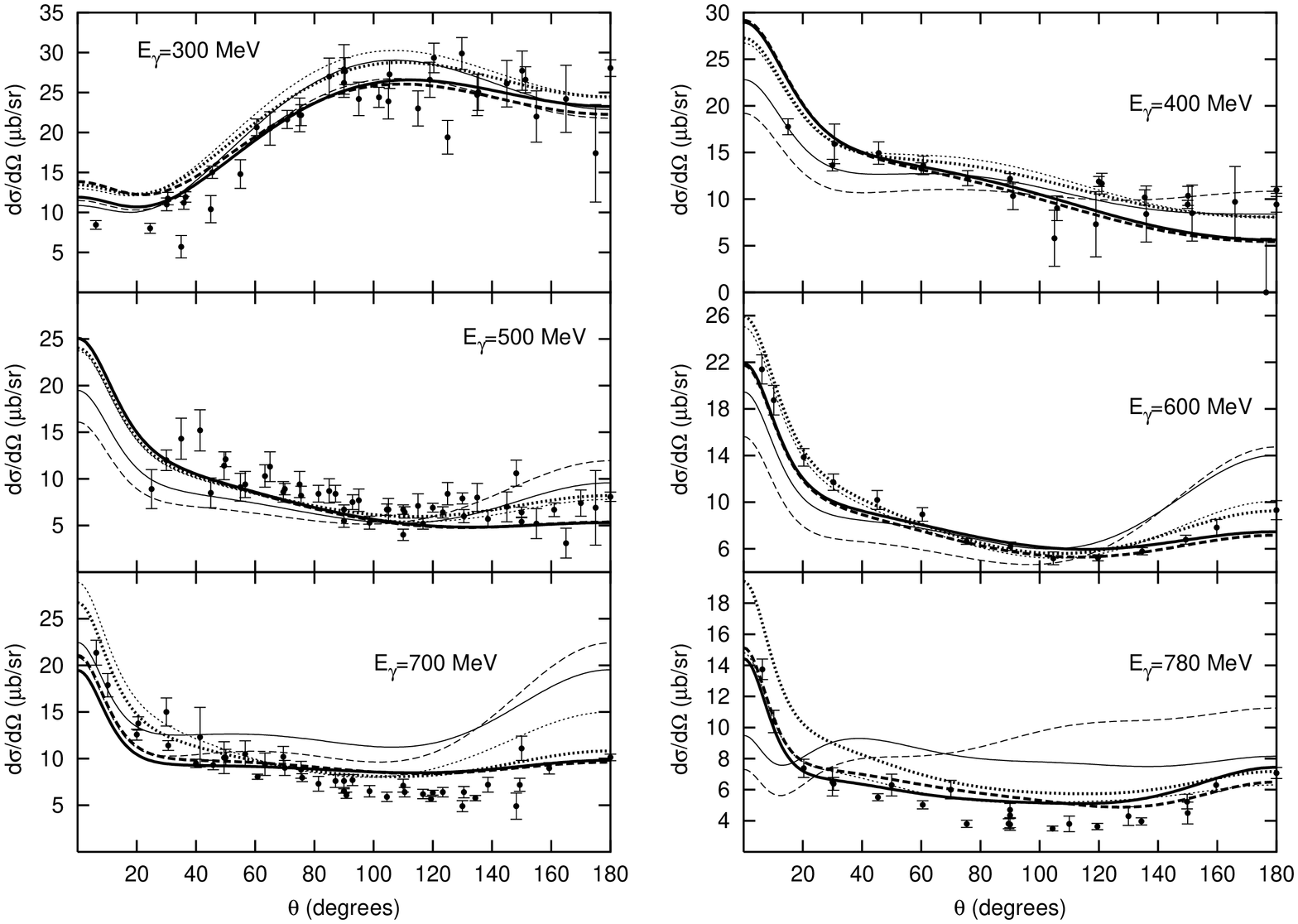}}}
\end{center}
\caption{Differential cross section of the 
$\gamma p \to \pi^- p$ reaction.
Experimental data are 
within the range $E_\gamma \pm 4$ MeV.
Same conventions as in Fig. \ref{fig:diffpi0p}.} 
\label{fig:diffpimp}
\end{figure}

\begin{figure}
\begin{center}
\rotatebox{-90}{\scalebox{0.52}[0.52]
{\includegraphics{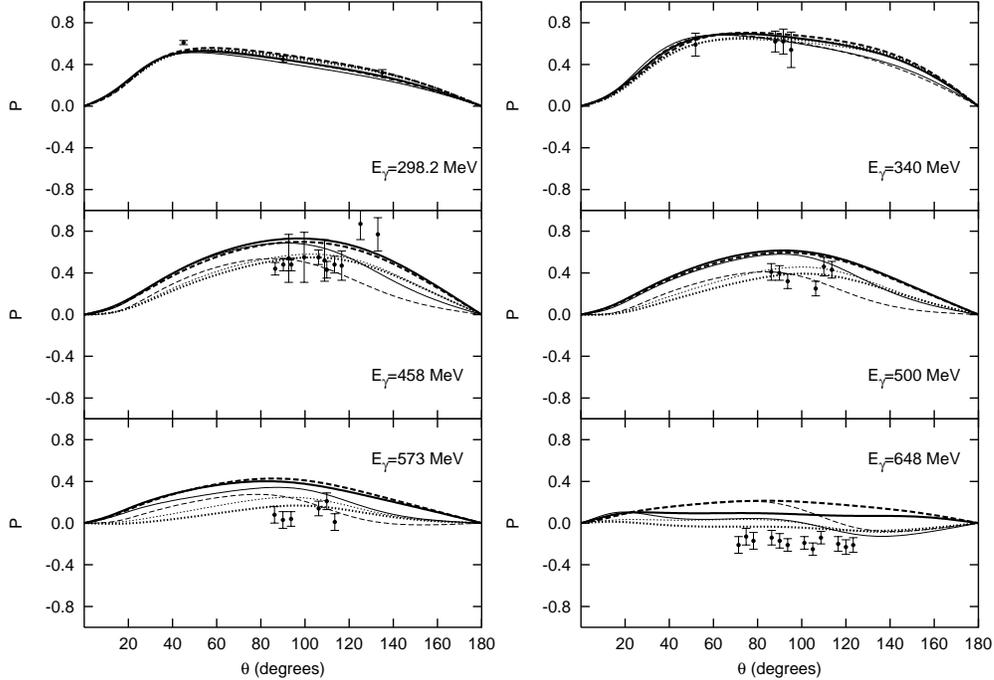}}}
\end{center}
\caption{Recoil nucleon polarization of the 
$\gamma p \to \pi^- p$ reaction.
Experimental data are 
within the range $E_\gamma \pm 1$ MeV.
Same conventions as in Fig. \ref{fig:Ppi0p}.} 
\label{fig:Ppimp}
\end{figure}

\begin{figure}
\begin{center}
\rotatebox{-90}{\scalebox{0.52}[0.52]
{\includegraphics{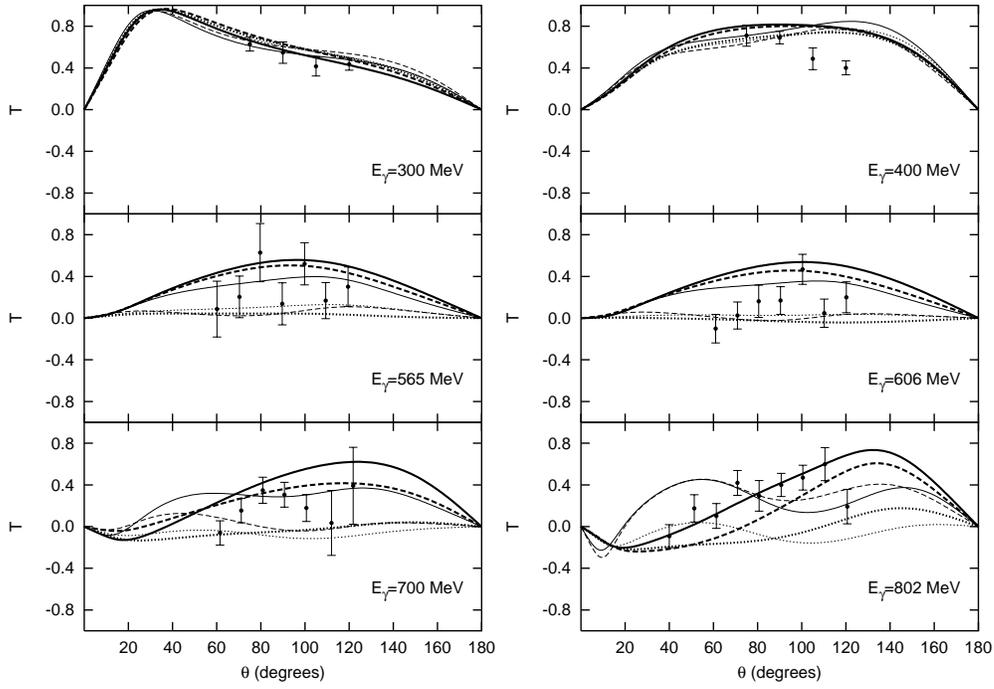}}}
\end{center}
\caption{Polarized target asymmetry of the 
$\gamma p \to \pi^- p$ reaction.
Experimental data are 
within the range $E_\gamma \pm 5$ MeV.
Same conventions as in Fig. \ref{fig:Ppi0p}.} 
\label{fig:Tpimp}
\end{figure}

\begin{figure}
\begin{center}
\rotatebox{-90}{\scalebox{0.52}[0.52]
{\includegraphics{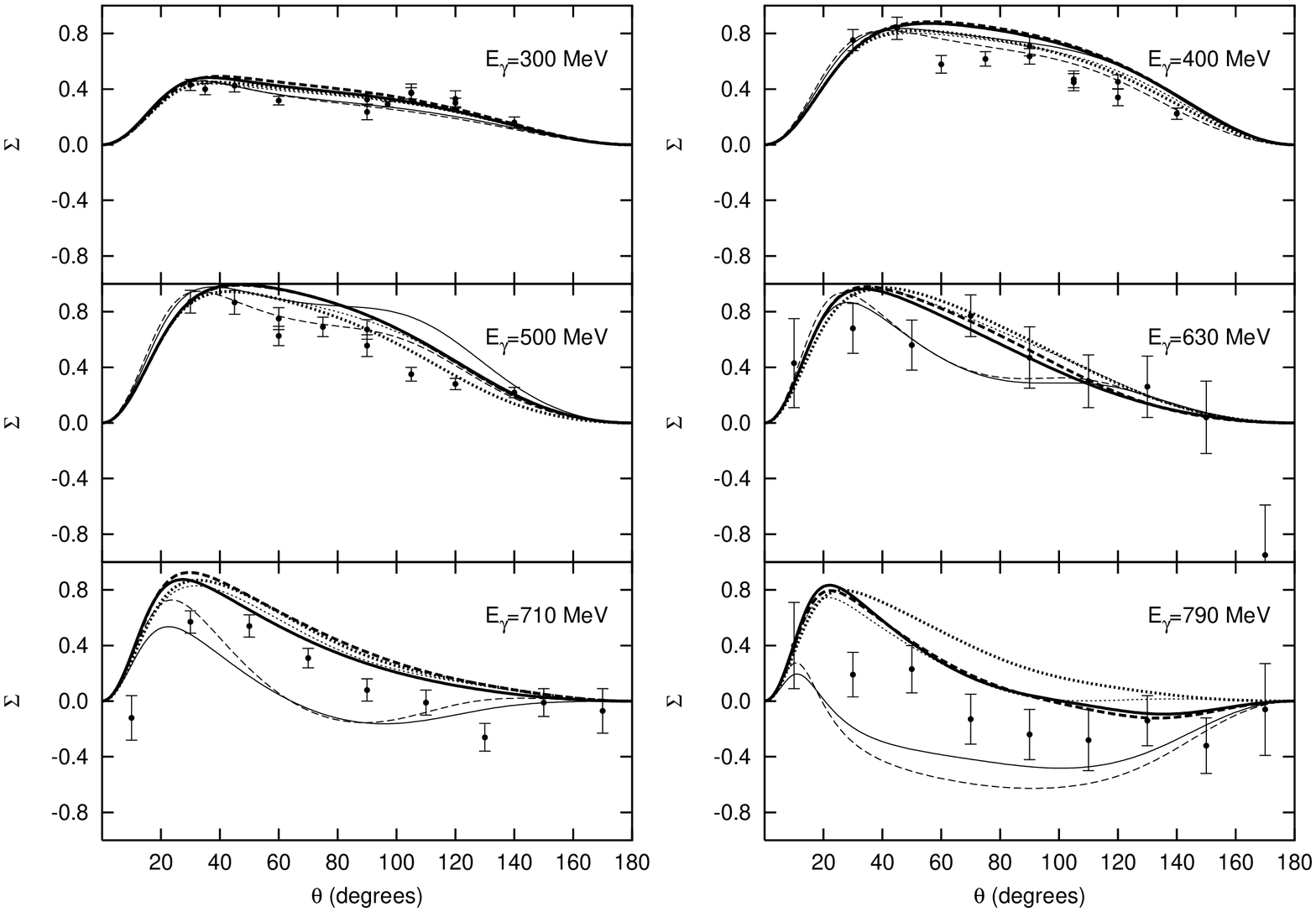}}}
\end{center}
\caption{Photon beam asymmetry of the 
$\gamma p \to \pi^- p$ reaction.
Experimental data are 
within the range $E_\gamma \pm 1$ MeV.
Same conventions as in Fig. \ref{fig:Ppi0p}.} 
\label{fig:Spimp}
\end{figure}

\begin{figure}
\begin{center}
\rotatebox{-90}{\scalebox{0.52}[0.52]
{\includegraphics{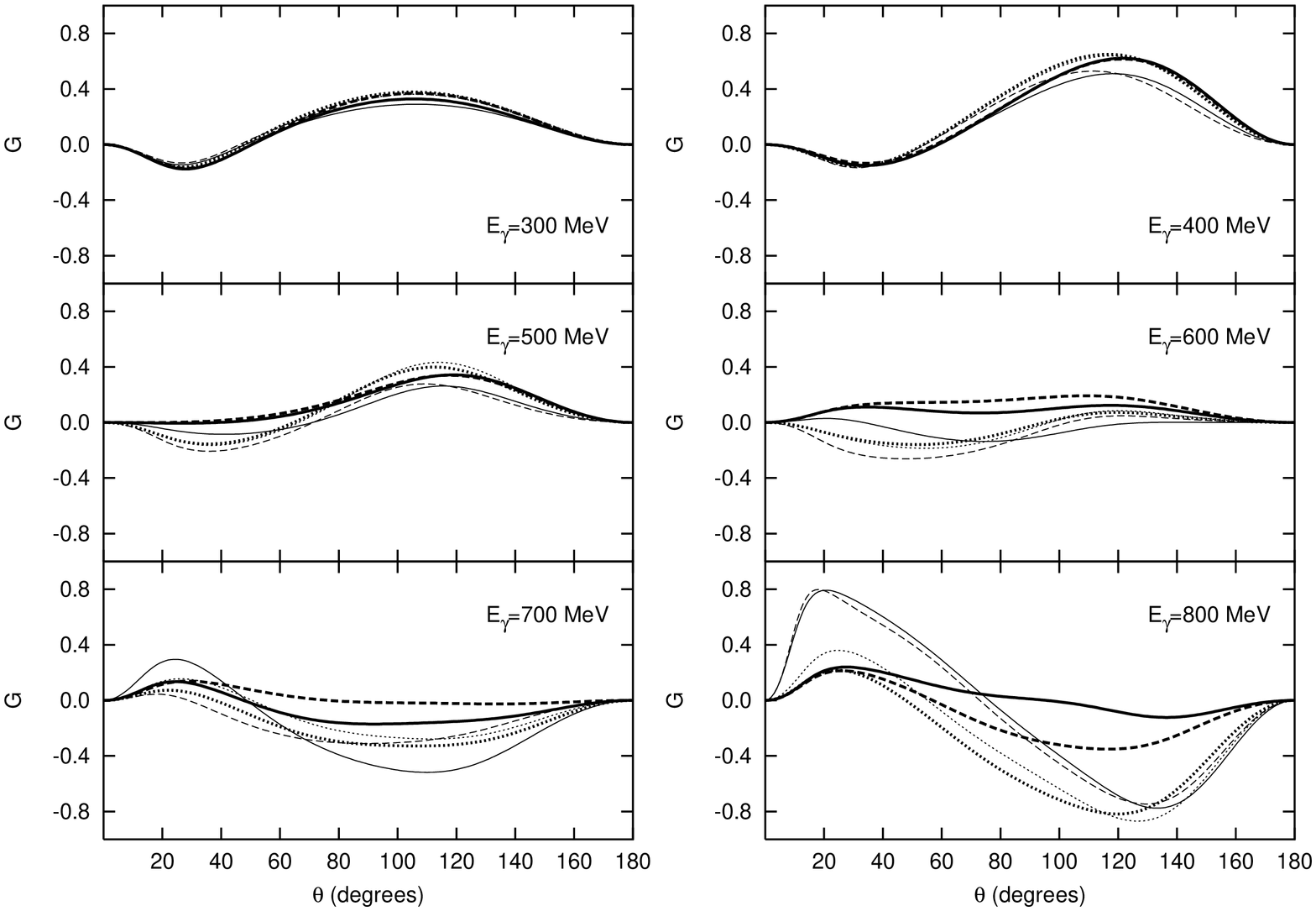}}}
\end{center}
\caption{$G$ asymmetry of the 
$\gamma p \to \pi^- p$ reaction.
Same conventions as in Fig. \ref{fig:Ppi0p}.} 
\label{fig:Gpimp}
\end{figure}

\begin{figure}
\begin{center}
\rotatebox{-90}{\scalebox{0.55}[0.52]
{\includegraphics{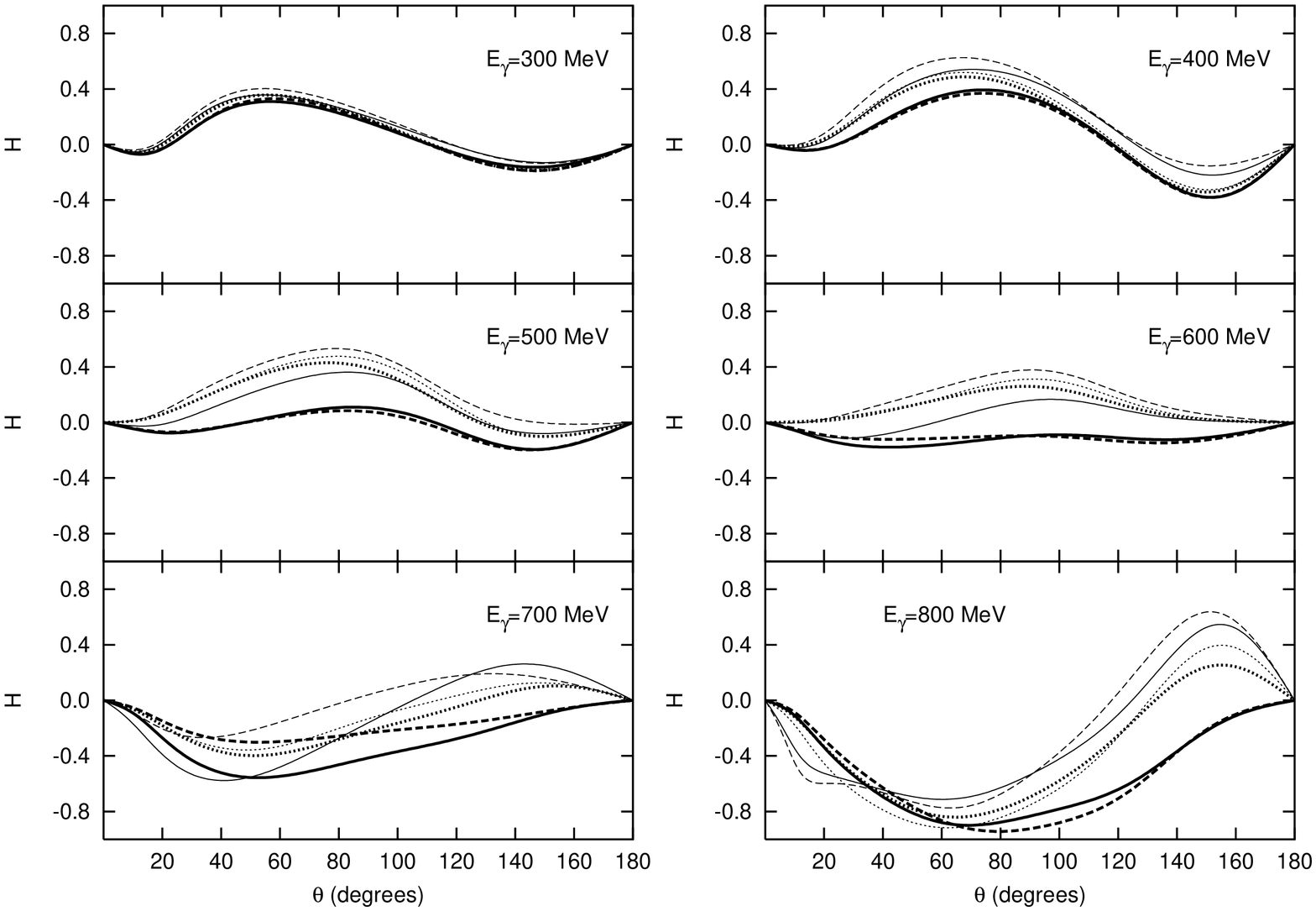}}}
\end{center}
\caption{$H$ asymmetry of the 
$\gamma p \to \pi^- p$ reaction.
Same conventions as in Fig. \ref{fig:Ppi0p}.} 
\label{fig:Hpimp}
\end{figure}

In the next paragraphs we go in detail through the predicted
differential 
cross sections and asymmetries
for charged pion processes, and compare them to available data
(Figs. \ref{fig:diffpipn}--\ref{fig:Hpimp}).

$\gamma p \to \pi^+ n$ differential cross sections 
(Fig. \ref{fig:diffpipn}) are well predicted by the model
in the whole energy range by all parameter sets. 
In the high energy regime (two last figures of the panels)
differential cross sections are not well predicted
by any of the parameters sets in the forward scattering region,
with the exception of set \#1 (PDG with $\delta_{FSI}$) 
which provides an impressively good agreement.
For the $P$ asymmetry (Fig. \ref{fig:Ppipn})
all curves are alike  and reproduce
data correctly up to $400$ MeV. 
As the energy is increased, sets \#1 and \#4 (PDG values)
provide the best results. The
$T$ asymmetry is qualitatively well predicted, 
but quantitative agreement is only achieved
up to $500$ MeV (Fig. \ref{fig:Tpipn}).
Sets with and without
FSI provide a good agreement with data for the $\Sigma$ 
asymmetry (Fig. \ref{fig:Spipn}). 
Only in the last figure of the panel ($700$ MeV)
we observe different qualitative behaviors 
from one set of constants to another.
Data are scant and not reliable for $G$ and $H$ asymmetries. 
As in previous asymmetries, in the low energy regime 
all the curves are alike, but as energy is increased 
their predictions become quite different.

Differential cross section data  for the
reaction $\gamma p \to \pi^- p$
are well predicted by the sets with FSI (\#1, \#2, and \#3).
All the curves are similar for the
$P$ asymmetry (Fig. \ref{fig:Ppimp}) and are close to data.
Overall agreement is good for the $T$ asymmetry (Fig. \ref{fig:Tpimp}). 
This agreement becomes excellent 
for the highest energy ($E_\gamma = 802$ MeV)
if we consider only curves \#2 and \#3.
$\Sigma$ asymmetry (Fig. \ref{fig:Spimp})
is very well predicted by curves \#2 and \#3 in the whole energy range.
All  predictions are qualitatively quite similar for the 
$G$ and $H$ asymmetries (Figs. \ref{fig:Gpimp} and \ref{fig:Hpimp}) 
except for $E_\gamma = 800$ MeV, where large differences are found.

The model works quite well for processes with 
charged pion. So, this is remarkable if we take 
into account that no $\delta_{FSI}$ have been included,
and indicates that FSI are not as important in the studied
energy region, for charged pions as they are 
for neutral pion channels.
Quantitatively, the model provides 
satisfactory results nearly in the whole energy range and in 
almost every observable.
Even in the cases where good quantitative result is not achieved,
at least the qualitative 
behavior of data
is well reproduced (i.e. Figs. 
\ref{fig:diffpipn} and \ref{fig:Tpipn}).

\subsection{Cross sections}\label{sec:crosssections}
\begin{figure}
\begin{center}
\rotatebox{-90}{\scalebox{0.55}[0.5]{
\includegraphics{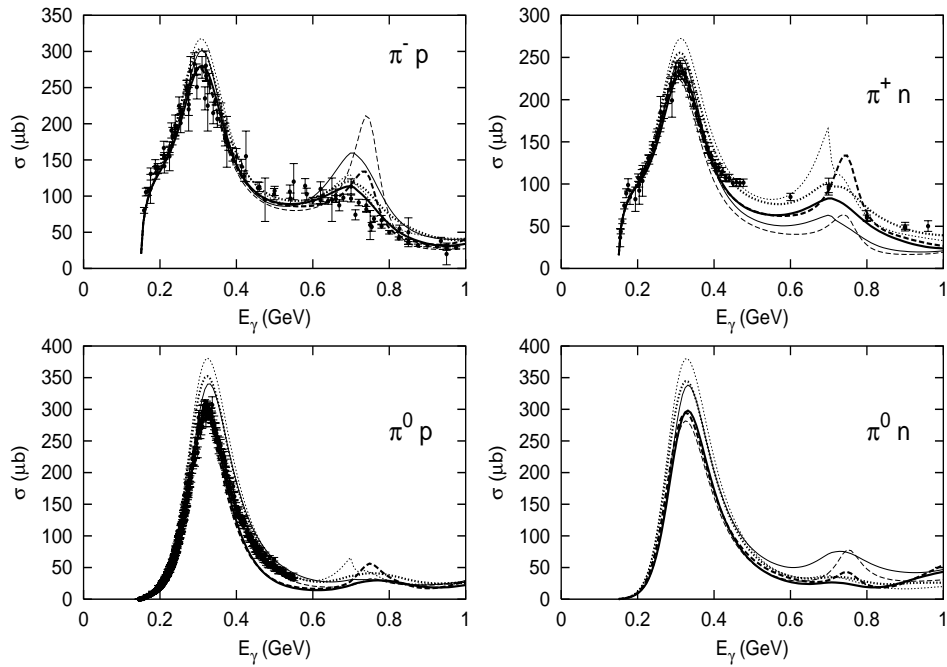}}}
\end{center}
\caption{Total cross section as a function of 
photon energy in laboratory frame.
Same curve conventions as in Fig. \ref{fig:diffpi0p}.} 
\label{fig:xsec}
\end{figure}

Finally, in Fig. \ref{fig:xsec} we present results for 
the total cross sections compared to available experimental data.
The two upper figures show the total 
cross section for charged pion channels and 
the two lower are for neutral pion photoproduction.

It is interesting to notice how some of the 
observed effects 
in the multipoles show up
in the cross sections.
For example, 
sets \#1, \#4, and \#5 overestimate 
the first resonance region
due to the overestimation of 
$M_{1+}^{3/2}$ peak. On the other hand,
set \#4 presents a cusp peak in 
multipole Im$E_{0+}^p$, that also
shows up in the cross section, 
specially so in the $\pi^+n$ channel.
The high energy behavior is well regularized. 
Nevertheless, it has to be considered 
that we do not take into account 
resonances D$_{15}$ and F$_{15}$ 
which may change the shape of the cross section in the 
second resonance region. 

The low enery behavior of the
charged processes is quite 
well reproduced by all the sets of parameters. 
Actually, 
curves obtained with coupling constants from 
sets \#1 and \#2 agree quite well with data 
in almost the whole energy range. 
Other sets do not provide good results: 
sets \#5 and \#6 overestimate greatly the second 
resonance region for $\pi^- p$ channel, and
set \#4 does the same in $\pi^+ n$ channel.
Overestimation of the second resonance region 
by set \#3 is due to the overestimation of 
multipoles related to resonance N(1520).

Concerning the $\gamma n \to \pi^0 n$ channel
we found several differences among sets either
in the region of the first or in the region 
of the second resonance.
As 
no data are available for $\pi^0 n$ total cross sections, 
we rely on results on differential cross section to infere
that: Up to 400 MeV, sets \#2 and \#3 may provide a good estimation
of the total cross section; and that beyond that energy,
there may be probably an underestimation of the total cross section.

In summary, we conclude that set \#2 is the most reliable one
because it provides the best results when all data are considered
globally. We weigh that this is so, regardless of the fact that
other sets may provide better fits to individual cases.
For instance, set \#6 provides the best fit to $\pi^0p$
total cross section and set \#1 is very good for charged pion channels.
As a matter of fact, set \#2 has the lowest $\chi^2$
for the electromagnetic multipoles.
With this set the only deviations from 
experimental data in the total cross section are the 
slight underestimation
of $\pi^+ n$ and $\pi^0 p$ processes beyond 400 MeV.

\section{Summary and final remarks} \label{sec:conclusions}
We have elaborated on a pion photoproduction model which is
based on an 
Effective Lagrangian Approach (ELA) and is 
guided by Weinberg's theorem,
fulfilling chiral symmetry, gauge invariance, and crossing symmetry.
We have included Born terms,
$\rho$ and $\omega$ mesons exchange, 
and seven nucleon resonances: 
$\Delta(1232)$, N(1440), N(1520), N(1535), 
$\Delta(1620)$, N(1650), and $\Delta(1700)$.
Under these premises, the model is independent of the underlying
subnuclear physics (quarks, gluons), which is embedded in
the parameters of the model, such as coupling constants,
masses, and widths.

With respect to former models along similar lines, this is the
first one that covers all the well established spin-1/2 and spin-3/2
resonances up to 1.7 GeV and, at the same time,
fulfills gauge invariance as well as
chiral and crossing symmetries. Crossing symmetry could not be achieved 
in previous models such as the one of Ref. \cite{EMoya} due, among other
things, to pathologies of former spin-3/2 Lagrangians.
This problem is fixed in the present work by:
\begin{enumerate}
\item [(a)] the use of a spin-3/2 Lagrangian due to Pascalutsa that 
contains no spurious spin-1/2 components in the direct channel, 
\item [(b)] the use of consistent, energy dependent, strong couplings and
widths, as well as form factors.
\end{enumerate}
One of the goals of this paper is to establish a reliable set of 
parameters for the model. In addition to the cutoff $\Lambda$ 
-- which is related to short-distance effects and can be 
considered as the only free parameter of the model --
we adjust electromagnetic coupling constants of the nucleon resonances
within the usually accepted ranges. The determination of the parameters
has been performed by fit to the experimental
$\gamma p \to \pi^0 p$ multipoles, through a minimization procedure.
In the minimization we have considered three different sets of masses
and widths:
\begin{enumerate}
\item[(a)] Masses and widths taken from PDG with electromagnetic
coupling constants within the PDG error bars.
\item[(b)] Masses and widths taken from the multichannel analysis
of Vrana et al. \cite{Vrana} with electromagnetic coupling constants
considered as free parameters.
\item[(c)] Masses and widths obtained by means of a speed plot
calculation  with the
electromagnetic coupling constants considered as free parameters.
\end{enumerate}
On the other hand we have considered final state interactions (FSI)
phenomenologically by adding an extra phase to the $\gamma p \to \pi^0 p$
multipoles, in order to match the current energy dependent 
solution of SAID phases \cite{SAIDdata}.
In all, we have derived six sets of parameters, one with 
and one without FSI
for each of the above mentioned sets of masses and widths:
\begin{enumerate}
\item[(a)] Sets \#1 and \#4. 
\item[(b)] Sets \#2 and \#5.
\item[(c)] Sets \#3 and \#6.
\end{enumerate}

Electromagnetic multipoles for $\gamma p \to \pi^0 p$ are globally well
reproduced by sets \#1, \#2, and \#3 that include FSI. 
The fits without FSI (\#4, \#5, and \#6) are also good in the low energy
regime.
Other experimental observables are surveyed such as differential cross
section, asymmetries, and total cross sections. At threshold we find
good agreement with experimental data. In our model almost all the
contribution at threshold comes from Born terms 
at variance with results in Ref. \cite{EMoya}.

For charged pion production, 
where we have no adjustable parameters, the agreement
is remarkably good for almost all the observables.
We note that FSI phases obtained for the $\gamma p \to \pi^0 p$ process
are not applicable to charged pion production. Thus, no FSI phases
have been included in these charged pion photoproduction calculations.
The fact that we get good agreement with data indicates that FSI are small
in $\gamma p \to \pi^+ n$ and $\gamma n \to \pi^- p$.

Although all the parameter sets are reasonable,
we favor set \#2 because of
its lowest $\chi^2$ to the multipole data 
and its better agreement with the
total cross sections for all processes. Set \#3, which also has 
a low $\chi^2$, is very similar to set \#2 and also 
yields similar helicity
amplitudes for all the resonances except for $\Delta (1700)$. This
resonance is poorly known and more precise information would be necessary.
A better experimental knowledge of multipole $M_{2-}^{3/2}$ would
improve the determination of the properties of 
$\Delta (1700)$ resonance.
Similarly better knowledge of the $M_{1-}^p$ multipole would help
to establish more reliably properties of N(1440).

For the future, it would be interesting to analyze contributions
from spin-5/2 nucleon resonances. 
Although they are not essential to the multipoles considered here,
they may contribute to the background and their effect can be sizeable
in the second resonance region of the total cross section and asymmetries.
The incorporation of the
spin-5/2 resonances will require to take into account 
higher order multipoles in the analysis.
The inclusion of
other resonances not considered here (three stars in PDG 
and \textit{misssing} resonances) could also improve the fits in 
some energy regions, but it is difficult to perform a reliable 
determination of the parameters without the aid of other 
physical processes where their contribution would be more sizeable.

The results obtained here are encouraging and estimulate 
the application of this model to other processes such as pion
electroproduction, two pion production in nucleons and nuclei,
as well as electro- and photo- production of other mesons.

\begin{ack}
One of us (C.F.-R.) thanks Prof. H. Garcilazo and
Prof. R.A. Arndt for useful comments, and
Dr. V. Pascalutsa
and Prof. W. Weise for useful discussions. 
C.F.-R. work is 
being developed under Spanish Government grant UAC2002-0009. 
This work
has been supported in part under contracts of 
Ministerio de Educaci\'on y Ciencia (Spain)
BFM2002-03562 and BFM2003-04147-C02-01. 
\end{ack}

\appendix
\section{Invariant amplitudes} \label{sec:invariantamplitudes}
In this appendix we show all the invariant amplitudes needed 
for the calculations in the isospin decomposition and the 
notation for kinematics of section \ref{sec:kinematics}. 
We name $v$ to the four momentum of the exchanged particle 
in each diagram.
\subsection{Born amplitudes}
\begin{itemize}
\item s-channel (diagram ($a$) in Fig. \ref{fig-diag1})
\begin{eqnarray}
A^0_s&=&-\frac{ef_{\pi N}}{2m_\pi}\bar{u}(p')\feyns{k}
\gamma_5\frac{\feyns{v}+M}{s-M^2}\left[\feyns{A}F_1^S-
\frac{F_2^S}{2M}A^\alpha \gamma_{\alpha \beta}q^\beta 
\right]u(p) \\
A^-_s&=&A^+_s=A^0_s\left(F_1^S\rightarrow F_1^V,F_2^S
\rightarrow F_2^V \right)
\end{eqnarray}
\item u-channel (diagram ($b$) in Fig. \ref{fig-diag1})
\begin{eqnarray}
A_u^0&=&-\frac{ef_{\pi N}}{2m_\pi}\bar{u}(p')\left[
\feyns{A}F_1^S-\frac{F_2^S}{2M}A^\alpha \gamma_{\alpha \beta}
q^\beta  \right]\frac{\feyns{v}+M}{u-M^2}\feyns{k}\gamma_5 
u(p) \\
A^+_u&=&-A^-_u=A^0_u\left(F_1^S\rightarrow F_1^V,F_2^S
\rightarrow F_2^V \right)
\end{eqnarray}
\item t-channel (diagram ($c$) in Fig. \ref{fig-diag1})
\begin{equation}
A^-_t=-eF_1^V\frac{f_{\pi N}}{m_\pi}\bar{u}(p')
\frac{A\cdot \left(v+k \right)}{t-m_\pi^2}\feyns{v}\gamma_5u(p) 
\end{equation}
\item Kroll-Rudermann (contact) term (diagram ($d$) in Fig. 
\ref{fig-diag1})
\begin{equation}
A_{KR}^-=eF_1^V\frac{f_{\pi N}}{m_\pi}\bar{u}(p')\feyns{A}
\gamma_5 u(p)
\end{equation}
\end{itemize}

\subsection{Vector meson amplitudes}
\begin{itemize}
\item $\rho$ meson (diagram ($f$) in Fig. \ref{fig-diag2})
\begin{equation}
A_{\rho}^0=-ie\frac{G_{\rho \pi \gamma} F_{\rho NN}}{m_\pi} 
\bar{u}(p')\frac{\epsilon_{\sigma \lambda \nu \mu}q^\sigma k^\nu 
A^\lambda g^{\alpha \mu}}{t-m^2_\rho}\left[\gamma_\alpha
+\frac{K_\rho}{2M}\gamma_{\alpha \beta}v^\beta \right]u(p) 
\end{equation}
\item $\omega$ meson (diagram ($f$) in Fig. \ref{fig-diag2})
\begin{equation}
A_\omega^+=-ie \frac{G_{\omega \pi \gamma} F_{\omega NN}}{m_\pi} 
\bar{u}(p')\frac{ \epsilon_{\sigma \lambda \nu \mu}
q^\sigma k^\nu A^\lambda g^{\alpha \mu}}{t-m^2_\omega}\left[ 
\gamma_\alpha +\frac{K_\omega}{2M}\gamma_{\alpha \beta}v^\beta 
\right]u(p)
\end{equation}
\end{itemize}

\subsection{S$_{11}$ resonance amplitudes}
\begin{itemize}
\item s-channel
\begin{eqnarray}
A^0_{s,\text{S}_{11}} &=&\frac{eg_Sh}{2M f_\pi}\bar{u}(p')
\feyns{k}G(v)A^\mu\gamma_{\mu \nu}q^\nu \gamma_5 u(p) \\
A^+_{s,\text{S}_{11}}& =& 
A^-_{s,\text{S}_{11}} = A^0_{s,\text{S}_{11}}(g_S
\rightarrow g_V)
\end{eqnarray}
\item u-channel
\begin{eqnarray}
A^0_{u,\text{S}_{11}}&=&-\frac{eg_Sh}{2M f_\pi}\bar{u}(p')
A^\mu\gamma_{\mu \nu}q^\nu \gamma_5 G(v)\feyns{k} u(p) \\
A^+_{u,\text{S}_{11}}&=&-A^-_{u,\text{S}_{11}}
=A^0_{u,\text{S}_{11}}(g_S \rightarrow g_V)
\end{eqnarray}
\end{itemize}
\subsection{S$_{31}$ resonance amplitudes}
\begin{itemize}
\item s-channel
\begin{equation}
A^+_{s,\text{S}_{31}}=-2A^-_{s,\text{S}_{31}}
=\frac{2}{3}\frac{egh}{M f_\pi}
\bar{u}(p')\feyns{k}G(v)A^\mu \gamma_{\mu \nu}q^\nu \gamma_5 u(p) 
\end{equation}
\item u-channel
\begin{equation}
A^+_{u,\text{S}_{31}}=2A^-_{u,\text{S}_{31}}=
-\frac{2}{3}\frac{egh}{M f_\pi}
\bar{u}(p')A^\mu \gamma_{\mu \nu}q^\nu \gamma_5 G(v)\feyns{k}u(p)
\end{equation}
\end{itemize}
\subsection{P$_{11}$ resonance amplitudes}
\begin{itemize}
\item s-channel
\begin{eqnarray}
A^0_{s,\text{P}_{11}}&=& \frac{eg_Sh}{2M f_\pi}\bar{u}(p')
\feyns{k}\gamma_5 G(v)A^\mu \gamma_{\mu \nu}q^\nu u(p) 
\\
A^+_{s,\text{P}_{11}}&=&A^-_{s,\text{P}_{11}}
=A^0_{s,\text{P}_{11}}(g_S \rightarrow g_V)
\end{eqnarray}
\item u-channel
\begin{eqnarray}
A^0_{u,\text{P}_{11}}&=&\frac{eg_Sh}{2M f_\pi}\bar{u}(p')A^\mu 
\gamma_{\mu \nu}q^\nu G(v)\feyns{k}\gamma_5u(p) \\
A^+_{u,\text{P}_{11}}&=&-A^-_{u,\text{P}_{11}}
=A^0_{u,\text{P}_{11}}(g_S \rightarrow g_V)
\end{eqnarray}
\end{itemize}
\subsection{P$_{33}$ resonance amplitudes}
\begin{itemize}
\item s-channel
\begin{eqnarray}
A^+_{s,\text{P}_{33}}&=&-2A^-_{s,\text{P}_{33}}
=\frac{-ihe}{f_\pi M^* M
\left(M^* +M \right)} \nonumber \\
& \times & \bar{u}\left( p'\right)
\epsilon_{\mu \nu \lambda \beta} v^\mu k^\lambda 
\gamma^\beta \gamma^5 G^{\nu \alpha}\left( v \right) 
\\
&\times& \left[  ig_1 \epsilon_{\omega \alpha \rho \phi}
v^\omega q^\rho A^\phi +g_2 \gamma^5 
\left(v\cdot q A_\alpha - v \cdot A q_\alpha  \right) \right] 
u\left( p \right) \nonumber
\end{eqnarray}

\item u-channel
\begin{eqnarray}
A^+_{u,\text{P}_{33}}&=&2A^-_{u,\text{P}_{33}}
=\frac{ihe}{f_\pi M^* M 
\left(M^* +M \right)} \nonumber \\
&\times& \bar{u}\left( p'\right) \left[ ig_1 
\epsilon_{\mu \nu \alpha \beta}v^\mu q^\alpha A^\beta 
+g_2\gamma^5\left(v\cdot q A_\nu - v \cdot A q_\nu \right)
\right] \\
&\times& G^{\nu \phi} \left( v \right) 
\epsilon_{\omega \phi \lambda \rho}v^\omega k^\lambda 
\gamma^\rho \gamma^5 u \left( p \right) \nonumber
\end{eqnarray}
\end{itemize}

\subsection{D$_{33}$ resonance amplitudes}
\begin{itemize}
\item s-channel
\begin{eqnarray}
A^+_{s,\text{D}_{33}}&=&-2A^-_{s,\text{D}_{33}}= 
\frac{ihe}{f_\pi M M^*\left( M+M^*\right)} \nonumber \\
&\times& \bar{u} \left( p' \right) 
\epsilon_{\mu \nu \lambda \beta}
v^\mu k^\lambda \gamma^\beta 
G^{\lambda \alpha}\left( v \right) 
\\
&\times&\left[ ig_1 \epsilon_{\omega \alpha \rho \phi} 
v^\omega q^\rho A^\phi \gamma^5 + g_2 
\left( v \cdot q A_\alpha - v \cdot A q_\alpha \right) 
\right] u\left( p \right) \nonumber
\end{eqnarray}
\item u-channel
\begin{eqnarray}
A^+_{u,\text{D}_{33}}&=&2A^-_{u,\text{D}_{33}}
=\frac{ieh}{f_\pi MM^*
\left(M+M^*\right)} \nonumber \\
&\times& \bar{u}\left( p' \right) \left[ ig_1 
\epsilon_{\mu \nu \alpha \beta}q^\alpha v^\mu A^\beta 
\gamma^5 +g_2 \left( v \cdot q A_\nu - v \cdot A q_\nu 
\right) \right] \\
&\times&G^{\nu \lambda}\left( v \right) 
\epsilon_{\omega \lambda \rho \phi} v^\omega k^\rho 
\gamma^\phi u \left( p \right) \nonumber 
\end{eqnarray}
\end{itemize}
\subsection{D$_{13}$ resonance amplitudes}
\begin{itemize}
\item s-channel
\begin{eqnarray}
A^0_{s,\text{D}_{13}}&=& \frac{3ihe}{4f_\pi M M^*
\left( M+M^*\right)}  \nonumber \\
&\times& \bar{u} \left( p' \right) 
\epsilon_{\mu \nu \lambda \beta}v^\mu k^\lambda 
\gamma^\beta G^{\lambda \alpha}\left( v \right) 
\\
&\times& \left[ ig_1^S \epsilon_{\omega \alpha \rho \phi} 
v^\omega q^\rho A^\phi \gamma^5 + g_2^S 
\left( v \cdot q A_\alpha - v \cdot A q_\alpha \right) 
\right] u\left( p \right) \nonumber \\
A^+_{s,\text{D}_{13}}&=&A^-_{s,\text{D}_{13}}
=A^0_{s,\text{D}_{13}} 
\left( g^S_{1,2} \to g^V_{1,2}\right) 
\end{eqnarray}
\item u-channel
\begin{eqnarray}
A^0_{u,\text{D}_{13}}&=&\frac{3ieh}{4f_\pi MM^*
\left(M+M^*\right)} \nonumber \\
&\times& \bar{u}\left( p' \right) \left[ ig_1^S 
\epsilon_{\mu \nu \alpha \beta}q^\alpha v^\mu A^\beta 
\gamma^5 +g_2^S \left( v \cdot q A_\nu - v \cdot A q_\nu \right) 
\right] \\
&\times&G^{\nu \lambda}\left( v \right) 
\epsilon_{\omega \lambda \rho \phi} v^\omega k^\rho 
\gamma^\phi u \left( p \right) \nonumber \\
A^+_{u,\text{D}_{13}}&=&-A^-_{u,\text{D}_{13}}
=A^0_{u,\text{D}_{13}} 
\left( g^S_{1,2} \to g^V_{1,2}\right) 
\end{eqnarray}
\end{itemize}

\section{Decay widths} \label{sec:decaywidths}
The coupling constants at the strong vertices
are related 
to the decay widths of the resonances. Given the 
following kinematical definitions
\begin{eqnarray}
k^*   &=& \frac{1}{2M^*}\left[ \left( M^{*2}-M^2
-m_\pi^2 \right)^2-4 m_\pi^2 M^2 \right]^\frac{1}{2} \: ,\\
E_\pi &=& \sqrt{k^{*2}+m_\pi^2} \: ,\\
E_N   &=& \sqrt{k^{*2}+M^2} \: ,
\end{eqnarray}
the decay widths related to the resonance Lagrangians 
of section \ref{sec:model} are
\begin{eqnarray}
\Gamma_\pi^{\text{S}_{11}} 
&=& 3\frac{k^*h^2}{2\pi M^* f_\pi^2}
\frac{\left[ E_\pi \left( E_N+M\right) +k^{*2}\right]^2}
{2\left( E_N+M \right)} \: ,\\
\Gamma_\pi^{\text{S}_{31}} &=& \frac{k^* h^2}{2\pi M^*f_\pi^2}
\frac{\left[E_\pi \left(E_N+M \right)+k^{*2} \right]^2}
{2\left(E_N+M \right) } \: ,\\
\Gamma_\pi^{\text{P}_{11}} &=& 3\frac{k^{*3}h^2}{2\pi M^* f_\pi^2}
\frac{\left( E_N+M+E_\pi\right)^2}{2\left( E_N+M \right) } \: ,\\
\Gamma_\pi^{\text{P}_{33}} &=& \frac{h^2}{3\pi f_\pi^2}
\frac{k^{*3}}{M^*}\left( E_N+M\right) \: ,\\
\Gamma_\pi^{\text{D}_{33}} &=& \frac{h^2}{3\pi f_\pi^2}
\frac{k^{*5}}{M^*}\frac{1}{E_N+M} \: ,\\
\Gamma_\pi^{\text{D}_{13}} &=& \frac{h^2}{\pi f_\pi^2}
\frac{k^{*5}}{M^*}\frac{1}{E_N+M} \: .
\end{eqnarray}

\section{Electromagnetic Multipoles} \label{sec:multipole}
The starting point for multipole analysis is to define 
the helicity spinors in the c.m. reference 
system for incoming,
\begin{equation}
u(p)^{\lambda=\frac{1}{2}}=\sqrt{\rho}
\begin{bmatrix} 0 \\ 1 \\ 0 \\ \zeta 
\end{bmatrix} \quad ,
\quad u(p)^{\lambda=-\frac{1}{2}}=\sqrt{\rho} 
\begin{bmatrix} 1 \\ 0 \\ -\zeta \\ 0 \end{bmatrix} \: ;
\end{equation} 
and outgoing nucleons,
\begin{eqnarray}
& &\bar{u}(p')^{\lambda=\frac{1}{2}}=\sqrt{\rho '}
\begin{bmatrix}
-\sin \frac{\theta}{2}, & \cos \frac{\theta}{2}, 
& \zeta ' \sin \frac{\theta}{2} ,& - \zeta ' 
\cos \frac{\theta}{2}
\end{bmatrix} \: ,\\
& &\bar{u}(p')^{\lambda=-\frac{1}{2}}=\sqrt{\rho '}
\begin{bmatrix}
\cos \frac{\theta}{2}, & \sin \frac{\theta}{2}, & 
\zeta ' \cos \frac{\theta}{2}, & \zeta ' \sin \frac{\theta}{2}
\end{bmatrix} \: ;
\end{eqnarray}
where $\rho=E^*+M$, $\rho '=E'^*+M$,  
$\zeta =\frac{q^*}{\rho}$, and $\zeta ' =\frac{k^*}{\rho '}$.

All formulae in this and forthcoming 
appendices are in c.m. reference system and with this 
spinors definition.

In order to build up the multipoles it is convenient to 
change the isospin basis from $(A^0,A^+,A^-)$ to 
$(A^{3/2},  \:  _pA^{1/2}, \:  _nA^{1/2})$. 
Both bases are related by means of
\begin{eqnarray}
A^{3/2}    &=&  A^+ - A^- \: ,\\
_p A^{1/2} &=&  \frac{1}{3}A^+ + \frac{2}{3}A^- + A^0 \: ,\\
_n A^{1/2} &=& -\frac{1}{3}A^+ - \frac{2}{3}A^- + A^0 \: .
\end{eqnarray}

Defining $\lambda = \lambda_\gamma - \lambda_1$, 
initial helicity state along the photon, and $\mu = -\lambda_2$, 
final helicity state along the pion, the spin and isospin 
projection of amplitudes can be written as
\begin{equation}
H_{\lambda \mu}^{I, j}(W)=\frac{1}{8W\pi}
\int^1_{-1} d\left( \cos \theta \right) d^j_{\lambda \mu} 
\left( \theta \right) A_{\lambda \mu}^I\left( \theta, W \right) \: ,
\end{equation}
where $W=\sqrt{s^*}$, $j$ is the spin of the resonance, 
and $d^j_{\lambda \mu}\left( \theta \right)$ 
are Wigner d-functions with the conventions of 
\cite{PDG2004}. The lowest order multipole amplitudes 
are \cite{Arndt90-2,Walker,Chew}:
\begin{eqnarray}
E_{0+}^I &=&  \frac{\sqrt{2}}{4} \left[ H_{1/2,1/2}^{I,1/2} 
+ H_{1/2,-1/2}^{I,1/2} \right] \: ,\\
M_{1-}^I &=& - \frac{\sqrt{2}}{4} \left[ H_{1/2,1/2}^{I,1/2} 
- H_{1/2,-1/2}^{I,1/2} \right] \: ,\\
E_{1+}^I &=& \frac{\sqrt{2}}{8} \left[ \left( H_{1/2,1/2}^{I,3/2} 
+ H_{1/2,-1/2}^{I,3/2} \right) -   \frac{1}{\sqrt{3}}
\left( H_{3/2,1/2}^{I,3/2} + H_{3/2,-1/2}^{I,3/2}  \right) 
\right] \: ,\\
M_{1+}^I &=& \frac{\sqrt{2}}{8} \left[ \left( H_{1/2,1/2}^{I,3/2} 
+ H_{1/2,-1/2}^{I,3/2} \right) + \sqrt{3}\left( H_{3/2,1/2}^{I,3/2} 
+ H_{3/2,-1/2}^{I,3/2} \right) \right] \: ,\\
E_{2-}^I &=&   \frac{\sqrt{2}}{8} \left[ \left( H_{1/2,1/2}^{I,3/2} 
- H_{1/2,-1/2}^{I,3/2} \right) + \sqrt{3}\left( H_{3/2,1/2}^{I,3/2} 
- H_{3/2,-1/2}^{I,3/2} \right) \right] \: ,\\
M_{2-}^I &=& - \frac{\sqrt{2}}{8} \left[ \left( H_{1/2,1/2}^{I,3/2} 
- H_{1/2,-1/2}^{I,3/2} \right) -  \frac{1}{\sqrt{3}}
\left( H_{3/2,1/2}^{I,3/2} - H_{3/2,-1/2}^{I,3/2} \right) \right] \: .
\end{eqnarray}

\section{Experimental helicity amplitudes} 
\label{sec:helicityamplitudes}
In this appendix we present the connection between our 
amplitudes and the helicity amplitudes of the resonances 
as they are found in Ref. \cite{PDG2004}  in order to 
relate the coupling constants to the usual partial wave 
analyses. To perform this connection the isospin decomposition
$\left(A^\Delta,A^p,A^n\right)$ is 
needed instead of the one in section \ref{sec:kinematics}. 
Both are related in the following way
\begin{eqnarray}
A^{\Delta} &=&\sqrt{\frac{2}{3}} \left( A^+-A^-\right) \: ,\\
A^p &=& -\frac{1}{\sqrt{3}} \left(A^+ + 2A^- +3 A^0\right) \: ,\\
A^n &=& \frac{1}{\sqrt{3}} \left( A^+ + 2A^- -3A^0\right) \: .
\end{eqnarray}

And the helicity amplitudes are given by \cite{EMoya,Arndt90-3}
\begin{equation}
A_\lambda^Id^j_{\lambda \mu}\left(\theta \right)
=\frac{i}{8\pi \left( 2j+1 \right)} \sqrt{ 
\left( 2j+1 \right) \frac{2\pi}{s^*} \frac{k^*}{q^*} 
\frac{M^*}{M} \frac{\Gamma^2}{\Gamma_\pi} } 
A_{\lambda_1 \lambda_2 \lambda_\gamma}^I \: ,
\end{equation}
where $\lambda$, $\mu$, $j$, and $d^j_{\lambda \mu}
\left( \theta \right)$ have the same meaning as in 
appendix \ref{sec:multipole}; $\Gamma$ is the total 
decay width and $\Gamma_\pi$ the pion-nucleon decay 
width of the resonance as defined in appendix 
\ref{sec:decaywidths}. $k^*$ and $q^*$ are the pion 
and the photon momenta in the c.m. system. We define 
the kinematical coefficients
\begin{eqnarray}
q^*&=& \frac{M^{*2}-M^2}{2M^*} \: , \\
\xi &=&\frac{q^*}{\sqrt{q^{*2}+M^2}+M} \: ,\\
T &=& \frac{1}{4}\frac{M^*}{M \left( M+M^* \right)} 
\frac{q^*}{\sqrt{M \xi}} \: ,
\end{eqnarray}
to obtain finally the following results:
\begin{itemize}
\item Resonance S$_{11}$
\begin{eqnarray}
A^{p,n}_{1/2}\left(\text{S}_{11} \right) 
&=&\frac{1}{\sqrt{2}}
\frac{eg^{p,n}}{M}
\sqrt{\frac{\xi}{M}}\left(M+M^*\right)
\end{eqnarray}

\item Resonance S$_{31}$
\begin{eqnarray}
A^{\Delta}_{1/2}\left(\text{S}_{31}\right)
&=&-\frac{1}{\sqrt{3}}
\frac{eg}{M}
\sqrt{\frac{\xi}{M}}\left( M+M^*\right)
\end{eqnarray}

\item Resonance P$_{11}$
\begin{eqnarray}
A^{p,n}_{1/2}\left(\text{P}_{11} \right)
&=&-\frac{1}{\sqrt{2}}
\frac{eg^{p,n}}{M}
\sqrt{\frac{\xi}{M}}\left( M+M^* \right)
\end{eqnarray}

\item Resonance P$_{33}$
\begin{eqnarray}
A_{1/2}^\Delta \left( \text{P}_{33} \right) 
&=& -eT
\sqrt{\frac{1}{2}}
\left[ g_{1}
-\xi g_{2} \right] \\
A_{3/2}^\Delta \left( \text{P}_{33} \right) 
&=& -eT
\sqrt{\frac{3}{2}}
\left[ g_{ 1}
+\xi g_{2} \right]
\end{eqnarray}

\item Resonance D$_{33}$
\begin{eqnarray}
A_{1/2}^\Delta \left( \text{D}_{33} \right) 
&=& eT
\frac{\sqrt{2}}{2} \left[g_{2}
-\xi g_{1} \right]  \\
A_{3/2}^\Delta \left( \text{D}_{33} \right) &=& eT
\sqrt{\frac{3}{2}}\left[g_{2}
+\xi g_{1} 
\right]
\end{eqnarray}

\item Resonance D$_{13}$
\begin{eqnarray}
A_{1/2}^{p,n} \left( \text{D}_{13} \right) &=& -eT
\frac{3}{2\sqrt{3}}\left[g_{2}^{p,n}
-\xi g_{1}^{p,n} \right]  \\
A_{3/2}^{p,n} \left( \text{D}_{13} \right) &=& -eT
\frac{3}{2}\left[g_{2}^{p,n}
+\xi g_{1}^{p,n} \right]
\end{eqnarray}
\end{itemize}

\end{document}